\documentclass[prx,twocolumn,aps,epsf,showpacs,superscriptaddress,longbibliography,nofootinbib,notitlepage]{revtex4-2}
\usepackage[pdftex]{graphicx}

\usepackage[normalem]{ulem}
\usepackage{xfrac}
\usepackage{dcolumn}
\usepackage{bm}
\usepackage{epsfig}
\usepackage{latexsym} 
\usepackage{amsmath,mathtools}
\usepackage{amssymb}
\usepackage{color}
\usepackage{array}
\usepackage{bbm}
\usepackage{color}
\usepackage{array}
\usepackage{cancel}
\usepackage[dvipsnames]{xcolor}
\usepackage{tikz}
\usepackage[colorlinks=true,allcolors=blue]{hyperref}%

\usepackage{array,multirow}
\newcolumntype{L}[1]{>{\raggedright\let\newline\\\arraybackslash\hspace{0pt}}m{#1}}
\newcolumntype{C}[1]{>{\centering\let\newline\\\arraybackslash\hspace{0pt}}m{#1}}
\newcolumntype{R}[1]{>{\raggedleft\let\newline\\\arraybackslash\hspace{0pt}}m{#1}}

\usepackage{titlesec}
\titleformat{\subsubsection}[runin]
        {\itshape}
        {\thesection.\thesubsection.\thesubsubsection)}
        {4pt}
        {}
        [~---]
\titlespacing*{\subsubsection}{12pt}{10pt}{4pt}

\usepackage[normalem]{ulem} 

\newcommand{\<}{\langle}

\newcommand{\up}{\uparrow}
\newcommand{\down}{\downarrow}

\renewcommand{\>}{\rangle}
\renewcommand{\(}{\left(}
\renewcommand{\)}{\right)}
\renewcommand{\[}{\left[}
\renewcommand{\]}{\right]}

\renewcommand{\d}{\partial}

\newcommand{\Z}{\mathbb{Z}}
\newcommand{\T}{\mathcal{T}}

\newcommand{\N}{\mathcal{N}}
\newcommand{\C}{\mathcal{C}}

\newcommand{\Guv}{G_\text{UV}}
\newcommand{\Gir}{G_\text{IR}}

\newcommand{\ir}{\underset{\text{\tiny IR}}{\approx}}

\makeatletter
\def\l@subsubsection#1#2{}
\makeatother
\renewcommand{\eqref}[1]{Eq.~\ref{#1}}

\newcommand{\red}[1]{\textcolor{red}{#1}}
\newcommand{\blue}[1]{\textcolor{blue}{#1}}

\definecolor{RW}{HTML}{5BC2E7}

\begin{document}
\title{Bulk-boundary correspondence for intrinsically-gapless SPTs from group cohomology}

\author{Rui Wen}
\affiliation{Department of Physics and Astronomy, and Stewart Blusson Quantum Matter Institute, University of British Columbia, Vancouver, BC, Canada V6T 1Z1}

\author{Andrew C. Potter}
\affiliation{Department of Physics and Astronomy, and Stewart Blusson Quantum Matter Institute, University of British Columbia, Vancouver, BC, Canada V6T 1Z1}
\begin{abstract}
Intrinsically gapless symmetry protected topological phases (igSPT) are gapless systems with SPT edge states with properties that could not arise in a gapped system with the same symmetry and dimensionality.
igSPT states arise from gapless systems in which an anomaly in the low-energy (IR) symmetry group emerges from an extended anomaly-free microscopic (UV) symmetry
We construct a general framework for constructing lattice models for igSPT phases with emergent anomalies classified by group cohomology, and establish a direct connection between the emergent anomaly, group-extension, and topological edge states by gauging the extending symmetry.
In many examples, the edge-state protection has a physically transparent mechanism: the extending UV symmetry operations pump lower dimensional SPTs onto the igSPT edge, tuning the edge to a (multi)critical point between different SPTs protected by the IR symmetry.
In two- and three-dimensional systems, an additional possibility is that the emergent anomaly can be satisfied by an anomalous symmetry-enriched topological order, which we call a quotient-symmetry enriched topological order (QSET) that is sharply distinguished from the non-anomalous UV SETs by an edge phase transition. 
We construct exactly solvable lattice models with QSET order.
\end{abstract}
\maketitle
{ \hypersetup{hidelinks} \tableofcontents }
%

\newpage
\section{Introduction}
Conventional topological phases of matter rely crucially on a bulk energy gap to ensure rigidly-quantized topological invariants and to protect topological edge states.
However, sharply-quantized topological properties can also arise in the far less well-understood realm of gapless quantum systems (including stable gapless phases or critical points).
Lieb-Shultz-Matthis (LSM) type constraints in which microscopic (UV) structure imposes strong constraints on the low-energy long-wavelength (IR) physics~\cite{lieb1961two,oshikawa2000commensurability,hastings2004lieb,cheng2016translational}, such as constraining the type of orders and excitations that can emerge. 
Moreover, non-local symmetry implementations, such as particle-hole symmetry of a Landau level, can lead to symmetry-protected topological features in gapless systems such as a quantized Berry phase of a particle-hole symmetric composite Fermi liquid~\cite{son2015composite,wang2016half,metlitski2016particle,potter2017realizing,sodemann2017composite}.
Gapless systems can also host new types of topological edge states such as the Fermi-arc surface states of Weyl semimetals~\cite{armitage2018weyl}, which have properties that would be fundamentally forbidden in a gapped system, and hence can be considered ``intrinsically-gapless" forms of topology.

While band topology of nodal semimetals inherently relies on both translation symmetry and absence of interactions, recently there is a growing litany of interacting systems with symmetry-protected topological (SPT) features~\cite{kestner2011prediction,fidkowski2011majorana,keselman2015gapless,scaffidi2017gapless,verresen2018topology,verresen2019gapless,thorngren2021intrinsically}.
These include examples where topological edge modes of a gapped SPT survive when the bulk gap closes, for example at a critical point between an SPT and a symmetry-broken phase~\cite{kestner2011prediction,fidkowski2011majorana,keselman2015gapless,scaffidi2017gapless,verresen2018topology,verresen2019gapless}, as well as interacting intrinsically gapless SPTs (igSPT's) with topological edge features that could not arise in a gapped system~\cite{thorngren2021intrinsically}. Recent work~\cite{ma2022edge,zhang2022exactly} highlights an intriguing connection between gapless SPTs and unconventional de-confined quantum critical points (DQCP), such as a direct transition between a quantum spin-hall state and a superconductor~\cite{ma2022edge} that has potential relations to graphene multilayers~\cite{lee2014deconfined}, and shows that certain DQCPs exhibit symmetry-protected edge modes that modify their boundary criticality.

Many forms of gapless symmetry protected topology share a common origin story: they stem from symmetry anomalies that emerge, in the renormalization group (RG) sense, at energies well below a characteristic energy scale, $\Delta$. Following standard quantum field theory terminology, we will refer to this low-energy regime as the infra-red (IR), and microscopic scales above $\Delta$ as the ultra-violet (UV).
 For example, the original LSM theorem applies to a spin-1/2 chain, which emerges as the low-energy description from an anomaly-free system of electrons and ions below the energy scale for crystallization of the material and below a Mott insulating gap for the electronic excitation. LSM anomalies can be interpreted as a mixed anomaly between the crystalline translation symmetry and spin-rotation or time-reversal symmetry of the spin-interactions~\cite{cheng2016translational}.
Such anomalies are discrete, and cannot be continuously deformed without closing the gap protecting their emergence. In this sense their topology is still protected by a gap in the system. Yet, they impact the  physics of a system that can ultimately be gapless and fluctuating in the IR, and hence have a mixed rigid yet gapless nature.

In a similar way, the $1+1d$ igSPT phase in the model of~\cite{thorngren2021intrinsically}, exhibits an IR symmetry anomaly that emerges at low-energies from an anomaly free lattice model with an extend UV symmetry.
The same anomaly could also arise at the surface of a higher-dimensional gapped SPT phase, in which case there is no possibility of asking about edge states of the anomalous system (since there is no edge of a boundary).
The igSPT construction realizes this anomaly without the higher-dimensional bulk, enabling one to expose an edge to an anomalous system. 
Strikingly, in the model of \cite{thorngren2021intrinsically}, this edge hosts SPT edge degrees of freedom (DOF) that could not arise in a gapped system with the same symmetry.
It is known~\cite{wang2018symmetric,tachikawa2020gauging} that various anomalous symmetries can emerge from an extended UV symmetry in this fashion, suggesting numerous possible igSPTs with various symmetries and dimensionality~\cite{thorngren2021intrinsically}.

In this paper, we introduce a systematic method for constructing examples of igSPT systems from symmetry-anomalies that are classified by group cohomology~\cite{chen2013symmetry}.
This framework generalizes the structure identified in $1d$ igSPT models introduced in Ref.~\cite{thorngren2021intrinsically,li2022symmetry}, to enable the construction of a large variety of igSPT lattice models in various dimensions. Furthermore, this formalism clarifies the relationship between the topological edge degrees of freedom (DOF), and the emergent IR symmetry anomaly that protects the bulk gaplessness.
Using this approach, we construct igSPT models for bosonic systems in one-, two-, and three- spatial dimensions ($1d$, $2d$, $3d$) with various symmetries.

\begin{figure}[]
\begin{centering}
	\includegraphics[width=1.0\columnwidth]{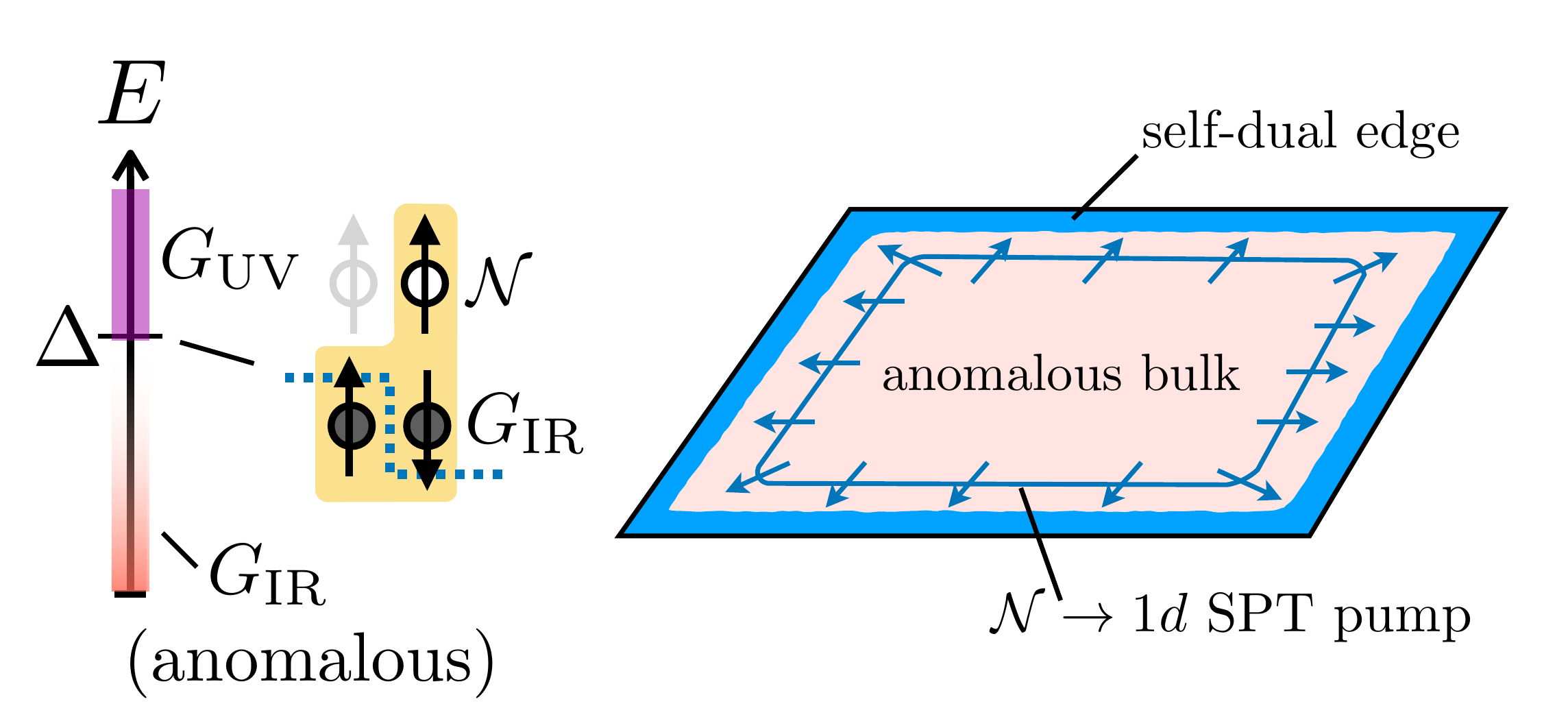}
\end{centering}
	\caption{{\bf Schematic of an igSPT -- } with symmetry group $\Guv$. 
	(Left) $\N$-rotors (spins with open circles) are locked to topological defects in $\Gir$-rotor (spins with filled circle) symmetry-breaking patterns (e.g. dashed blue line depicts $1d$ Ising domain wall) at high-energy (UV) scale $\Delta$. This imprints an anomalous $\Gir = \Guv/\N$ symmetry action at low energies (IR). (Right) Schematic of a $2d$ igSPT. The igSPT bulk $\Gir$ anomaly forbids a trivially gapped symmetric bulk. In the IR $\N$ symmetry operations act trivially in the bulk, and pump a lower-$d$ gapped $\Gir$ SPT onto the boundary. This pumping action forces the edge to a self-dual critical point, leading to symmetry-protected gapless edge states.
	}
	\label{fig:schematic}
\end{figure}

Denoting the number of spatial dimensions as $D$, our construction takes as input a group-cohomology anomaly specified by an element of $H^{D+2}(\Gir,U(1))$ for the low-energy symmetry group $\Gir$, and outputs a lattice model of an igSPT in $Dd$ with an enlarged UV symmetry group $\Guv$ that realizes this anomalous $\Gir$-symmetry action in the IR.  
Specifically, we define a lattice model with $\Gir$-rotors on vertices of the lattice and $\N$ rotors on plaquettes, with a microscopic onsite $\Guv$ symmetry action. Then, we design a Hamiltonian, $H_\Delta$, that locks the $\N$ rotors on each plaquette to topological defects of the $\Gir$ rotors such that, for energy scales $\ll \Delta$ (``the IR"), the $\N$-rotors are gapped, and an emergent anomaly is exactly imprinted on the low-energy sector. 
We show that, if the symmetry $\Gir$ remains unbroken and the system does not develop fractionalized topological order, then the emergent anomaly protects nontrivial bulk and edge modes. The connection between emergent anomalies and igSPTs was pointed out in the original work on this topic~\cite{thorngren2021intrinsically}, and the structure of group extensions was explored in~\cite{li2022symmetry}. 
This work extends these concepts to lattice models in dimensions higher than one, constructs a number of new examples, and elucidates the connection between the emergent anomaly and the igSPT edge physics.
Specifically, we give a general argument for a bulk-boundary correspondence connecting the emergent $\Gir$ anomaly and the structure of the group extension, based on gauging the extending $\N$ symmetry to form an anomalous $\Gir$ symmetry enriched topological order (SET).
In many cases, the edge states can also be understood via a physically-intuitive mechanism of SPT pumping.  Namely, for these cases, the extended symmetry operations $\N$ act trivially in the igSPT bulk, but pump a $(D-1)d$ $\Gir$-SPT onto the edge. 
This lower-dimensional SPT pumping action forces the igSPT edge to reside at a self-plural~\footnote{The generalization of self-dual to potentially more than two phases.} critical point among different $(D-1)d$ $\Gir$-SPT phases, which prevents the edge from being in a trivial symmetric state.

In $1d$, the emergent anomaly forces the igSPT bulk to either be gapless or spontaneously break symmetry. By contrast, higher dimensions $D\geq 2$, an additional possibility arises that the emergent bulk anomaly bulk can be satisfied by certain gapped and symmetric states with anomalous symmetry-enriched topological order (SET). 
Using techniques introduce in~\cite{tachikawa2020gauging,wang2018symmetric}, we construct exactly solvable igSPT-type models with an emergent anomalous SET ground state. In this context, the moniker ``intrinsically gapless" is no longer appropriate, and we instead refer to these phases as quotient-symmetry enriched topological phases (QSETs) following the language of quotient-symmetry protected topological phases (QSPTs) introduced in~\cite{verresen2021quotient}. QSETs have an anomalous implementation of the IR symmetry, which is lifted to an anomaly-free symmetry action in the UV, a notion that requires a separation of scales, $\delta \ll \Delta$, between the gap, $\delta$, to creating anyonic excitations in the QSET, and the scale $\Delta$ at which the anomaly emerges. As for QSPTs, we argue that the QSET orders are sharply distinguished from ordinary $\Guv$-SETs by an edge phase transition where the gap to the extending $\N$-DOF closes at the edge. Namely, if the gap to $\N$ charges remains open, then the anomalous lower-dimensional SPT pumping symmetry ensures that the QSET edge is either gapless or symmetry breaking ($D=2$). In $3d$, the edge is itself $2d$, and it is also possible that both the bulk and edge form a QSET order. We construct an explicit example of this in Section~\ref{sec:3d}, which takes the form of a $3d$ toric-code topological order in which gauge magnetic flux lines are decorated with $1d$ igSPT states.

The paper is organized as follows. 
In Section~\ref{sec:taxonomy}, we review the taxonomy of known types of gapless SPTs.
In Section~\ref{sec:1d}, we review the $1d$ igSPT models with $\Guv=\Z_4$ symmetry introduced in~\cite{thorngren2021intrinsically,li2022symmetry} in the language of the group-cohomology framework, and show how this perspective connects their edge states to the emergent anomaly. Here, we also lay the ground-work for constructing lattice models with fractionalized anomalous QSET orders in higher dimension. 
In Section~\ref{sec:general}, we formally generalize this structure holds to igSPTs in various dimensions and symmetry classes with emergent anomalies classified by group-cohomology. 
We then illustrate this general formalism by constructing lattice models for a $2d$ time-reversal symmetric igSPT (Section~\ref{sec:2d}) and a $3d$ Ising igSPT (Section~\ref{sec:3d}). We close with a brief discussion of prospects for realizing igSPTs with beyond group cohomology models and as Mott insulators of realistic electron systems. 
Since there are few reliable theoretical tools for studying strongly coupled field theories in $2d$ and $3d$, throughout, we will not attempt to analyze the ultimate low-energy (deep-IR) field theory description of the lattice models we construct. Rather, we will use the presence of a gapped sector to make sharp topological statements about the emergent anomalous symmetry, and use these to deduce information about the possible structure of higher-dimensional igSPT bulk and edge modes.
Finally, the Appendices contain additional formal details, and construct several additional igSPT model examples in $1d$, $2d$, and $3d$ with various combinations of $\Z_N$, $U(1)$, and time-reversal symmetries. 

\section{Gapless SPTs: Definitions and Taxonomy \label{sec:taxonomy}}
Colloquially speaking, a gapless symmetry protected topological (SPT) state is a scale-invariant gapless system that possesses symmetry protected topological edge states that cannot be removed without undergoing a phase transition where the bulk changes its universal scaling properties.
Previous studies~\cite{scaffidi2017gapless,verresen2018topology,verresen2019gapless,thorngren2021intrinsically,li2022symmetry} have identified four distinct categories of gapless SPTs defined by two binary characteristics:
\begin{enumerate}
\item whether or not the gapless SPT edge states can be trivialized by stacking it with a gapped SPT with the same symmetry,
\item whether or not the edge states are exponentially well-confined to the edge by a gapped sector.
\end{enumerate}
Gapless SPTs that fail to have the first property are called intrinsically gapless SPTs (igSPTs)~\cite{thorngren2021intrinsically,li2022symmetry}, and will be the focus of this work. Those that fail to have the second property have been called ``purely gapless" SPTs, and have edge states that are not exponentially localized to the edge. Rather, their influence decays as a power of the distance into the bulk~\cite{verresen2019gapless}. It remains an open question whether there are gapless SPTs that are both purely and intrinsically gapless~\cite{RubVPC}.

Despite their nom de plume, ``gapless SPTs" are perhaps better thought of as a form of symmetry-enriched criticality (SEC)~\cite{verresen2021gapless}~\footnote{Unlike gapped phases, which are stable to generic perturbations, gapless states may have one or more relevant perturbations that change their universality class. We will remain agnostic about the number of relevant perturbations, i.e. whether we are discussing a stable gapless phase or (more commonly) a fine-tuned critical or multi-critical point, and use the terms ``critical" and ``gapless" as synonyms.}. Adding a symmetry to gapped systems with a fixed type of intrinsic topological order leads to distinct symmetry enriched topological (SET) phases. Unlike gapped SPT phases, SET phases cannot be smoothly connected to a trivial atomic insulator via a gapped path of Hamiltonians, even when the protecting symmetry is broken due to the underlying long-range entanglement of the intrinsic topological order. Similarly, for a given universality class of a gapless system, there may be multiple inequivalent implementations of symmetry, which cannot be smoothly interpolated between along a path of Hamiltonians whose ground-states have the same universality class~\cite{verresen2021gapless}. 

Gapless SPTs are examples of SECs, in which one cannot locally distinguish between a ``trivial" critical point and a symmetry-enriched one from local bulk measurements. Rather, the differences between different gapless SPTs are only evident in non-local bulk probes, or local edge probes.
One can make this notion more specific by analogy to gapped SPTs. Different gapped $G$-SPT ground-states can be connected by a finite-depth local unitary (FDLU), $U$, that is overall symmetric ($[U,g]=0~\forall g\in G$) but which is not symmetrically generated ($U\neq e^{-iH}$ for any local $G$-symmetric $H$). This definition cannot be directly ported to the gapless setting as even different instances of a gapless state with the same universal scaling properties, e.g. two instances of a conformal field theory (CFT) perturbed by different irrelevant operators, cannot be connected by an FDLU. However, we can generalize the notion of symmetric FDLU-(in)equivalence by defining i) that two ground-states are in the same universality class if they flow to the same RG fixed point after applying an overall symmetric FDLU, ii) ground-states with the same universality class are distinct gapless SPT classes if they cannot be connected in this way by any symmetrically-generated FDLU.
An immediate corollary of this definition is that local scaling operators in distinct gapless SPTs of the same type of criticality have the same symmetry properties as conjugating a local operator with definite symmetry quantum number with an overall-symmetric $U$ preserves its symmetry quantum number. However, the symmetry properties of non-local scaling operators, such as a disorder operator that inserts a domain wall, may change under such an overall symmetric FDLU, leading to distinct classes of gapless SPTs.  These notions can be made more precise for $1+1d$ conformal field theories (CFTs)~\cite{verresen2021gapless}, for which the data specifying a universality class is well understood and characterized by the spectrum and fusion rules for primary scaling operators. 

\section{Low dimensional igSPTs \label{sec:1d}}
Given the formal nature of our constructions, we begin by warming up with an (almost trivial) example of how emergent anomalies can arise in a $0d$ system.
We then, review the $\Guv=\Z_4$ symmetric $1d$ igSPT with emergent $\Gir=\Z_2$ anomaly previously constructed in~\cite{thorngren2021intrinsically}, in a language that is amenable to generalization to other symmetry groups and dimensions.

\subsection{Warmup: Emergent anomalies in $0d$}
To see a simple example of how an IR anomaly can emerge from an anomaly-free UV system, consider the $0d$ spin-1/2 edge state of a Haldane/AKLT spin-chain~\cite{haldane1983nonlinear,affleck2004rigorous} with spin-rotation symmetry ($\Gir=SO(3)$). While this $1d$ G-SPT phase is typically discussed as a spin-1 chain, in any physical realization, it arises only as an emergent description of spin-1/2 electrons in a Mott phase, so that only spin-1 DOF are active below the Mott gap, $\Delta$. The spin-1/2 DOF transform under a larger group $\Guv=SU(2)$: rotations that add up to $2\pi n$ around any spin-axis, which are identity operations in $\Gir$, give a Berry phase of $(-1)^n$ with $n\in \N=\Z_2$.  Formally $\Guv=SU(2)$ is a (central) extension of $\Gir=SO(3)$ by the normal subgroup $\N=\Z_2\vartriangleleft \Guv$. The spin-1/2 representation of $SO(3)$ is a projective representation characterized by a nontrivial element of the second group-cohomology $\omega_2(g,h)\in H^2(SO(3),U(1)) = \Z_2$, where the $\Z_2$ group structure indicates that two spins-1/2 form an ordinary linear representation of $SO(3)$. Namely, for $\Gir,h \in SO(3)$, the spin-1/2 representation, $R_{g},R_h$ satisfy multiplication rule $R_gR_h = \omega_2(g,h) R_{g\cdot h}$ where an explicit representation of $\omega_2$ is the Berry phase for sweeping the  spin around a loop starting from up in the z-direction, rotating by $g$ then $h$, then $(gh)^{-1}$. 

One can realize the system with a spin-1/2 ground-space trivially in a single spinful fermion site, simply by considering a single-site Hubbard model: $H=-\mu c^\dagger_\sigma c^{}_\sigma+\frac{U}{2}n(n-1)$ with onsite repulsion $U$ and chemical potential $0<\mu<U/2$ so that there is a single occupied electron in the ground-space. This system has a (Mott) gapped sector with even fermion parity consisting of the empty and doubly-occupied states with energy $\Delta = \text{min}(\mu,U-2\mu)$, and a doubly-degenerate ground-space consisting of the singly-occupied spin-up or down states. 
In this low-energy ground-space there is an emergent $SO(3)$ anomaly: the states transform projectively under $SO(3)$ as an ordinary spin-1/2 representation of $SU(2)$. This anomaly is stable: it is protected by the Mott gap to fermion excitations, and can only be removed by closing the gap separating the fermion parity even and odd ground-spaces.

While this example may seem merely an overly complex re-interpretation of a completely trivial single-site problem, it lays the ground-work for more complicated higher dimensional examples. Specifically, it suggests that ingredients for realizing an emergent anomaly are: i) a (central) group extension that lifts the anomaly, ii) an interaction term that locks the extending degrees of freedom away at high energies in such a way that imprints the anomaly on the low-energy subspace.

\subsection{Review of $1d$ igSPT with $\Z_4$ symmetry}
With this nearly-trivial $0d$ example in hand, we next review the constructions of~\cite{thorngren2021intrinsically,li2022symmetry} for $1d$ igSPTs. Our goal will be to adapt the notation such that generalizations to higher dimension and other symmetry groups become obvious.
We also reveal additional structure about the connection of igSPT edge states, and the group extension, via a symmetry under pumping lower dimensional $\Gir$ SPTs onto the boundary.
Additionally, we introduce various notions of gauging the symmetry that serve as useful tools for characterizing the igSPT topology, and constructing QSET phases in higher dimensions.

In this section, we focus on the example where the low-energy IR symmetry group is $\Gir=\Z_2$ and the UV symmetry group is $\Guv=\Z_4$. Additional examples for other $\Gir$ are detailed in Appendix~\ref{app:1d} and summarized in Table~\ref{tab:results}. The emergent $\Gir=\Z_2$ anomaly is the same one that protects the edge of the $2d$ G-SPT constructed by Levin and Gu~\cite{levin2012braiding}. However, realized as a pure $1d$ igSPT with extended $\Guv=\Z_4$ symmetry, we can interrogate the $0d$ ends of an open chain with this bulk anomaly.

Ref.~\cite{thorngren2021intrinsically} proposed an elegant physical model realizing this igSPT phase in terms of an extended Ising-Hubbard model with discrete Ising symmetry corresponding to $\pi$-rotations around a specific ($x$) spin-axis. The electronic degrees of freedom form a spin-1/2 projective representation of this group for which a $2\pi$ rotation results in a $(-1)$ phase, and hence only a $4\pi$ rotation is trivial corresponding to symmetry group $\Guv=\Z_4$. 
Denoting $\Z_4 = \{0,1,2,3\}$, the fermion parity, $(-1)^{N_f}$ where $N_f$ is the number of fermions, forms a normal sub-group $\N=\{0,2\} = \Z_2^F$.
The igSPT phase corresponds to a Mott insulator, in which the fermion excitations have an energy gap, so that the only IR degrees of freedom are bosonic spins that transform under a quotient group $\Gir=\Guv/N = \{0,[1]\} = \Z_2$ where $[\dots]$ denotes the equivalence class under the quotient.
The IR igSPT phase has an emergent anomaly of $\Gir$, corresponding to the anomaly of the edge of a $2d$ gapped $\Z_2$-SPT~\cite{levin2012braiding}.
While the extension of the spin-model by fermionic spinor DOF is natural for physical realizations, Ref.~\cite{li2022symmetry} showed that the same IR theory can arise from a purely bosonic $\Z_4$ spin chain.
We will follow the latter all-boson approach, since it meshes nicely with the group cohomology formalism, but will comment on cases where the igSPT might alternatively arise from a Mott insulator of fermions.
While all the main points of this $1d$ $\Z_4$ igSPT were previously explained in~\cite{thorngren2021intrinsically,li2022symmetry}, we give additional arguments that clarify the structure of edge states, and use a notation and framework that readily generalizes to other dimensions and symmetries.

\subsection{Anomalous Ising spin-chain}
Consider a spin-1/2 chain with on-site $\Z_2$ spin DOF $g_i\in \Gir = \Z_2 = \{0,1\}$. Specifically, we can define the standard Pauli operator $\sigma^z_i = (-1)^{g_i}$.
Further, define the conjugate operator as $\hat{g}_i\in \{0,1\}$, via $\sigma^x_i = (-1)^{\hat{g}_i}$.
To start consider an infinite chain or with periodic boundary conditions (PBCs). We will then consider the effect of open-boundary conditions and the edge.
An anomalous non-onsite symmetry action is:
\begin{align}
U_{1}^A &= \prod_i \omega_3(g_i-g_{i-1},-g_i-1,1) U_1^\text{os} 
\nonumber\\
&\equiv \prod_i (-1)^{g_i(g_i-g_{i-1})} \prod_i (-1)^{\hat{g}_i}
\nonumber\\
\label{eq:U1A1dZ2}
\end{align}
where $\omega_3(g,h,k) =(-1)^{ghk}$, and the second factor in $U_1^A$ is just the ordinary on-site action of $\Gir=\Z_2$ on each site which maps $|g_i\>\rightarrow |g_i+g \mod 2\>$.
Physically, the non-onsite phases, $(-1)^{g_i(g_i-g_{i-1})}$
 means that each domain wall (DW) between $\sigma^z_i = \up$, $\sigma^z_{i-1} = \down$ carries a $\Z_2$ charge (gives a phase of $(-1)$ in $U_1^A$)~\footnote{This phase can be more symmetrically apportioned such that $\up\rightarrow \down$ and $\down \rightarrow \up$ DWs each contribute a factor of $i$, which is related by a finite-depth unitary or equivalently redefining $\omega_3$ by a coboundary. However, we prefer that the non-onsite phases take values in the IR symmetry group ($\Gir=\Z_2$).}.
Roughly-speaking, this symmetry action is anomalous because it causes non-trivial (semionic) statistics for Ising domain walls (DWs)~\cite{levin2012braiding},
 which intuitively presents an obstacle for reaching a trivial gapped, symmetric state from a spontaneous symmetry broken one by ``condensing" DW defects, as the non-onsite factors lead to destructive interference in the DW dynamics preventing them from condensing~\cite{kawagoe2021anomalies}.

\begin{figure}[]
\begin{centering}
	\includegraphics[width=1.0\columnwidth]{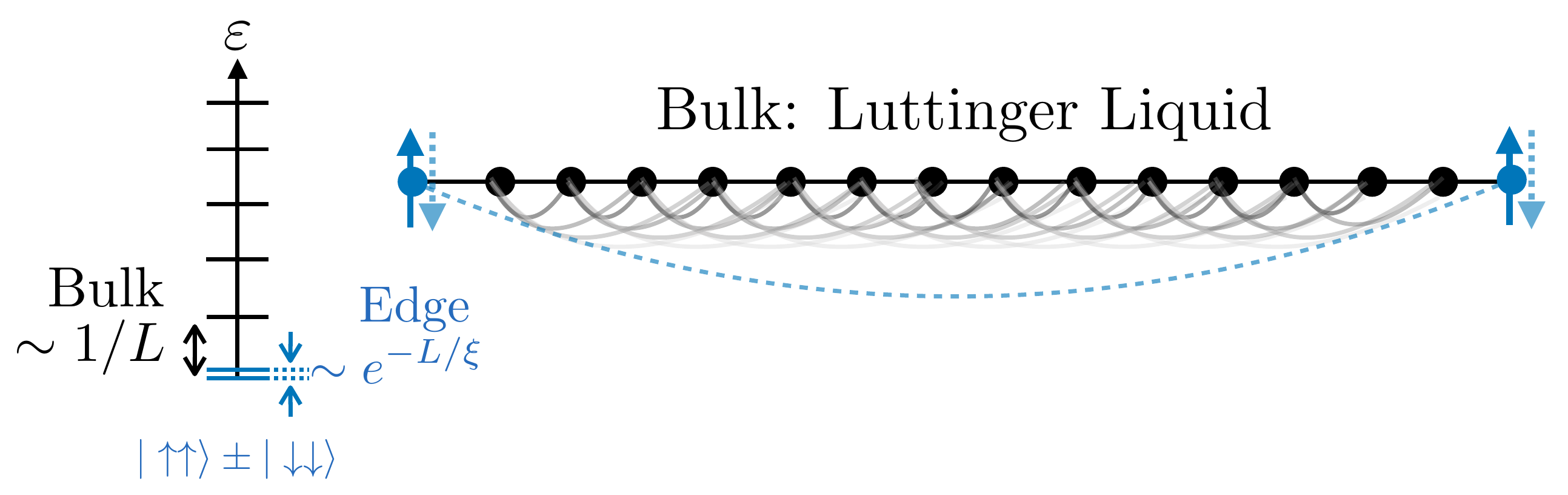}
\end{centering}
	\caption{{\bf Anatomy of a $1d$ igSPT -- } Schematic depiction of the structure of $1d$ $\Guv=\Z_4$ intrinsically gapless SPT as originally explained in~\cite{thorngren2021intrinsically}. The bulk has an emergent anomalous $\Z_2$ symmetry, which protects a gapless Luttinger liquid state (gray arcs schematically denote power-law range correlations) with finite size gap $\sim 1/L$ for chain length $L$ or requires spontaneous symmetry breaking (SSB). With open boundaries, there is a twofold ground-space degeneracy with much smaller finite-size splitting $\sim e^{-L/\xi}$. Crucially, the edge state space does not have a local tensor product structure, or equivalently, there is long-range entanglement between the edge modes in the exact ground-states (long dashed blue line).}
	\label{fig:1digspt}
\end{figure}

\subsection{Onsiteing the symmetry}
Following~\cite{thorngren2021intrinsically,li2022symmetry}, this $\Gir=\Z_2$-anomaly can emerge as the low-energy sector of an igSPT with an enlarged symmetry $\Guv=\Z_4$ which is a central extension of $\Gir$ by $\N=\Z_2$. 
Introduce auxiliary qubit DOF with Pauli operators
$\tau^x_i = (-1)^{\hat{n}_i}$, $\tau^z_i = (-1)^{n_i}$
with ordinary on-site (extended) symmetry generated by:
\begin{align}
U_1^\text{os} = \prod_i (-1)^{g_i\hat{n}_i} (-1)^{\hat{g}_i}.
\label{eq:U1hatG1dZ2}
\end{align}
Then, energetically lock $\hat{n}_i$ to $(g_i-g_{i-1})$ via a Hamiltonian:
\begin{align}
H_\Delta = -\Delta \sum_i \delta_{\hat{n}_i,g_i-g_{i-1}} = -\Delta \sum_i \frac12\(1+\tau^x_i\sigma^z_i\sigma^z_{i-1}\),
\end{align}
such that the symmetry action restricted to the ground-space of $H_\Delta$ is equivalent to the anomalous symmetry action:
\begin{align}
U_1^\text{os} \ir  \prod_i (-1)^{g_i(g_i-g_{i-1})} \prod_i (-1)^{\hat{g}_i} =U_1^A
\end{align}
We will denote this as $U_1^\text{os}\underset{IR}{\approx} U_1^A$.
To determine the structure of the UV symmetry group $\Guv$ implemented by \eqref{eq:U1hatG1dZ2}, note that $(U_1^\text{os})^2 = \prod_i(-1)^{\hat{n}_i}$, and $(U_1^\text{os})^4=1$, from which we see that the extended group structure is $\Guv=\Z_4$.

\subsubsection{Transforming into the IR space}\label{sec:transintoir}
An alternate perspective on the emergent anomaly is obtained by starting with the on-site $\Z_4$ symmetry action \eqref{eq:U1hatG1dZ2}, and performing a local unitary transformation: 
\begin{align}
V = \prod_i (-1)^{n_i(g_i-g_{i-1})}
\label{eq:v1d}
\end{align} 
that  maps $H_\Delta$ to a trivial $\N$-paramagnet:
\begin{align}
V^\dagger H_\Delta V = 
-\Delta \sum_i \delta_{\hat{n}_i,0}
\end{align}
and converts the onsite symmetry to the anomalous one:
\begin{align}
V^\dagger U_1^\text{os}V = \prod_i (-1)^{g_i(g_i-g_{i-1})} \prod_i (-1)^{g_i\hat{n}_i} \prod_i (-1)^{\hat{g}_i}
\end{align}
which coincides with the anomalous symmetry transformation, \eqref{eq:U1A1dZ2} if we restrict to the ground-space of $H_\Delta$ where $\hat{n}_i=0$.

\subsubsection{Zero-correlation length (ZCL) IR Hamiltonians} We can write down an idealized model for the emergent anomaly by adding generic local interactions: $H=H_\Delta+H_\text{IR}$, where $H_\text{IR}$ is symmetric, and commutes with $H_\Delta$, i.e. exactly preserves the IR subspace with emergent anomaly. For such Hamiltonians, the emergent IR symmetry operation is precisely \eqref{eq:U1A1dZ2}, for which the non-onsite phases are strictly local, acting only on neighboring sets of three sites. We will refer to such models as having zero-correlation length (ZCL), with the understanding that this refers to the spatial range of the symmetry action (and, as we will see shortly, the localization length of igSPT edge states) but not to the correlation length of local correlation functions which is infinite in a gapless system.

A general prescription to construct a ZCL $H_\text{IR}$ is to take any local interaction involving only $\Gir$-rotors, conjugate it by $V$ to generate an interaction term that commutes with $H_\Delta$, but may not respect symmetry. Then explicitly symmetrize the term by superposing it with its symmetry conjugate (assuming this sum does not vanish). For example, starting with a simple transverse field term $H_1=-J\sum_i (-1)^{\hat{g}_i}=-h\sum_i \sigma^x_i$, conjugation by $V$ gives: $V^\dagger H_1 V = -J\sum_i \tau^x_{i=1}\tau^x_i\sigma^x_i$ which commutes with $H_\Delta$ but is not symmetric. Then, we can add its symmetry conjugate, to obtain $H_\text{IR} = V^\dagger H_1 V+U_1^\dagger V^\dagger H_1 V U_1$:
\begin{align}
H_\text{IR} = -J\sum_i \(\tau^x_{i-1}\tau^x_{i}+\tau^y_{i-1}\tau^y_{i}\)\sigma^x_i
\end{align}  
which satisfies the fixed-point properties. This Hamiltonian can further be exactly solved by fermionization, and results in a gapless Luttinger liquid with topological edge states~\cite{li2022symmetry}.
We mention in passing that an alternative prescription for constructing fixed-point $H_\text{IR}$'s would be to consider including a generic local symmetry-preserving term $H_1$ with coefficient $\ll \Delta$, but which may not commute with $H_\Delta$. Then, perform (degenerate) higher-order perturbation theory with $H_1$ to approximately compute its interaction projected into the low-energy subspace. 

Clearly there are many different possible options for $H_\text{IR}$, resulting in a potentially rich phase diagram in the deep IR. This phase diagram is constrained by the emergent anomaly, which prevents trivial (symmetric and non-fractionalized) gapped states. Rather than confronting the (generally hard) task of working out this phase diagram for specific choices of $H_\text{IR}$, we focus instead on deducing sharp topological features arising from the emergent anomaly.

\subsection{Edge states via lower-$d$ SPT pumping} 
To explore the edge of this Ising igSPT, restrict the system to an open chain of $\Z_4$ rotors, with sites $i=1\dots L$, and microscopic (UV) symmetry action \eqref{eq:U1hatG1dZ2}. 
Further, restrict $H_\Delta$ to this open chain by simply omitting term that spill past the boundaries, and add the edge term $-\Delta\delta_{n_1,0}$ to remove the dangling $\N$ rotor that does not interact with any $\Gir$ domain walls on the links.
The numerical analysis and analytic arguments of~\cite{thorngren2021intrinsically,li2022symmetry} show that the $\Z_4$ igSPT has gapless DOF corresponding to operators localized to the edge of the igSPT chain with open boundaries. Specifically, in a length-$L$ chain, the bulk topological Luttinger liquid has a finite size gap $\delta_B \sim 1/L$. Within this gap, there is a near-exact two-fold degeneracy when the system has open boundary conditions, with the two near-ground-states being split by an amount $\delta_E \sim e^{-L/\xi}$, where $\xi$ is proportional to the inverse of the energy gap $\Delta$ to the UV $\N$ DOFs.

Here we give a general argument that clarifies this structure, and which readily generalizes to other examples with various symmetries and dimensions.
Specifically, consider the anomalous symmetry action $U_g^A$.
In the ground-space (IR) of $H_\Delta$, the extending $n_i$ spins are locked to DWs of the $g_i$ spins, and we can write a description of the action just in this IR sector:
\begin{align}
U_1^\text{os}\underset{IR}{\approx} \prod_{i=2}^L (-1)^{g_{i}(g_i-g_{i-1})} \prod_{i=1}^L (-1)^{\hat{g}_i}
\end{align}
In the IR symmetry group $\Gir=\Z_2=\{0,1\}$ the group operation is $1+1=0$.
One might therefore expect that $U_{g=2}$ would yield the identity in the IR. Instead, one finds:
\begin{align}
	U_2^\text{os}=\(U_1^\text{os}\)^2\underset{IR}{\approx} \prod_{i=2}^L (-1)^{g_{i}-g_{i-1}} = (-1)^{g_L-g_1} = \sigma^z_1\sigma^z_L.
\end{align}
 We can interpret $U_2$ as a unitary that pumps a $0d$ $\Gir$-SPT onto the edges of the chain. $0d$ G-SPTs are classified by representations of $\Gir$ on $U(1)$ -- i.e. different $0d$ SPTs correspond to different possibly symmetry quantum numbers or ``charges" of the ground-state (which cannot be changed without closing the gap to an excited state with a different charge). 
For $\Z_2$ symmetry there are two possible symmetry charges, $\pm 1$, and $U_2$ flips the local symmetry charge of each edge, toggling this $0d$ SPT invariant. 
Specifically, the local action of $U_2$ on one end of the chain, say $x=1$, is $(-1)^{g_1}$ which anticommutes with the $\Gir$-symmetry generator $\prod_i (-1)^{\hat{g}_i}$.
As a result, the edge is tuned to a degenerate critical point where there are two degenerate edge configurations.

We now connect this lower-$d$ SPT pumping symmetry to the ground-space degeneracy of the open igSPT chain, and also to the non-local string order identified in~\cite{thorngren2021intrinsically}.
By assumption, $U_{\Guv=0,1,2,3}$ are symmetries of the full bulk and edge Hamiltonian.

Moreover, since in the IR $U_2$ acts nontrivially only on the edge of the chain, $(U_2)_{[1,L]}=(-1)^{g_1-g_L}=\sigma^z_1\sigma^z_L$, one can define the restriction to one end of the chain by $U_2|_{1,L}:=\sigma^z_{1,L}$. This local action of the pumping actions must commute with any local Hamiltonian that's invariant under the $\Guv$ symmetry. Furthermore, each of them anti-commutes with the symmetry generator $U_1$. This non-commuting algebra of symmetry acitons $\{(U_2)|_{i=1},U_1\}=0$ immediately leads to an exact ground state degeneracy that is at least 2-fold. Moreover, this degeneracy is directly associated with edge-local operators and does not arise with periodic boundary conditions.

We refer to the edge-local version of the pumping operators as topological edge modes.
However, a key point that distinguishes them from ordinary topological edge states of gapped systems is that the degenerate ground-space does not have a tensor-product structure. Specifically, unlike for the Haldance/AKLT chain~\cite{haldane1983nonlinear,affleck2004rigorous}, the degeneracy cannot be decomposed into a local ``spin-1/2" living at each end of the chain (which would result in 4-fold GS degeneracy). 
Formally, the distinction is that the edge pumping modes, $(U_2)_{i=1,L}$ share a common global conjugate operator $U_1$, whose action cannot be localized to one or the other edge because of the gapless bulk degrees of freedom (in contrast to the Haldane/AKLT chain where there are independent edge mode $S^{x}$ and $S^z$ operators for each end of the chain). 
We note that a similar non-tensor product structure edge degeneracy was observed at the critical point between a Haldane chain (gapped $\Z_2^2$-SPT) and a reduced symmetry magnet gapped with only a single $\Z_2$ symmetry~\cite{scaffidi2017gapless}. 
The fractional quantum dimension of the edges harkens that of Majorana zero modes at the end of a topological superconductor. 
However, a more apt analogy is that of (boundary) spontaneous symmetry breaking (SSB) since: the true ground-state of a finite-size chain is a Schrodinger-cat/GHZ like state of opposite edge magnetizations which has long-range entanglement that can be collapsed by a local measurement of the edge spin or which can be removed by an infinitesimal boundary field. Ordinarily, a $0d$ system could not spontaneously break symmetry on its own, however, here the $0d$ edge is stabilized by the $1d$ bulk, despite that the bulk itself does not break symmetry.
An equivalent interpretation of these edge modes is as different symmetry-breaking conformal boundary conditions on the bulk CFT~\cite{scaffidi2017gapless,duque2021topological}.

\subsubsection{Effect of finite correlation length}
Next consider moving away from the ZCL limit, where $H_\text{IR}$ is symmetric, but does not commute with $H_\Delta$.
The allowed emergent anomalies are discrete and cannot be continuously altered by $\Gir$-symmetric perturbations that do not close the gap to the $\N$ charge degrees of freedom. 
Namely, consider perturbing the Hamiltonian to: $H=H_\Delta + \lambda V$, where $\lambda$ is a dimensionless coupling constant and $V$ is $\Guv$-symmetric. Then, the stability of the gapped $\N$ degrees of freedom to local perturbations~\cite{bravyi2010topological} implies there is a critical $\lambda_c>0$, such that for $\lambda<\lambda_c$ the $\N$-charged DOF remain gap, and the emergent anomaly remains stable.
In this regime, one can define an IR Hilbert space by adiabatically continuing the ground-space of $H_\Delta(s) = H_\Delta +\lambda(s) V$ with $s\in [0,1]$, $\lambda(0)=0$, and $ \lambda(1)=\lambda<\lambda_c$.
The resulting adiabatic evolution is accomplished by a $\Guv$-symmetric finite-depth local unitary circuit (FDLU): $F = \lim_{t_f\gg V/\Delta^2} \mathcal{T}\{e^{-i\int_0^{t_f}H_\Delta(t/t_f) dt}\}$. 
Properties in the FDLU-transformed frame, $F$, then differ from those in the lab-frame by exponentially-well-localized symmetric dressing~\cite{hastings2010locality}. In particular, the IR action of symmetry is equivalent to the ideal nearest-neighbor anomalous action, $U_g^A$, above, up to this FDLU transformation, For example, in the perturbed model, the pumping operation $F^\dagger U_1^2 F$ is no longer precisely localized to the edge, but has exponentially decaying tails into the bulk that decay with distance $x$ as $e^{-x/\xi}$ where the correlation length $\xi$ is bounded by the Lieb-Robinson distance for $F$.

From this we deduce that, away from the ZCL limit, we can define quasi-locally dressed operators $W_{1,L} = F^\dagger \sigma^z_{1,L}F$ that approximately commute with $H$ to accuracy $e^{-L/\xi}$, and this edge-state degeneracy is split by this exponentially small amount, but however, remains sharply distinct from the bulk finite-size gap $\sim 1/L$ for $L\gg \xi$.

\subsubsection{The $\N$-charge gap does not close at the edge}
Since the SPT pumping operation is implemented by an $\N$-symmetry operation, and ends up acting nontrivially at the edge of the igSPT, one might be tempted to conclude that the gap to the $\N$ charged DOF closes at the edge. We emphasize that this is \emph{not} the case. 
Specifically, the unitary $V$ in \eqref{eq:v1d} explicitly transforms $H_\Delta$ to a paramagnet that fully gaps all the $\N$ charges.
In the ZCL limit, one can readily check that there are no local $\N$-charged operator that acts within this 2-fold degenerate ground-space associated with these edge model (a feature that is preserved up to local dressing by $F$ away from the ZCL limit).
Physically, the effect of the $\N$ symmetry transformation on the IR theory with emergent anomalous $\Gir$ symmetry, arises due to a non-trivial entanglement between $\N$ charges and $\Gir$ charges that is locked in by $H_\Delta$ at UV scale $\Delta$. However, it costs finite energy, $\Delta$, to unlock an $\N$ rotor.

\subsubsection{Connection to string order}
We note that the lower-$d$ SPT pumping operation is closely connected to the nonlocal string order identified by TVV in~\cite{thorngren2021intrinsically}. In TVV's Ising/Hubbard model exhibited a string order $S^z_i\prod_{i<x<j} (-1)^{N_x} S^z_j$ where $N_i$ was the fermion number on site $i$ and $S^z_i$ was the electron spin operator at site $i$. In that example, the group extension that lifted the anomaly was to extend a $\Gir=\Z_2$ $x$-axis spin rotation symmetry by fermion parity $\N=\Z_2^F$. Hence we can understand the structure of the string operator as follows: the bulk of the string $\prod_{i<x<j} (-1)^{n_x}$ where $n_x$ is the number of fermions at position $x$, is simply the $\N=\Z_2^F$ symmetry generator restricted to an interval $(i,j)$. The $S^z$ operators terminating the ends of the string add a symmetry charge, i.e. add a $0d$ SPT onto the end of the string. This cancels the $0d$ SPT pumping action of the symmetry operation, such that the string operator has a finite expectation value for an arbitrarily long string. In principle, this picture coupled with the SPT-pumping perspective enables one to identify non-local membrane ``order-parameters" that detect higher dimensional igSPTs which have an analogous form of acting with an $\N$ symmetry operation restricted to $Dd$ region, multiplied by $(D-1)d$ boundary operators that add an appropriate $\Gir$ SPT to their boundaries. A subtlety (common to all higher-$d$ analogs of non-local string order), is that, away from the ZCL, long-range membrane order is characterized by exponential decay of membrane correlations with the perimeter/surface-area of the \emph{membrane-boundary} (an $\mathcal{O}(1)$ constant suppression for $1d$ string order parameters) rather than with the bulk membrane area (similar to the perimeter vs. area law for diagnosing confinement in $2d$ gauge theories). For this reason, we do not pursue this non-local order perspective in higher-dimensions. 

\subsection{Gauging the symmetry}
Gauging the symmetry often provides non-perturbative insights into its topology. Gauging a $G$-symmetry can either be done by exploring the response of the system to an external, background $G$-gauge field, or by promoting a global $G$ symmetry to a local gauge redundancy coupled to dynamical gauge fields. For our purposes, it will be most useful to consider gauging the central extension, $\N$, that lifts the anomaly, which results in an anomalous $\Gir$-SET.
While this gauging perspective will be most useful in $2d$ and $3d$, we use the $1d$ version as a warmup.

\subsubsection{Coupling to a background gauge $\N$-gauge field} 
For a gapped system, the response theory to an external gauge field is a local topological quantum field theory (TQFT). For an igSPT, the presence of gapless bulk modes can lead to non-local responses to a general $\Guv$-gauge field. To avoid this non-locality, one must be selective in what gauge-fields or gauge configurations we couple into the system.
Since the $\N$ DOF are gapped in the IR, gauging only the normal sub-group $\N$ of the extended symmetry $\Guv$ results in a local response theory.
As explained in~\cite{tachikawa2020gauging}, gauging the extending symmetry $\N$ results in a theory with an anomalous $\Gir=\Guv/\N$ symmetry.
Here, we explore the relation between this and the igSPT edge states.

To explicitly gauge the lattice model, introduce a non-dynamical (backgrond) $\Z_2$ gauge link variable $A_{i,i+1}\in \N=\{0,1\}$ associated with link $(i,i+1)$, and conjugate gauge electric field $E_{i,i+1}\in \widehat{\mathcal{N}}=\{0,1\}$ satisfying $\{(-1)^{A_{i,j}},(-1)^{E_{i,j}}\} = 0$, and enforce the Gauss' law gauge constraints: $\hat{n}_i = E_{i,i+1}-E_{i-1,i}$, which generate the gauge transformations: 
$n_i \rightarrow n_i+\chi_i$ and $ A_{i,j} \rightarrow A_{i,j}+\chi_j-\chi_i$
where $\chi_i\in \{0,1\}$.
We note that the following discussion is equivalent to the discussion of the effect of symmetry-twisted boundary conditions on $1d$ igSPTs in~\cite{li2022symmetry}. 

%

The unitary $V$ in \eqref{eq:v1d}, which converts the anomalous symmetry to an on-site one, is not gauge invariant as written. However, we can remedy this by minimally coupling $n_i$ to $A_{i,i+1}$. With periodic boundary conditions (PBCs) note that: $\prod_i (-1)^{n_i(g_i-g_{i-1})} = \prod(-1)^{g_i(n_{i+1}-n_i)}$. Hence, we can minimally couple $(n_{i+1}-n_i)\rightarrow (n_{i+1}-n_i - A_{i,i+1})$ to make this operator gauge invariant resulting in:
\begin{align}
\mathcal{V} = \prod_{i=1}^{L-1} (-1)^{g_i(n_{i+1}-n_{i}-A_{i,i+1})}
\end{align}
where the script font indicates that the $\N$ symmetry subgroup has been gauged.
This transformation effects $\mathcal{V}^\dagger H_\Delta \mathcal{V} = -\Delta\sum_i \delta_{\hat{n}_i,0}$ and (with PBCs):
\begin{align}
\mathcal{V}^\dagger U_1 \mathcal{V} &\ir (-1)^{\sum_i A_{i,i+1}} \prod_{i}(-1)^{g_i(g_i-g_{i-1})} \prod_i (-1)^{\hat{g}_i}
\end{align}
We see that the $U_1$ symmetry eigenvalues $\pm 1$ are toggled by an $\N$-gauge flux $(-1)^{\sum A} = (-1)$.
Hence, the $\N$ gauge flux carries a $\Gir$-symmetry charge. This is consistent with our above discussion of the edge states -- since inserting an $\N$ symmetry flux with PBCs is equivalent to cutting open the chain, acting locally on one end with $U_1^2 = W$ (which we have seen changes the $U_1$ charge), and then gluing the chain back together. Hence, the structure of the igSPT edge states can also be deduced by gauging the extending symmetry, $\N$.

\subsubsection{Fractionalizing the anomalous $\Gir$-symmetry} 
In higher dimensions, it is possible that the bulk igSPT anomaly is satisfied by a gapped symmetric, but topologically ordered state. Following~\cite{wang2018symmetric}, we can view this topological order as arising from fractionalizing $\Gir$-rotors into $\Guv$ DOF, and projecting out the unphysical fractional states by coupling the $\N$-subgroup of the $\Guv$ rotors to a dynamical $\N$ gauge field. Here, it will be crucial to assume a hierarchy of scales between the UV gap, $\Delta$ that imprints the anomaly, and the (assumed much smaller) energy gap, $\delta$, to the fractionalized excitations: $\delta \ll \Delta$.

Specifically, consider a microscopic lattice model with $\Guv$ symmetry, imposing $H_\Delta$ so that in the IR at scales $\ll \Delta$ the symmetry action is effectively the anomalous one given by \eqref{eq:U1A1dZ2} acting on $\Gir$-rotors. Then, fractionalize each $\Gir$-rotor into $\Guv$ rotor, with onsite basis $|\nu_i,g_i\>$ with $\nu_i \in \N_g = \Z_2$, but where the $\nu_i$ DOF is coupled to a dynamical $\N_g$-gauge field, $a_{i,i+1}$ (we use lower- and upper-case letters to distinguish emergent and external/background gauge fields). 
We caution that one should distinguish the gapped $n_i$ DOF from the fractionalized $\nu$ DOF: the latter are not gauge invariant, i.e. are not microscopic excitations, but may only emerge as fractionalized IR excitations if $a$ is in a deconfined phase (similar to spinons in a more typical parton or slave-particle description of electronic models). For this purpose, we include a $\Gir$ subscript on $\N_g$, to distinguish the gauge group from the global $\N$ symmetry acting on the gapped UV sector.

In higher dimensions, this dynamical gauge theory could emerge from a parton description where a microscopic boson operator $B$ with $\Gir$ charge $(-1)$, is fractionalized into a pair of $\frac12$-$\Gir$ charged operators: $B=b^2$ (i.e. $U_g:b\rightarrow i^g b$ acts as a $\Z_4$ symmetry on the partons). In $1d$, any such gauge theory will generically become confined if there are any quantum fluctuations in the gauge field. Nevertheless, we find it useful to study a fine-tuned fluctuation-less gauge model as  preparation for higher-dimensional examples.

The gauge invariant sector satisfies a Gauss law $\hat{\nu}_i = e_{i,i+1}-e_{i-1,i}$ where $\hat{\nu}$ and $e$ are conjugate to $\nu$ and $a$ respectively. This Gauss' law generates gauge transformations $\nu_i \rightarrow \nu_i+\chi_i$ and $a_{i,i+1}\rightarrow a_{i,i+1}+\chi_{i+1}-\chi_i$ for any local gauge transformation $\chi_i$. Further, the IR symmetry action on the fractionalized degrees of freedom is:
\begin{align}
U_1^\text{frac} = \prod_{i=1}^{L} (-1)^{g_i(g_{i}-g_{i-1})} \prod_{i=1}^L (-1)^{g_i\hat{\nu}_i}(-1)^{\hat{g}_i}
\end{align}

As for the external gauge response, the finite-depth unitary $\mathcal{V}$ (with $n\rightarrow \nu$ and $A\rightarrow a$) simplifies the symmetry action to be almost on-site: 
\begin{align}
\mathcal{V}^\dagger U_1^\text{frac} \mathcal{V} \ir (-1)^{\nu_L-\sum_{i=1}^{L-1} a_{i,i+1}-\nu_1}\prod_i (-1)^{g_i\hat{\nu}_i}(-1)^{\hat{g}_i}.
\end{align}
This differs from an ordinary onsite $\Z_4$ symmetry only by a gauge-string with charged ends that terminate on the edge. 
As a result, one can write add a trivial $\Z_4$ bulk-paramagnet Hamiltonian in the $\mathcal{V}$-transformed frame:
\begin{align}
\mathcal{V}^\dagger H_{\text{$\Z_4$-PM} }\mathcal{V} &= -\delta\sum_{i=2}^{L-1}\delta_{\hat{\nu}_i,0}\delta_{\hat{g}_i,0} \nonumber\\
&=  -\delta \sum_{i=2}^{L-1}
\sum_{(\nu_i,g_i),(\nu_i',g_i')\in \Z_4}\frac14|(\nu_i,g_i)\>\<(\nu_i',g_i')|
\end{align}
that gaps out all the bulk DOF (except the total gauge flux).
In additional to the topological edge degeneracy, this leaves an additional accidental edge degeneracy that can be partially removed by further adding:
 \begin{align}
 H_\text{edge}=-\mathcal{V}\(\delta\sum_{i={1,L}} (-1)^{g_i+\hat{\nu}_i}\)\mathcal{V}^\dagger,
 \end{align}
  which commutes with $U_1^\text{frac}$ and locks $g_i{1,L}=\hat{\nu}_{1,L}$. Then the remaining DOF are simply $\hat{\nu}_{1,L}$ which naively leaves a fourfold GSD. However, total gauge charge vanishes, $\sum_i \hat{\nu}_i=0$, for any physical states, which (considering the bulk ground-state of $H_{\text{$\Z_4$-PM} }$ has no charge) reduces the physical edge-GSD of gauge invariant states to two (for fixed gauge-flux sector). 
Again, the pumping operation pumps $0d$ $\Gir$-SPTs onto the edge: $\(\mathcal{V}^\dagger U_1^{\text{frac}} \mathcal{V}\)^2 = \(U_1^\text{frac}\)^2 = (-1)^{g_L-g_1}$, so that, just as for the gapless Luttinger-Liquid bulk, there is no local symmetric perturbation one can further reduce this two fold degeneracy since it is protected by a non-trivial local anti-commutation between $U_1$ and (the edge restriction of) $U_2$.

In addition there is a pair global topological supers-election sectors labeled by total gauge flux $\sum_i a_{i,i+1}=0,1$. As mentioned above adding quantum fluctuations in the gauge field would immediately lead to a confined theory in $1d$. However, this construction will prove useful in producing lattice models of bona-fide anomalous topological orders in higher-dimensional examples.

\section{Constructing igSPTs from group cohomology data \label{sec:general}}
We next formally show how this $1d$ example generalizes for emergent anomalies classified by group cohomology. To begin let us fix some notation.

\subsection{Notation \label{sec:notation}}

\paragraph*{General groups -- } For equations involving a general, unspecified group, $G$ we will use multiplicative notation for group operations: $(\cdot): G\times G \mapsto G$, denoted by $g\cdot h$ (we will occasionally omit the $\cdot$ for brevity). 
We denote the representation of symmetry element $g\in G$ on quantum states by $U_g$.

\paragraph*{Abelian groups -- } For specific examples, we focus on groups that are either Abelian or semidirect products of Abelian groups and time-reversal. 
For Abelian groups we will typically use additive notation for the group operation $+:G\times G\mapsto G$, for example we write $\Z_{N} = \{0,1,\dots N-1\}$ with group operation addition modulo $N$, and denote $U(1)$ elements by a phase $\alpha \in [0,2\pi)$.
For direct product of cyclic groups, such as $\Z_{N_1}\times \Z_{N_2}$, denote $N_{12} = \text{GCD}(N_1,N_2)$ where GCD denotes the greatest-common divisor.

For Abelian groups, define the ``vorticity indicators" $v$ as follows: for $g,h\in \Z_N, 0\le g,h \le N-1$ define $v_N(g,h)$ to be $1$ if $g+h\geq N$ and $0$ otherwise, where, here, we regard $g,h$ as integers and and $+$ as integer-addition (i.e. not modulo $N$). Similarly for $U(1)$ define $v_{2\pi}(\alpha,\beta)$ to be $1$ if $\alpha+\beta\geq 2\pi$, and $0$ otherwise.  When clear from context, we will drop the subscripts on $v$.
This function will frequently appear in cocycles, where it can physically be interpreted as detecting vortices on triangular plaquettes.

We denote the Pontryagin dual of a group $G$ by $\widehat{G}\equiv H^1(G,U(1))=\text{Hom}(G,U(1))$, which consists of linear representations of $G$ on $U(1)$, i.e. $\hat{g}\in\widehat{G}$ is a homomorphism $\hat{g}: G\mapsto U(1)$, satisfying property $\hat{g}(g_1)\cdot \hat{g}(g_2) = \hat{g}(g_1\cdot g_2)~\forall g_{1,2}\in G$.
For Abelian groups the Pontryagin dual is given by the group Fourier transform. All the examples we will consider involve the relations: $\widehat{\Z_N}=\Z_N, \widehat{\Z}=U(1)$.
As an example, when $G=\Z_N$, we  have $\widehat{G}=\Z_N$, and an element $\hat{g}\in \widehat{G}$ is the homormphism: 
$\hat{g}(g) = e^{2\pi i \hat{g}g/N}$.
%

\begin{figure}[t]
\begin{centering}
	\includegraphics[width=1\columnwidth]{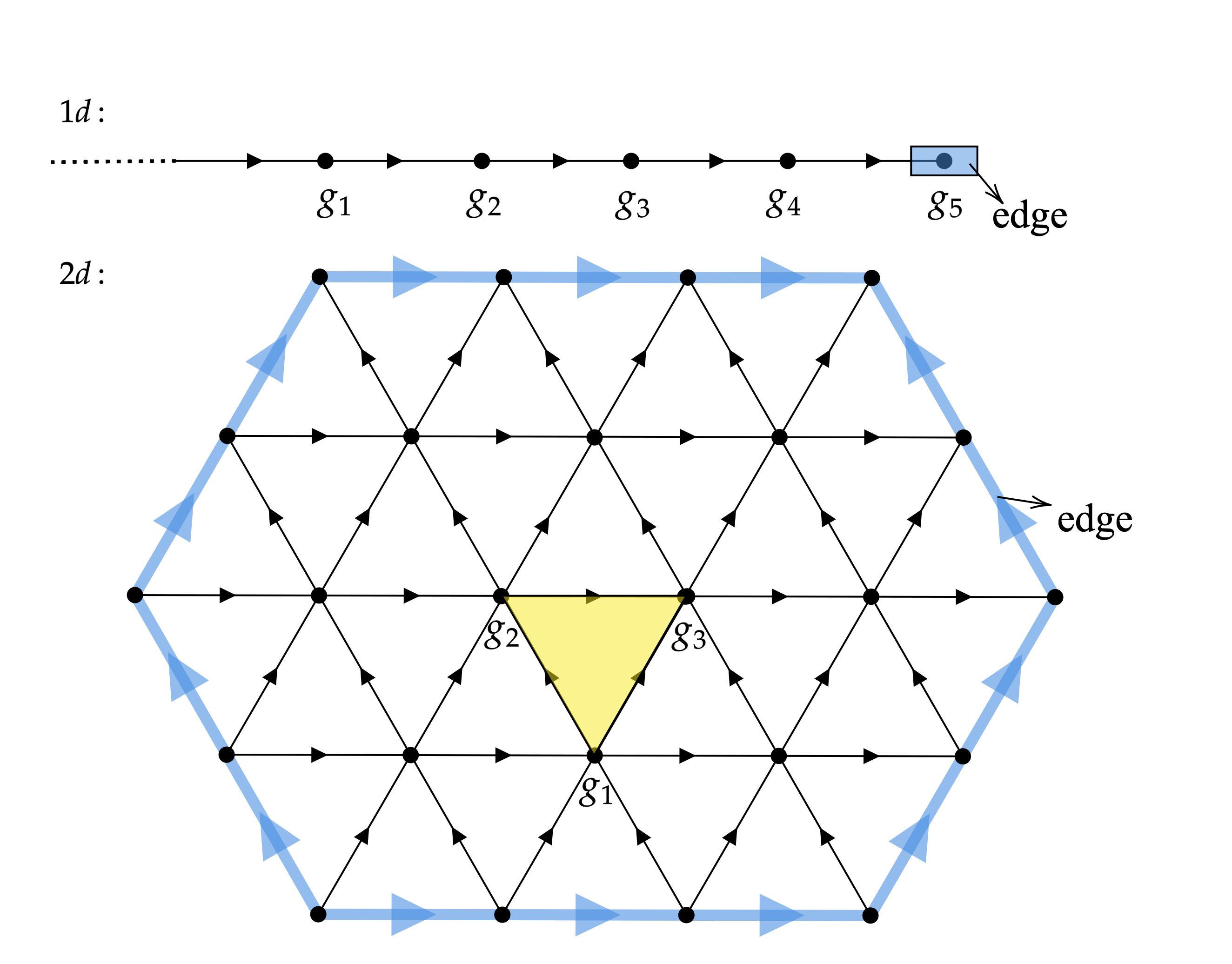}
\end{centering}
	\caption{{\bf Lattice models with anomalous $\Gir$-symmetry from group cohomology -- } are defined on a simplicial complex: a generalized triangulation with orientation $s=\pm$ for each $D$-simplex (link in $1d$, triangle in $2d$, tetrahedron in $3d$, etc...). The on-site Hilbert space for the anomalous model consists of $|g_a\in G\>$ for each vertex $a$. The anomalous  $\Gir$-symmetry action is specified by an element $[\omega_{D+2}]$ of the cohomology group $H_{D+2}(G,U(1))$, it differs from the trivial onsite action by a product of phase factors given by the cocycle $\omega_{D+2}$ and field configuration on each simplex. E.g. in 2d the phase factor associated with the yellow triangle is $\omega_4(g_1^{-1}g_2,g_2^{-1}g_3,g_3^{-1}g^{-1},g)^{-1}$, the overall inverse is due to the clock-wise orientation of the triangle.}
	\label{fig:cohomologymodel}
\end{figure}

\paragraph*{Cup Product -- } We will make use of the cup product of two co-cycles, defined as follow. For $\omega_m\in H^m(G,M)$ and $\eta_n\in H^n(G,N)$, the cup product of them is an element of $H^{m+n}(G,M\otimes_{\Z} N)$ defined as:
\begin{align}
	\omega_m\cup\eta_n&(g_1,\cdots,g_{m+n})\nonumber\\
	&\equiv \omega_m(g_1,\cdots,g_m)\otimes_{\Z} \eta(g_{m+1},\cdots,g_{m+n}). 
\end{align}
We will only be concerned with the case where $M$, $N$ are Pontryagin dual of each other: $M=\widehat{N}$, in which case we can identify the tensor product $M\otimes_{\Z} N$ as $U(1)$ by identifying the element $m\otimes_{\Z}n$ as $m(n)\in U(1)$, this identification will be made implicitly throughout this paper. 
For example when $M=\widehat{N}=\Z_N$, both $\omega_m$ and $\eta_n$ are integers $ \in \{0,1,\dots N-1\}$, and accroding to our definition we have: $\omega_m\cup\eta_n(g_1\dots g_{m+n})  = \exp\[\frac{2\pi i}{N} \omega_m(g_1,\dots g_n)\eta_n(g_{m+1},\dots g_{m+n})\]$.
Therefore $\omega_m\cup \eta_n$ is a $m+n$ cocycle with $U(1)$ coefficients.

\paragraph*{$G$-rotor models -- } We will consider quantum lattice models with onsite Hilbert space built from quantum rotors, i.e. $\mathcal{H}_i=\text{Span}\{|g_i\>, g_i\in G\}$ for each site $i$ in the lattice. With a slight abuse of notation, we define the operator $g$ which is diagonal in the $g$-basis with eigenvalue $g$, and the dual operators $\hat{g}$ whose exponential raise the eigenvalues of $g$, for example:
\begin{align}
\begin{cases}
	e^{\frac{2\pi i}{N}j\hat{g}}|g\>=|g+j\>,&~~G=\Z_N,\\
	e^{i\alpha\hat{g}}|g\>=|g+\alpha\>,&~~G=U(1)
\end{cases}.
\end{align}
We will refer to such DOF as $G$-rotors.
 Note the eigenvalues of the dual operators take values in the Pontryagin dual group, thus the notation here is consistent with the notation for the Pontryagin dual.



\subsection{Review: Anomalous lattice models from cohomology data}
We will exploit several constructions from the group cohomology (partial) classification of bosonic SPT phases in $(D+1)d$ with symmetry $G$.
Refs.~\cite{chen2013symmetry,chen2014symmetry} provided a general lattice model for the anomalous $Dd$ surface of a $(D+1)d$ gapped SPT with symmetry $G$. These models are $G$-rotor models on a lattice with simplicial structure, i.e. a suitable triangulation of space with sites indexed by $a\in \Z$, and a $G$-rotor $|g_a\>$ at each site $a$. 
The order of site labels induces an orientation to edges, which is conventionally drawn as arrows pointing from smaller to larger site number. 
We will focus on cases where the simplicial structure forms a regular lattice: a chain of sites in $1d$, a triangular lattice in $2d$, or cubic lattice of face-sharing tetrahedra in $3d$. We label elementary $D$-simplices (links in $1d$, triangles in $2d$, or tetrahedrons in $3d$) by letters $i,j,k\cdots$, and vertices by $a,b,c\cdots$. The $D+1$ vertices of a fixed simplex $i$ are labeled as $i_1,i_2,\cdots, i_{D+1}$. Each simplex is assigned an orientation $s_i=\pm$. For example, in 2d with the simplicial structure shown in Fig.\ref{fig:cohomologymodel}, upwards pointing triangles have $s=1$ and downwards pointing triangles have $s=-1$.


An anomalous symmetry action is locally generated but not strictly onsite. It differs from the ordinary onsite $G$-symmetry action $|g_a\>\rightarrow |gg_a\>$ by non-onsite phases that depend on topological defect configurations on each simplex $i$.
Specifically, for $g\in G$, the anomalous symmetry can be chosen (up to a symmetric finite-depth local unitary transformation) to act as~\cite{chen2013symmetry,chen2014symmetry}:
\begin{align}
&U^A_g| \{g_a\}\> = \nonumber\\&\prod_{i} \[\omega_{D+2}(g^{-1}_{i_1}g^{}_{i_2	},\dots, g^{-1}_{i_{D}}g_{i_{D+1}}^{},g_{i_{D+1}}^{-1}g^{-1},g)\]^{s_i} |\{gg_a\}\>
\label{eq:UAgeneral}
\end{align}
where $\omega_{D+2}\in Z^{D+2}(G,U(1))$ is a representative co-cycle for the anomaly, the product is over all simplices. 
For group elements that act anti-unitarily (i.e. time-reversal), the phases above should additionally be complex conjugated.

The general form of this equation is that the ordinary onsite symmetry action $g_a\rightarrow g g_a$ is modified by phases, $\omega_{D+2}$, that depend on symmetry domain wall configurations $g_{i_{\alpha}}^{-1}g_{i_{\alpha+1}}$ on the links $(i_\alpha i_{\alpha+1})$ of each simplex $i$.
Physically, these phases give rise to destructive quantum interference between different domain wall rearrangements~\cite{kawagoe2021anomalies}, that prevent the system from reaching a gapped symmetric state (which would be a quantum superposition of all possible domain wall configurations).
Group cohomology implicitly assumes both a tensor product structure (i.e. the anomaly is in the symmetry action not the Hilbert space) and that all topological defects in symmetry breaking order are gappable. These assumptions are known to breakdown for selected ``beyond-group-cohomology" SPTs that are characterized by symmetry defects that carry ungappable chiral modes or unpaired Majorana zero modes. In the following we focus on the group-cohomology anomalies, and comment only briefly at the end about possible generalizations to beyond cohmology igSPTs.


\subsection{\label{la} Lifting the anomaly}
Starting from an anomalous $\Gir$-rotor model, we now show a general prescription for obtaining an lattice model with an extended and anomaly-free onsite symmetry $\Guv$ in the UV, and emergent $\Gir$-anomaly in the IR. 
The idea will be to introduce additional $\N$-rotors for some appropriate Abelian group $\N$, $n_i$ (with dual operators $\hat{n}_i$), on each plaquette, such that the central extension:
\begin{align}\label{eq:extension}
0\rightarrow \N \xhookrightarrow{~~\iota~~} \Guv \xrightarrow{\Guv/\N} \Gir \rightarrow 0
\end{align}
lifts the $\Gir$ anomaly in $\Guv$, meaning that a nontrivial cocycle $[\omega]\in H^{D+2}(\Gir,U(1))$ becomes trivial when viewed as a cocycle in $H^{D+2}(\Guv,U(1))$. 
Extensions of of $\Gir$ by $\N$ are classified by $H^2(G,\N)$: denoting elements of the extended $\Guv$ symmetry group by $(n\in\N,g\in G)$, the extended group corresponding to an element $e_2\in H^2(G,\N)$ has group operation: $(n_1,g_1)\cdot(n_2,g_2) = \(n_1+n_2+e_2(g_1,g_2),g_1\cdot g_2\)$, and the associativity of this group operation, $\gamma_1\cdot(\gamma_2\cdot\gamma_3) = (\gamma_1\cdot \gamma_2)\cdot\gamma_3$ for $\gamma_i\in\Guv$, is guaranteed by the cocycle condition $d e_2=1$.

It was shown that group-cohomology anomalies can be lifted by such a central extension~\cite{tachikawa2020gauging,wang2018symmetric} including a constructive proof in~\cite{tachikawa2020gauging}. 
Moreover, it was shown in~\cite{tachikawa2020gauging} that existence of such a group extension guarantees the $D+2$ cocycle $\omega_{D+2}$ that specifies the $Dd$ $\Gir$-anomaly has a representative in the decomposed form:
\begin{align}
\omega_{D+2}= b_D\cup e_2,
\label{eq:bcupe}
\end{align}
with $b_D\in Z^D(G,\widehat{\N}), e_2\in Z^2(G,\N)$. 
We will show below that this decomposition implies a bulk-boundary correspondence for igSPTs tying the emergent group-cohomology anomaly to the presence of SPT edge states. In general, $b_D$ will characterize the symmetry transformation of $\N$ gauge fluxes when one gauges the extending $\N$ symmetry which we will argue guarantees non-trivial igSPT edge states. In many cases, $b_D$ will manifest as a lower-dimensional SPT pumping symmetry that generalizes the $1d$ $\Guv=\Z_4$ case discussed above, and gives a physically transparent mechanism for igSPT edge states.



Second, since $b_D$ has coefficients that are dual to those of $e_2$, referring to the discussion in Section~\ref{sec:notation} above, their cup product $[b_D\cup e_2]$ can be identified with an element of $H^{D+2}(G,U(1))$ by ``feeding" the $e_2(g_{D+1},g_{D+2})\in \N$ output into $b_D(g_1,\dots g_D)\in \widehat{\N}=\text{Hom}(\N, U(1))$.
As a concrete example, the $1d$ $\Guv=\Z_4$ igSPT discussed above cocycle $\omega_3(g,h,k) = (-1)^{ghk}$ which we can decompose into $b_1\cup e_2$ with $b_1(a)=a\in \widehat{\N}=\Z_2$ and $e_2(g,h) = g\cdot h\in \N=\Z_2$.

\begin{figure}[t]
	\begin{centering}
		\includegraphics[width=1.1\columnwidth]{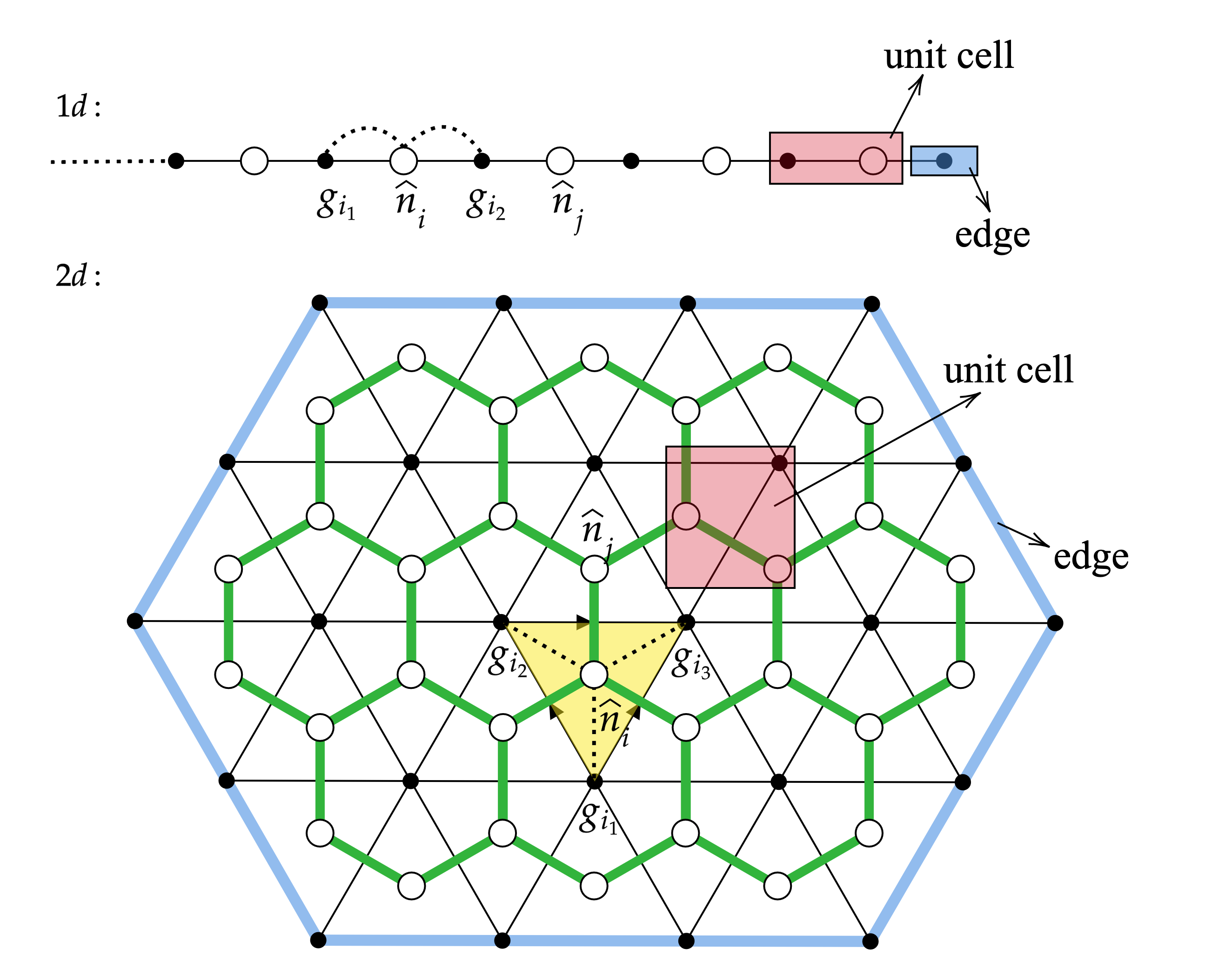}
	\end{centering}
		\caption{{\bf The igSPT lattice.} The $\N$ rotors $\hat{n}_i$ live on the sites of the dual lattice(white circles), or equivalently, the center of every simplex of the simplicial lattice. An $\N$ rotor in a simplex is grouped with the vertex of that simplex with largest site label to form a unit cell. The $\Guv$ symmetry action becomes onsite with this definition of a unit cell. The dashed lines in the figure indicate the locking of $\N$ rotors to $\Gir$ rotors in the IR.}
		\label{fig:uv_lattice}
\end{figure}

\begin{widetext}
\subsection{Extending and onsiteing the symmetry} 
The decomposition structure permits us to define an ordinary onsite $\Guv$ symmetry action that can be reduced to the anomalous $\Gir$ action in the IR. We introduce an $\N$-rotor $n_i\in \N$ for each simplex $i$. We define a single unit cell of the model as follows: associate each simplex, $i$, with the vertex $a_i\in i$ that has the largest site number in that simplex $n_i$. This results in groups of $D$ simplices for each vertex. Define a single unit cell of the lattice model to consist of associated $D$ $\N$-rotors and one $\Gir$ rotor. Then, we may  define the $\Guv$ symmetry action as:
\begin{align}\label{eq:ugos}
	U^\text{os}_{g}|\{\hat{n}_i\},\{g_i\}\>=\prod_i \hat{n}_i\otimes_{\Z}e_2(g_{i_{D+1}}^{-1}g^{-1},g)|\{\hat{n}_i\},\{gg_a\}\>.
\end{align}
which is a tensor products of local unitaries acting only on a single unit-cell of the lattice model, i.e. is an on-site symmetry action. See Fig.\ref{fig:uv_lattice} for an illustration of 1d and 2d lattices.


Then, to imprint the IR anomaly, we introduce an interaction:
\begin{align}\label{eq:HDelta}
	H_\Delta = -\Delta \sum_{i}\delta_{\hat{n}_i,b^{s_i}_D(g_{i_1}^{-1}g_{i_2}^{},\dots, g_{i_{D}}^{-1}g_{i_{D+1}}^{})},
\end{align}
In the ground-space of $H_\Delta$, $\hat{n}_i$ gets locked to $b^{s_i}_D(g_{i_1}^{-1}g_{i_2}^{},\dots,  g_{i_{D}}^{-1}g_{i_{D+1}}^{})$, which we denote as: $\hat{n}_i\ir b_D$, and we obtain the low energy anomalous action 
\begin{align}
    U_g^{\text{os}}|\{g_a\}\rangle&\ir\prod_{i}b_D^{s_i}(g^{-1}_{i_1}g_{i_2},g^{-1}_{i_2}g_{i_3},\cdots,g_{i_{D}}^{-1}g_{i_{D+1}})\otimes_{\mathbb{Z}}e_2(g^{-1}_{i_{D+1}}g^{-1},g)|\{gg_a\}\rangle\\
    &=\prod_{i}\omega_{D+2}^{s_i}(g^{-1}_{i_1}g^{}_{i_2},g^{-1}_{i_2}g_{i_3}^{},\cdots,g^{-1}_{i_{D+1}}g^{-1},g)|\{gg_a\}\rangle\\
    &=U^A_g|\{g_a\}\>.
\end{align}
\end{widetext}

The result is that, at energy scales $\ll \Delta$ (henceforth referred to as ``the IR"), the extending $\N$-rotors are gapped out by $H_\Delta$. Importantly, it is possible to fully gap all the $\N$ rotors in this way, even in open geometries with boundaries. Specifically, with open boundaries one needs to add additional terms $-\Delta \sum_{i\in  S} \delta_{\hat{n}_i,0}$ where $0$ is the identity element of $\widehat{\N}$ and $S$ is the set of boundary sites that are not the terminal site of some complete simplex, which fully gaps all $\N$-rotor DOF not involved in $H_\Delta$.
Importantly, this gapping is done in such a way as imposes a non-trivial entanglement between the $\Gir$ and $\widehat{N}$ rotors that imprints an anomalous $\Gir$-symmetry action on the $\Gir$ degrees of freedom in the IR.

The unitaries $U_g^{os}$ do not yet form a group, because the multiplication is not closed:
\begin{align}
	U^{\text{os}}_{g_1}U^{\text{os}}_{g_2}=\left(\prod_i \hat{n}_i\otimes_{\Z}e_2(g_1,g_2)\right)U^{\text{os}}_{g_1g_2}.
\end{align}
To obtain a closed group, we enlarge our set of symmetry actions by defining a symmetry action for each pair $(n,g),n\in \N,g\in G$ as 
\begin{align}
	U^{\text{os}}_{(n,g)}:=\left(\prod_i \hat{n}_i\otimes_{\mathbb{Z}}n\right) U_g^\text{os}.
\end{align}
The original symmetry actions \eqref{eq:ugos} are identified with $U^\text{os}_{0,g}$. It's easy to check that the Hamiltonian \eqref{eq:HDelta} is symmetric under $U^{\text{os}}_{(n,g)}$ actions. The multiplication rule is now
\begin{align}
    U^{\text{os}}_{(n_1,g_1)}U^{\text{os}}_{(n_2,g_2)}=U^{\text{os}}_{(n_1+ n_2+ e_2(g_1,g_2),g_1g_2)},
\end{align}
which is the group law of the extended group $\Guv$. In conclusion, we have constructed a lattice model with onsite $\Guv$-symmetry in the UV, and emergent $\Gir=\Guv/\N$-anomaly in the IR.

\subsection{Edge states: perspective from gauging $\N$}\label{sec:edge states}
How is the emergent anomaly related to the presence of SPT-edge states of the igSPT? To answer this question, consider the gedanken experiment of gauging the extending $\N$ symmetry of the igSPT model described above. 

\subsubsection{General picture}
Gauging $\N$ leaves an anomalous $\Gir$ symmetry enriched topological (SET) order. Due to this anomaly there must be no way to confine this SET order without either closing the gap to the $\N$ gauge charges (i.e. driving a phase transition out of the igSPT phase), or breaking the $\Gir$ symmetry. In particular, it must not be possible to condense $\N$-flux excitations to result in a trivial gapped, confined, and $\Gir$ symmetric theory. 
In general there can be three different types of obstacles to condensing the $\N$ flux excitations. i) The $\N$-flux may carry gapless modes, for example with gaplessness protected by a $\Gir$ anomaly. In this case, their condensation would result in a gapless state. ii) The $\N$ flux may carry a fractional $\Gir$ charge which cannot be screened by any local excitations. Here, condensing the $\N$ flux would necessarily break the $\Gir$ symmetry. iii) In $2d$ and $3d$ the $\N$ flux could also have non-trivial self-statistics, such that it could not be directly condensed. 

In our construction, the $\N$ flux is always a boson, but may have non-trivial $\Gir$ symmetry properties, so only possibilities i) and ii) are realized in our models.
We note that $\N$-fluxes are co-dimension two [$(D-2)$-dimensional] objects: instantons in $1d$, point-particles in $2d$, and line or loop excitations in $3d$.
In Appendix~\ref{app:gauged}, we show that the $\Gir$ symmetry properties of fluxes are characterized by $[b_D]\in H^D(\Gir,\widehat{N})$ which arises in the decomposition of the anomaly cocycle $\omega_{D+2}=b_D\cup e_2$.
Gapless fluxes [case i) above] corresponds to models where $[b_D\otimes_{\Z} n] \in H^{D}(\Gir,U(1))$ is non-trivial for some $n\in \N$. In this case, we will show below that the $n$-flux carries the gapless edge states of a $\Gir$-SPT, and $[b_D\otimes_\Z n]\in H^{D}(\Gir,U(1))$ characterizes the anomaly of those SPT edge states. 
Fractional $\Gir$-charged fluxes [case ii) above] correspond to situations where $[b_D\otimes_\Z n]=[0]$ is trivial $\forall n\in \N$, so that there is no obstruction to having a gapped $\N$-flux, but where $[b_D]\in H^D(\Gir,\N)$ is non-trivial. For example $H^2(\Z_2,U(1))=0$ (there are no projective representations of $\Z_2$) whereas $H^2(\Z_2,\Z_2)=\Z_2$ (it is possible to define a fractional half-charge for a $\Z_2$-gauge flux). We note that for an igSPT with nontrival emergent anomaly, the cocycles $b_D,e_2$ that constitute the anomaly cocycle $\omega_{D+2}$ must be both nontrival. For instance, if $b_D$ is trivial, i.e. $b_D=d c_{D-1}$, then $d(c_{D-1}\cup e_2)=dc_{D-1}\cup e_2+(-1)^{D-1}c_{D-1}\cup de_2=b_D\cup e_2=\omega_{D+2}$, making $\omega_{D+2}$ trivial. Therefore given an igSPT with nontrivial emergent anomaly, it is guaranteed that the fluxes and charges will have nontrivial $\Gir$ symmetry properties. 

The non-trivial properties of $\N$-gauge fluxes are directly related to the local action of symmetry on the edge of the igSPT. Namely, consider an interface between a trivial $\N$ gauge theory, ``the vaccum" (with no emergent anomalies) and the $\N$-gauged igSPT (with emergent $\Gir$ anomaly). Then 
the $\N$ flux transforms trivially in the vacuum. If we drag the $\N$ flux into the system, it must change its $\Gir$ symmetry action to $[b_D]\neq [0]$. In order that the overall system obey an ordinary linear representation of $\Gir$, there must be a compensating codimension two excitation that transforms as $-[b_D]$. Since the flux's symmetry property changes immediately upon passing through the vacuum--igSPT interface, this compensating excitation must reside on the igSPT edge. Hence, the igSPT faces the same obstacle to forming a trivial, $\Gir$-symmetric, confined state as the bulk vortex. Pulling this behavior back to the ungauged theory (e.g. by fixing to a flat $\N$-gauge configuration), this implies an obstacle to forming a trivial $\Guv$-symmetric edge of the igSPT -- establishing a bulk-boundary-correspondence between the emergent anomaly and the edge states.

\subsubsection{Special Cases: Edge states from SPT pumping symmetry}
Having elucidated a general bulk-boundary correspondence from the $\N$-gauging arguments presented above, we now restrict to the special case where $[b_D\otimes_\Z n]$ is a non-trivial element of $H^{D}(\Gir,U(1))$ for some $n\in \N$. In these cases, it is possible to understand the igSPT edge states from a physically intuitive picture of an emergent lower-dimensional SPT-pumping symmetry.
Specifically, consider the extending symmetry operation(s):
\begin{align}
&U^\text{os}_{(n,0)}=\prod_i\hat{n}_i\otimes_{\Z} n\nonumber \\
&\ir \prod_{i} b_D^s(g^{-1}_{i_1}g_{i_2}^{},g^{-1}_{i_2}g_{i_3}^{},\cdots,g_{i_{D}}^{-1}g_{i_{D+1}}^{})\otimes_{\mathbb{Z}}n.
\label{eq:sptpumping}
\end{align}
Since $b_D(g_1,\dots g_D)\in \widehat{\N}$, for fixed $n\in \N$, $[b_D(\circ,\dots ,\circ)[n]]\in H^{D}(\Gir,U(1))$ can be identified with a $(D-1)d$ $\Gir$-SPT. This suggests a close relation betwteen the decomposition $\omega=b\cup e$ and the SPT pumping action noted in the the $1d$ example.
Indeed, in Appendix~\ref{app:sptpump}, we show that \eqref{eq:sptpumping} is precisely a unitary that converts a trivial product state of the edge (with symmetry action $U^\text{trivial}$)  into an edge $\Gir$-SPT (with symmetry action given in \eqref{eq:modsymmetryedge}). More generally starting from an SPT with invariant $[\nu] \in H^{D}(\Gir,U(1))$, the pumping operation $U_n^{\text{os}}$ converts this into a state with SPT invariant $[\nu+b_D\otimes_{\Z}[n]]$.
Since these pumping operations are symmetries of any $\Guv$-symmetric Hamiltonian, any symmetry-respecting edge must be invariant under toggling its $\Gir$-SPT invariant. This SPT-pumping symmetry forbids the igSPT edge from being trivially gapped without breaking symmetry.  Heuristically, the SPT pumping symmetry forces the edge of the igSPT to sit at a self-plural\footnote{generalizing the notion of self-dual, self-trial, etc... to arbitrary number} (multi)critical point between the different possible $(D-1)d$ $\Gir$-SPT phases.
When this self-plural critical point is a continuous phase transition, this results in symmetry-protected gapless edges. If this represents a discontinuous, first order, transition, then this results in an edge that spontaneously breaks the $\N$-symmetry. A third possibility that arises only in $D\geq 3$ is that this self-dual point could be satisfied by a gapped-symmetry SET order (we construct an example in Sec.~\ref{sec:3d}).
Formally, the SPT-pumping symmetry forces the edge to have an anomaly that has an LSM-type obstruction to forming a trivial, gapped, symmetric state. Specifically, the igSPT edge with symmetry $\Guv$ is the same as the anomalous edge of a gapped $Dd$ SPT with a \emph{different} symmetry $\Gir\times \N$ in which $\N$ symmetry-breaking domain walls are decorated~\cite{chen2014symmetry} by $(D-1)d$ $\Gir$ SPTs with cocycle $b_D$. We remark that a related construction was used in~\cite{tsui2015quantum} to establish anomaly constraints on self-dual transitions between SPT and trivial phases.

We emphasize the fact that the $\Guv$-igSPT edge states have the same anomaly as a $\Gir\times \N$ gapped SPT does not imply that there is a gapped $\Guv$-SPT with the same edge states since $\Guv\neq \Gir\times \N$ for any non-trivial extension. As we prove in Appendix.~\ref{app:intrinsic}, and illustrate through specific examples below, the non-trivial extension required to lift the anomaly always forms an insurmountable obstacle to creating a gapped SPT with the igSPT edge properties -- i.e. the igSPT models with emergent group-cohomology anomalies are indeed \emph{intrinsically} gapless.

We also remark that the SPT pumping mechanism for igSPT edge states can be directly related to the anomaly-cancellation argument used by TVV~\cite{thorngren2021intrinsically} to deduce the existence of edge states. In Appendix~\ref{app:anomalycancel}, we explain how the edge-SPT-pumping picture reproduces the edge-anomaly-cancellation picture while revealing additional structure about the edge-anomaly.

\color{black}

\begin{table*}[]
\setlength{\tabcolsep}{6pt}
\renewcommand{\arraystretch}{2.5}

\begin{tabular}{C{2.2in}|C{0.8in}|C{1.4in}|C{2.0in}}

{\bf Symmetry group extension} $\N\rightarrow \Guv\rightarrow \Gir$ &  { \bf Anomaly (sub)group} $ H^{D+2}(G,U(1))$ & {\bf Representative Cocycle} $\omega_{D+2}(g,h,k,\dots)$ & {\bf SPT-pumping/Symmetry fractionalization} 
\\
\hline
\multicolumn{4}{c}{\bf $1d$ igSPTs}
\\
\hline
$\Z_N\rightarrow \Z_{N^2}\rightarrow \Z_N$ & $\Z_N$ & 
 $e^{\frac{2\pi i}{N}g\cdot v_N(h,k)}$ & $U_N$ pumps a $\Z_{N}$ rep.
\\
$\Z\rightarrow \mathbb{R}\rightarrow U(1) $ & $\Z$ & 
 $e^{i g \cdot v_{2\pi}(h ,k)}$ & $U_{2\pi}$ pumps a $U(1)$ rep.
\\
${\Z_{N_{12}}\rightarrow \Z_{N_1}}\times \Z_{N_2N_{12}}\rightarrow \Z_{N_1}\times\Z_{N_2}$ & $\Z_{N_{12}}$ & 
 $e^{\frac{2\pi i}{N_{12}}g^1\cdot v_{N_2}(h^2,k^2)}$ & $U_{(0,N_2)}$ pumps a $\Z_{N_1}$ rep.
\\
$\Z_{N_{123}} \rightarrow { \Z_{N_2}\ltimes (\Z_{N_1}\times \Z_{N_3}\times \Z_{N_{123}})} \rightarrow \Z_{N_1}\times\Z_{N_2}\times\Z_{N_3}$ 
& $\Z_{N_{123}}$ & $e^{\frac{2\pi i}{N_{123}}g^1h^2k^3}$ & $\scriptstyle \[U_{(0,0,N_2)},U_{(0,0,N_3)}\]_{\mathfrak{g}} $ pumps a $\Z_{N_1}$ rep. 
\\
$\Z\rightarrow \mathbb{R}\ltimes \Z_2^T\rightarrow U(1)\ltimes \Z_2^T$ & $\Z_2$ &$e^{ig^1h^2k^2}$ &$\T^2$ pumps a $U(1)$ rep. 
\\
\hline
\multicolumn{4}{c}{\bf $2d$ igSPTs}
\\
\hline
$ \(\Z_{2}\)_\text{boson}\rightarrow \Z_{4}^T \rightarrow \Z_{2}^T$ & $\Z_2$ 
& $(-1)^{ghkl}$ & $\T^2 $ pumps a Haldane chain.
 \\
$\Z_{N_{12}}\rightarrow \Z_{N_1}\times \Z_{N_2N_{12}} \rightarrow \Z_{N_1}\times \Z_{N_2}$ & $\Z_{N_{12}}^2$ &
 $e^{\frac{2\pi i}{N_{12}}g^1h^2 v_{N_2}(k^2,l^2)}$
  &$U_{(0,N_2)}$ pumps a  $1d~\Z_{N_1}\times \Z_{N_2}$-SPT
\\
$\Z_{N_{12}}\rightarrow \Z_{N_1}\ltimes \Z_{N_2N_{12}} \rightarrow \Z_{N_1}\times \Z_{N_2}$ & $\Z_{N_{12}}^2$ &
$e^{\frac{2\pi i}{N_{12}}g^1h^1k^1l^2}$ & 
{edge $\Z_{N_1}$ domain walls have fractional $\Z_{N_1}$-charges}
\\
$\Z_2 \rightarrow \frac{\mathbb{R}}{4\pi \Z}\times \Z_2^T\rightarrow \frac{\mathbb{R}}{2\pi \Z}\times \Z_2^T$ & 
$\Z_2^2$  &
$\begin{cases}
(-1)^{v_{2\pi}(g^1,h^1)v_{2\pi}(k^1,l^1)}\\
(-1)^{g^2h^2v_{2\pi}(k^1,l^1)}
\end{cases}
$ & 
$U_{2\pi} \text{ pumps a 1d } U(1)\times \Z_2^T \text{-SPT}$
\\
\hline
\multicolumn{4}{c}{\bf $3d$ igSPTs}
\\
\hline
$\Z_N\rightarrow \Z_{N^2}\rightarrow \Z_N$ & $\Z_N$ &  $e^{\frac{2\pi i}{N}g\cdot v_N(h,k)\cdot v_N(l,m)}$ &
$U_N$ pumps a 2d $\Z_N$-SPT
\\
\end{tabular}
\caption{{\bf  Examles of igSPTs from group-cohomology -- } 
Topological data for igSPTs in $1d$, $2d$, and $3d$ with Abelian symmetries and/or time-reversal. An anomalous IR $\Gir$ symmetry emerges from a UV symmetry $\Guv$  symmetry which is a central extension of $\Gir$ by $\N$, where the $\N$ DOF are gapped out.
The data is listed only for ``root" phases (from which other phases can be trivially obtained by stacking). For product groups we only list mixed anomalies that do not follow from previous table entries. E.g. for the third row ($1d$ $\Gir=\Z_{N_1}\times\Z_{N_2}$) we do not repeat the pure $\Z_{N_1}$ or $\Z_{N_2}$ anomalies described in the first row).
For product groups $G_1\times G_2 \ltimes G_3 \dots$ we denote elements by $(g^1,g^2,g^3,\dots)$.
The explicit cocyles $\omega_{D+2}\in Z^{D+2}(G,U(1))$ represent the anomalies. For on-site symmetries, we can read off the the cocycle from the higher-d SPT response theory by replacing $A^i\rightarrow g^i$ and $dA \rightarrow v(g,h)$.
We list the SPT pumping or the symmetry fractionalization that protect the non-trivial edge for each igSPT. 
For unitaries $U,V$, $[U,V]_{\mathfrak{g}} = UVU^\dagger V^\dagger$ denotes the group commutator.
\vspace{2in}
}
\label{tab:results}
\end{table*}

\subsection{Stability and bulk phase diagram} So far, we have not fully specified the bulk igSPT Hamiltonian, which is necessary to describe a particular igSPT state, or explore the bulk and boundary phase diagram of the system with the emergent anomaly.
To remedy this we should introduce a local $\Guv$ symmetric Hamiltonian, $H=H_\text{IR}+H_\Delta$ such that: i) $H_\text{IR}$ is $\Guv$ symmetric, and ii) $H_\text{IR}$ is sufficiently weaker than $H_\Delta$ that it does not close the gap that imprints the IR anomaly.
As argued in the $1d$ igSPT discussion above, the latter assumption implies that the anomalous IR symmetry action is adiabatically connected, by a  $\Guv$-symmetric finite-depth local unitary (FDLU) circuit, to the ideal zero-correlation length one in \eqref{eq:UAgeneral}, and similarly for the lower-dimensional SPT pumping action of $\N$-symmetries.
Hence, under these conditions, the edge states will be stable, and are at most spread out by virtual quantum dressing within the Lieb-Robinson lengthscale of this FDLU.
Therefore, as is common in gapped topological systems, for generic igSPTs, the edge states will not be confined strictly to the edge of the system, but will have an exponentially decaying envelope $\sim e^{-x/\xi}$ with distance $x$ from the edge, where $\xi$ is the (finite) correlation length induced by the gap to $\N$-rotors.
Throughout most of the remainder of the paper, we will not attempt to solve the (generally hard) problem of deciding what happens for a particular $H_\text{IR}$, but rather will deduce sharp, topological anomaly based constraints on the possible outcomes.

\paragraph*{Examples -- }
Table~\ref{tab:results} compiles a summary of igSPTs in various dimensions and with various symmetry groups obtained from this general framework. In the remainder of the main text we focus on a single representative example in $2d$ and $3d$ respectively. 
In higher dimensions, it is possible to have stable QSET phases consistent with the emergent $\Gir$ anomalies. Straightforward generalizations of the gauging-by-fractionalization procedure outlined for $1d$ above yield general formulas for solvable lattice models of these QSETs. Rather than displaying these results for general unspecified symmetry groups and dimensions, which follows closely the results of~\cite{wang2018symmetric} and results in somewhat unwieldy notation we instead apply this framework to specific examples $2d$ and $3d$ in the following sections.

\section{A $2d$ Time-Reversal Symmetric igSPT \label{sec:2d}}
We now apply this general framework to construct a $2d$ time-reversal symmetry igSPT, which has IR symmetry $\Gir=\Z_2^T$ where the $T$ superscript indicates that the symmetry operation is represented as an antiunitary operator.
Following the group-cohomology cook book, one can construct an anomalous action of $\Z_2^T$ on a lattice model with spin-1/2 DOF $\sigma^z_a = (-1)^{g_a}$ on site $a$ of a $2d$ triangular lattice, $\Sigma_2$, with a simplicial structure. There is a single anomalous $2d$ action of time-reversal symmetry, $H^4(\Z_2^T,U_T(1))=\Z_2$ which can be implemented as:
\begin{align}
\T^A = \prod_{i}(-1)^{g_{i_1i_2}g_{i_2i_3}g_{i_3}}\prod_a (-1)^{\hat{g}_a} K
\label{eq:TA}
\end{align}
where we have defined the shorthand $g_{i_1i_2} = g_{i_1}-g_{i_2}$, $K$ denotes global complex conjugation in the $g_a$ eigenbasis.
We can read from this anomalous action $\N=\widehat{\N}=\Z_2$, $b_1(g)=g$ and $e_2(g_1,g_2)=g_1 g_2$. To onsite this symmetry, we introduce additional $\widehat{N}=\Z_2$ rotors, $\hat{n}_i\in \{0,1\}$ on each triangle, 
with onsite symmetry action:
\begin{align}
\T^\text{os} = \prod_{i} (-1)^{\hat{n}_ig_{i_3}}\prod_a(-1)^{\hat{g}_a}K.
\end{align}
where $\hat{g},n$ denote the conjugate operators to $g,\hat{n}$ respectively, and $K$ is now global complex conjugation in the $g,n$-eigenbasis.
To imprint the anomaly in the IR, we then energetically lock $\hat{n}_i\ir g_{i_1i_2}g_{i_2i_3}$ on each triangle $i$ via the Hamiltonian:
\begin{align}
H_\Delta = -\Delta\sum_{i} \delta_{\hat{n}_i,g_{i_1i_2}g_{i_2i_3}}.
\end{align}
In the ground-space of $H_\Delta$, we then have an emergent time-reversal anomaly: $\T^\text{os}\ir \T^A$.

Similar to $1d$, we see that the extension which lifts the anomaly is $\N=\Z_2$, and $\Guv=\Z_4^T = \{1,\T,\T^2,\T^3\}$ where $\T^{2n+1}$ are antiunitary, $\T^{2n}$ are unitary, and $\T^4=1$, the extending group $\N$ is the normal subgroup $\{1,\T^2\}$. 
This clearly yields an intrinsically gapless SPT, as there are no gapped bosonic SPT phases with $\Z_4^T$ symmetry in $2d$. Noting that, microscopically, $\T^2 = (-1)^{\sum_i\hat{n}_i}$, we see that physically, we have to extending the Hilbert space to include Kramers doublet bosons~\footnote{We will later argue that extending by Kramers doublet fermions (such that $\T^2 = (-1)^{N_F}$ where $N_F$ is the fermion number) would not lift this anomaly, but rather would instead lift a more complicated beyond-cohomology $\Z_2^T$ anomaly.}.

For completeness, we note a unitary $V$ that maps the on-site symmetry to the anomalous one, and equivalently maps $H_\Delta$ to a simple paramagnet for the extending spins, $n$, $V^\dagger H_\Delta V = -\Delta\sum_{i}\delta_{\hat{n}_i,0}$:
\begin{align}
V = \prod_{i} (-1)^{g_{i_1i_2}g_{i_2i_3}n_i}.
\end{align}
This mapping makes it manifest that $H_\Delta$ (augmented with $-\Delta \sum_i \delta_{\hat{n}_i,0}$ terms for incomplete boundary simplices) fully gaps out the extending DOF, even in the presence of a boundary. As in $1d$, this means that the $\N$-rotors are fully gapped in both the bulk and the edge, and there are no local $\N$-charged operators (i.e. Kramers doublet boson operators) that create low energy excitations.

\subsection{Edge states} As in $1d$, we can expose the structure of the edge states by considering a composition of symmetry operations that produces an $\N$-element microscopically, and would act trivially on a system with non-anomalous IR $\Z_2^T$-symmetry. Here there is a single pumping operator given by $\T^2$ which gives a unitary pumping operation:
\begin{align}
\T^2 \ir \prod_{i\in \Sigma_2} (-1)^{g_{i_1i_2}g_{i_2i_3}} = \prod_{i\in \d\Sigma_2} (-1)^{g_{i_1}g_{i_1i_2}}.
\end{align}
where $\d\Sigma_2$ is the boundary (edge) of $\Sigma_2$.

Based on the $1d$ examples, we now expect that this pumping operation, $\T^2$ should toggle the $1d$ $\Z_2^T$-SPT invariant of $\d\Sigma_2$. In $1d$, there is a single nontrivial $\Z_2^T$-SPT: the Haldane/AKLT phase~\cite{haldane1983nonlinear,affleck2004rigorous} which is characterized by Kramers doublet edge states. This can be explicitly confirmed by verifying that $(-1)^{g_{ij}g_{jk}}=\omega_2(g_{ij},g_{jk})$ is a non-trivial $2$ cocycle in $H^2(\Z_2,U_T(1))$.
Alternatively, note that restricting the boundary string operator to an open $1d$ chain $x\in \{1,2,\dots L\}$ gives: $\T_{[0,L]}^2 \ir \prod_{i=2}^L (-1)^{g_i(g_i-g_{i-1})}$. Conjugating an ordinary on-site TR symmetry on this chain: $\T^{(0)} = \prod_{i=1}^L (-1)^{\hat{g_i}}K$, by $\T^2$ gives: 
\begin{align}
\(\T^2_{[1,L]}\)^\dagger \T^{(0)} \T^2_{[1,L]} &= (-1)^{g_L-g_1} \prod_i (-1)^{\hat{g}_i} K
\end{align}
and the local action of this twisted time-reversal on site $i=1$ is: $(-1)^{g_1}(-1)^{\hat{g}_1}K$, which squares to $\T^2_\text{edge}=(-1)$, which is the signature of anomalous edge action of a Haldane chain. 
From these considerations, we confirm that acting on the $2d$ igSPT with $\T^2$ pumps a Haldane/AKLT chain onto it s $1d$ edge.

The symmetry of the system under the SPT-pumping action of $\T^2$ forces the edge to reside at a self-dual point between the trivial and Haldane phases. 
As explained in~\cite{tsui2015quantum}, this self-duality enforces an anomaly constraint that forbids the edge from being trivially gapped and symmetric. 
Specifically, if we write down a $1d$ family of edge Hamiltonians:
\begin{align}
H_\text{edge}(\lambda) = -\frac12\sum_{i\in \d\Sigma_2} \[(1-\lambda)\sigma^z_{i-1} \sigma^x_i\sigma^z_{i+1}+(1+\lambda)\sigma^x_i\]
\label{eq:Hselfdual}
\end{align}
with $-1\leq\lambda\leq 1$.
which realizes a trivial paramagnet for $\lambda>0$, the Haldane/AKLT phase for $\lambda<0$, and a gapless CFT with central charge $c=1$ at the self-dual point $\lambda=0$.
The pumping symmetry maps $\lambda\rightarrow -\lambda$, and forces the edge to sit at the gapless point if this symmetry is intact.
We note that another possibility (which does not arise for this particular edge Hamiltonian) is that there could be a first order phase transition between the trivial and SPT states. This would correspond to spontaneously breaking the $\N$-symmetry of the system, since it would result in a degenerate pair of ground-states for the edge that are interchanged by $\N$.

While $H_\text{edge}$ in Eq.~\ref{eq:Hselfdual} with $\lambda=1$ operates in the IR subspace, and is symmetric under $\T^2_\text{IR}$, it is \emph{not} invariant under $\T_\text{IR}$ due to the non-onsite phases in this anomalous time-reversal symmetry operator which depend on the bulk spin configuration as well as the edge. 
Hence, in the true igSPT the Luttinger liquid described by Eq.~\ref{eq:Hselfdual} can potentially become intertwined with the bulk critical modes in a manner that is hard to calculate in general. Below, we will discuss two instances where this bulk-edge coupling can be studied in a controlled fashion, and the gapless self-dual Luttinger liquid described by Eq.~\ref{eq:Hselfdual} with $\lambda=1/2$ dynamically decouples from bulk modes.

\subsection{On the bulk phase diagram}
Our construction of the time-reversal symmetric $2d$ igSPT so far simply engineers a gapped sector that protects an emergent IR anomaly. However, $H_\Delta$ still has an extensively degenerate ground-space. To create a bona fide model, we should then add generic local time-reversal symmetry interactions, $H_\text{IR}$, that do not close the $\Delta$-gap. As in $1d$, any terms in $H_\text{IR}$ that do not commute with $H_\Delta$ can be adiabatically eliminated, so we may restrict our attention to interactions that commute term-by term the terms of $H_\Delta$.
We may follow either of the two procedures that were outlined for the $1d$ example for generating ZCL $H_\text{IR}$ models. 
In this case, the resulting models involve complex several body interactions: generically, in $2d$, single site terms map to seven-body interactions under $V$ involving products of operators over a site and its nearest neighbors on the surrounding hexagon of the triangular lattice).
Moreover, there are few non-numerical tools for studying $2d$ interacting lattice models.
Therefore, instead of confronting the difficult task of mapping out the phase diagram for a particular family of $H_\text{IR}$, we instead use the emergent anomaly data to deduce general anomaly constraints on the possible bulk and boundary phases that might arise.

\subsubsection{Symmetry breaking bulk}
A trivial option for the bulk is that it could be gapped and spontaneous symmetry breaking. An example Hamiltonian for this would be a nearest neighbor Ising ferromagnetic interaction: $H_\text{IR} = -J\sum_{\<ij\>}(-1)^{\hat{g}_i+\hat{g}_j}$ where $\<ij\>$ denote nearest neighboring sites.

\subsubsection{Gapless bulk}
One possible fate of the igSPT bulk that is consistent with the emergent anomaly is a symmetric gapless state. This problem has been previously studied in the context of $3d$ bosonic $\Z_2^T$-SPT surface physics~\cite{vishwanath2013physics}, and one candidate gapless state is an $SO(5)_1$ Wess-Zumino-Witten model~\cite{bi2015classification}, which has various lower-symmetry and dual fermionic incarnations~\cite{wang2017deconfined}. A closely related field theory also appears in descriptions of deconfined quantum critical points~\cite{senthil2004deconfined} between antiferromagnet and valence bond solids, or between quantum spin-Hall states and superconductors in fermionic systems~\cite{lee2014deconfined} (though both of those involve different emergent anomalous symmetries at the critical point). The fate of the interaction between gapless Luttinger liquid edges and bulk DQCP modes was recently studied in a large-$N$ limit~\cite{ma2022edge}, where the edge and bulk naturally decouple. Understanding the ultimate fate of these $2d$ DQCP-type field theories and their interplay with edge modes remains an outstanding challenge. For example, it remains an open question whether these DQCPs are continuous or weakly first order (which in the igSPT context with an emergent anomaly would signal spontaneous symmetry breaking).


\subsection{Quotient-symmetry-enriched topological (QSET) order}
For $1d$ igSPTs the only ground-states consistent with the emergent anomaly are either gapless or symmetry breaking. In $2d$, there is an additional possibility that the bulk of the igSPT is gapped and symmetric, but with an anomalous symmetry enriched topological order (SET). For the $2d$ $\Z_2^T$ in-cohomology that emerges in this model, a known anomalous SET is a $\Z_2$ (toric-code) topological order with anyons $\{1,e,m,f=e\times m\}$ such that $e$ and $m$ are both Kramers doublets under time-reversal, denoted as $e_Tm_T$~\cite{vishwanath2013physics,wang2013boson}. 

In this section, we construct an explicit lattice model in which $e_Tm_T$ SET order arises in the bulk of the $\Z_4^T$ igSPT model using the ``gauging by fractionalization" procedure introduced for the $1d$ igSPT. Analyzing the edge of this model shows that the edge-SPT-pumping symmetry and associated self-dual edge gapless states survive when the bulk becomes gapped to form the anomalous SET order. 
Here, some care is needed to properly define what one means by an anomalous SET, since the anomaly is lifted deep in the UV.
To see this, note that extending the symmetry to $\Z_4^T$ by adding microscopic Kramers bosons, $b_T$ clearly lifts the $e_Tm_T$ anomaly in the UV, as the SET can be trivially confined to a symmetric paramagnet by condensing the composite $b_T\times m_T$ (or $b_T\times e_T$)~\footnote{We note that, extending the symmetry by adding local Kramers doublet fermions, $c_T$, would not lift the $e_Tm_T$ anomaly, but rather would lift a different beyond-cohomology anomaly which supports a representative $e_{fT}m_{fT}$ SET in which the $e$ and $m$ particles are both Kramers doublets under time-reversal and also have fermionic statistics.}.
However, if there is a hierarchy of scales, such that the gap, $\Delta$, to $b_T$ excitations greatly exceeds the gap $\delta$ to the anyonic excitations, we can meaningfully talk about an emergent anomaly even with a gapped bulk.
Borrowing a term from an analogous situation for SPTs~\cite{verresen2021quotient}, we will call such a state a quotient-symmetry enriched topolgoical order (QSET).
The sharp distinction between a QSET and a $\Guv$-SETs is that the QSET will have symmetry-protected edge states that cannot be removed without undergoing an edge phase transition in which the gap, $\Delta$, to $\N$-DOF closes.

Following the $1d$ approach described above, we can construct a lattice model of this QSET by fractionalizing the on-site $g_i\in \Z_2$ DOF into $(\nu_a\in\{0,1\},g_a)\in \Z_4$ DOF, and couple the fractional $n$-DOF to an emergent $\Z_2$ lattice gauge field, $A_{a,b}$ so that the physical (gauge invariant) degrees of freedom are still $\Gir=\Z_2$ rotors. The anomalous time-reversal is implemented as
\begin{align}
\T^\text{frac} \ir  \prod_{i} (-1)^{g_{i_1i_2}g_{i_2i_3}g_{i_3}} \prod_a (-1)^{g_a\hat\nu_a}(-1)^{\hat{g}_a}K
\end{align}

\begin{figure}[]
\begin{centering}
	\includegraphics[width=1.0\columnwidth]{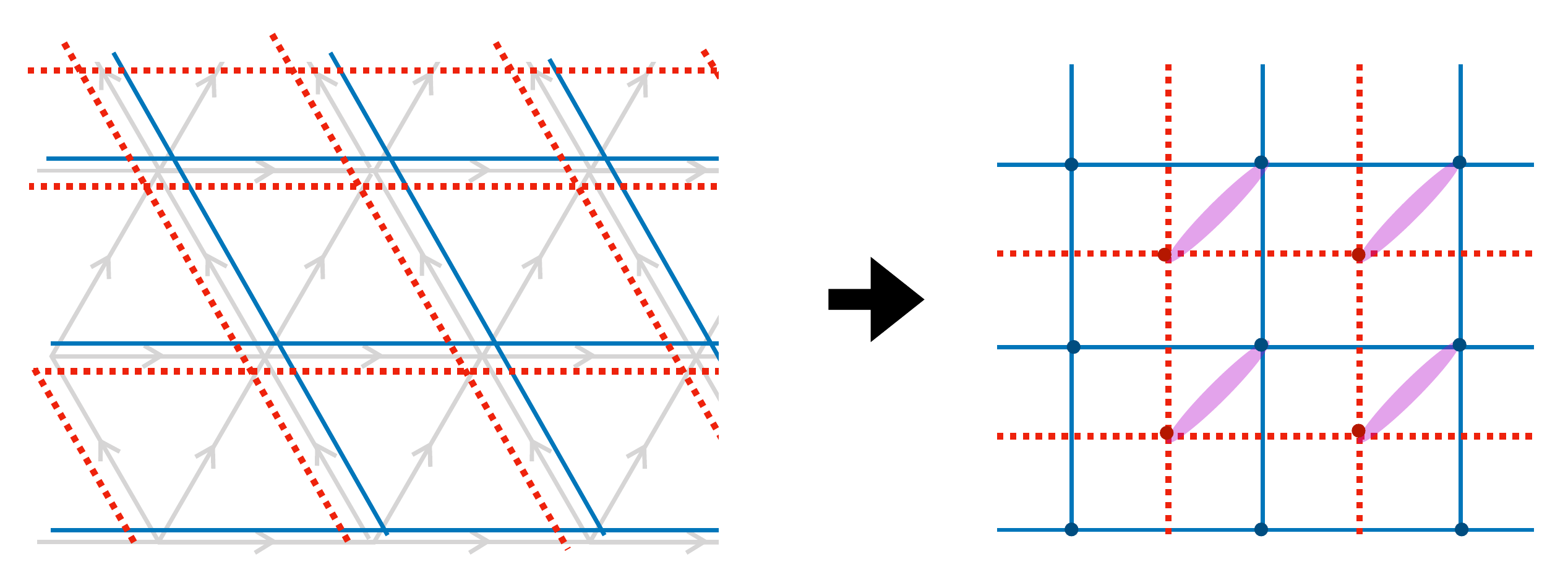}
\end{centering}
	\caption{{\bf Decorated domain wall picture} As described in the text, the triangulated simplicial complex (left) can be redrawn as an inter-penetrating pair of dual square lattices (right), and the anomalous symmetry can be interpreted as decorating domain walls on the blue lattice with $1d$ igSPT phases on the dashed red sublattice.
	}
	\label{fig:ddw}
\end{figure}

To interpret this expression it is useful to redraw the triangular lattice as shown in Fig.~\ref{fig:ddw}. Here, for each 2-simplex $(i_1i_2i_3)$, we color the $i_1i_2$ bond blue and the $i_2i_3$ bond dashed red. The blue and red bonds form a pair of (skewed) square lattices. We can straighten the plaquettes and shift the red square lattice relative to the blue one so that the red and blue square lattices are dual to each other. Here, we must remember that the purple-circled sites should be regarded as the same site.  Then, we can interpret the non-onsite part of the symmetry as follows: along TR DWs ($g_{ab}=1$) on the blue sub-lattice,  the non-onsite aspect of the symmetry action is essentially the same as that of the $1d$ igSPT model with $\Gir=\Z_2$ described above.
We can make this decorated domain wall (DDW) picture precise by temporarily extending the IR symmetry group to $\Z_2^T\times \Z_2 = \{1,\T\}\times \{1,g\}$, with $\Z_2^T$ rotors living on the vertices of the blue lattice and $\Z_2$ rotors living on the vertices of the dashed red lattice, and decorating $\Z_2^T$ DWs with $1d$ $\Gir=\Z_2$ ($\Guv=\Z_4$) igSPTs. Then, we can break the enlarged symmetry back down to the diagonal subgroup generated by $(\T,g)$, by locking the $\Z_2$ and $\Z_2^T$ rotors together in pairs as indicated by the purple ovals. 

With this DDW picture, we can readily extend the other $1d$ constructions, to obtain gauge-invariant a unitary $\mathcal{V}$ that transforms this fractionalized symmetry into an almost on-site one (up to edge terms):
\begin{align}
\mathcal{V} = \prod_{i} (-1)^{\blue{g_{i_1i_2}}\red{g_{i_2}(\nu_{i_2i_3}-a_{i_2i_3})}}.
\end{align}
which performs the same transformation as the $1d$ $\Z_4$ igSPT along blue DWs. We can write the transformed symmetry schematically as:
\begin{align}
\T' &= \mathcal {V}^\dagger \T^\text{frac} \mathcal{V} 
\nonumber\\
&=  \prod_{\blue{\mathcal{DW}}} (-1)^{\int_{\blue{\mathcal{DW}}} D_A\nu}  \prod_a (-1)^{g_a\hat\nu_a}(-1)^{\hat{g}_a}K
\end{align}
where $\blue{\mathcal{DW}}$ denotes the blue domain walls, and $\int_{\blue{\mathcal{DW}}} D_A\nu \equiv \sum_{(ab)\in \Guv} \(\nu_b-\nu_a - A_{a,b}\)$ denotes the lattice version of a Wilson line. These Wilson lines terminate only on the boundary, and so there are no $\nu$ operators in the bulk, only gauge connections $A_{a,b}$.

Using $\mathcal{V}$ we can write down a Hamiltonian that fully gaps the bulk: 
\begin{align}
\mathcal{V}^\dagger H \mathcal{V} = 
-\delta\sum_{a\in \Sigma_2^{\mathrm{o}}}
(-1)^{\hat{\nu}_a+\hat{g}_a} -\delta \prod_{i} (-1)^{A_{i_1i_2}+A_{i_2i_3}+A_{i_3i_1}}
\end{align}
where the first term makes a trivially-gapped paramagnet of the bulk matter ($\Sigma_2^{\mathrm{o}}$ denotes the interior of the space) and the second term produces a gap for gauge flux excitations.

This produces a lattice model of a $\Z_2$ (toric-code) QSET order.
Denote the topological super-selection sectors of the $\Z_4$-rotor excitations [$(\hat\nu_a,\hat{g}_a)\neq (0,0)$] as $e$ and the gauge flux as $m$.
The $\Z_4$ rotors transform under $\Guv=\Z_4^T$, and hence the $e$ particles are Kramers doublets.

\subsubsection{Deducing the QSET anomalous symmetry properties from the cohomology data} 
\begin{figure}[]
\begin{centering}
	\includegraphics[width=1.0\columnwidth]{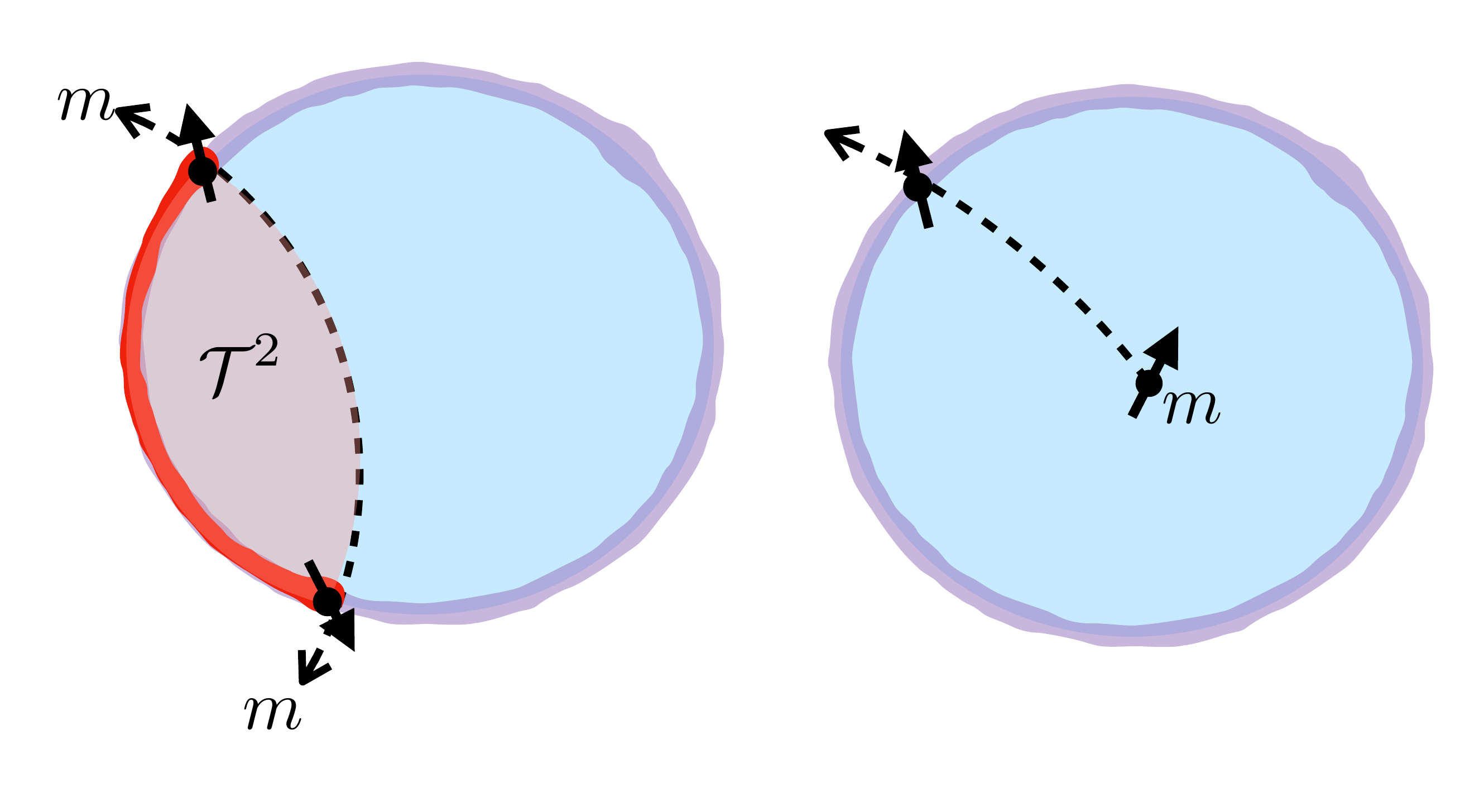}
\end{centering}
	\caption{{\bf Symmetry properties of $2d$ QSETs} constructed through the symmetry fractionalization approach can be deduced by the SPT-pumping symmetry action. Consider a $2d$ disk geometry where the bulk is a gapped time-reversal symmetric QSET. (Left) Creating a pair of gauge fluxes ($m$ particles)  and dragging them out of the sample is equivalent to acting locally on the shaded red region by $\T^2$, which by the SPT-pumping property of the bulk anomaly adds a Haldane/AKLT chain with spin-1/2 (Kramer's doublet) edge states to the boundary. (Right) Considering a similar process where only one $m$ particle is dragged out of the system, and noting that projective time-reversal DOF can only be created in pairs, we deduce that the $m$ particle in the bulk must carry a Kramer's doublet. This argument can be readily generalized to various other symmetry groups.}
	\label{fig:2dqset}
\end{figure}

We can deduce the symmetry properties of $m$ from the properties of the pumping symmetry operations through the following argument applied to the case where $\Sigma_2$ is an open disk (illustrated in Fig.~\ref{fig:2dqset}).  Since $m$ is a $\pi$-flux for $\nu$-DOF, braiding an $m$ flux around a $\nu$ excitation is equivalent to acting locally on that excitation by $\T^2$. Hence, creating a pair of $m$ particles in the QSET bulk dragging them around a loop $\Gamma$ and re-annihilating them is equivalent to locally acting with $\T^2$ in the interior of $\Gamma$, which we have seen pumps a Haldane chain onto $\Gamma$ due to the anomalous symmetry action. Thus, if we create and separate a pair of $m$'s, each $m$ must behave like the Kramers-doublet edge of a Haldane chain. This shows that the resulting QSET bulk is an anomalous $e_Tm_T$ SET.

The symmetry properties of $m$ can also be directly confirmed by examining the action of $\T'$ on a gauge flux. With PBCs, the Wilson loops are all closed, and the modified symmetry, $\T'$ differs from the on-site $\Z_4^T$ symmetry by a phase of $(-1)$ whenever the blue DWs enclose a flux -- and we can choose a gauge where we consider the interior of the blue DW to be regions with $g_i=1$. Suppose we have a gauge flux ($m$ particle) in an initially $g_i=0$ domain. Then acting with $\T$ flips $g_i:0\rightarrow 1$. Acting again with $\T$ then yields a $(-1)$ phase since the flux sits in an $1$ domain. This shows that the local action of $\T^2$ on $m$-particles gives a $(-1)$ phase. 

\subsubsection{Edge states of the QSET} 
The SPT-pumping symmetry prevents the edge of the QSET from being trivially gapped. We can deduce the structure of the QSET edge states from the fractionalized lattice model construction. 
This follows analogously to the $1d$ case discussed above: wherever a blue DW intersects the boundary one can lock $g_i$ to $\hat{\nu}_i$, and applying the trivial paramagnetic term on other sites without blue DWs:
\begin{align}
&\mathcal{V}H_\text{edge}\mathcal{V}^\dagger\nonumber \\
&= - 
\delta\sum_{(i_1i_2i_3)\cap\d\Sigma_2} \[(-1)^{\blue{g_{i_1i_2}}\red{(g_{i_3}+\hat{\nu}_{i_3})}} +(-1)^{\blue{(1-g_{i_1i_2})}\red{(1-\hat\nu_{i_1})(1-\hat{g}_{i_1})}}\]. 
\end{align}
The first term in this equation is the same as the $H_\text{edge}$ for the $1d$ igSPT from the prior section wherever a blue DW intersects the edge ($\blue{g_{ab}}=1$), and the second term simply implements a trivial paramagnet away from the DWs ($\blue{g_{ab}}=0$).
Eliminating the $\nu$ variables to satisfy this edge Hamiltonian, we obtain a $1d$ closed chain of free $\Gir$-rotors with the anomalous SPT-pumping symmetry described above in \eqref{eq:TA}.
Consequently, a valid symmetric edge Hamiltonian is \eqref{eq:Hselfdual} tuned to the self-dual point ($\lambda=0$) where it realizes a $c=1$ CFT. 
In contrast to the complicated case of a gapless bulk and edge, this edge CFT is obviously stable to symmetric coupling to the gapped QSET bulk.

\subsubsection{UV interpretation of the QSET} The QSET phase is sharply separated from an ordinary $\Guv$-SET by an edge phase transition where the edge states are lost by closing the $\N$ gap. One can ask what the resulting $\Guv$-SET is in this case. For the time-reversal symmetric model we have been considering, the $\Guv$-SET is trivial, since we can freely relabel the $e$ and $m$ excitations by binding them to microscopic Kramers-doublet bosons to make them Kramers singlets.
In the $\Guv=\Z_2\times \Z_4$ example studied in the appendix, the resulting $\Guv$-SET is actually a non-trivial one where the $m$ particle carries a projective representation of $\Guv$. Whether the resulting $\Guv$-SET is trivial or nontrivial seems to depend on how far the symmetry is extended to lift the $\Gir$ anomaly. Namely, we could lift the anomalous $\Gir=\Z_2^2$ anomaly in this example to $\Guv = D_4$, (the dihedral group with four elements), which is not the minimal lift, but would result in the QSET order being trivial if the $\N$-gap is closed at its edge.

\section{A $3d$ Ising igSPT \label{sec:3d}}
By now the pattern is clear, and we can readily construct a $3d$ igSPT model with $\Gir=\Z_2,\N=\Z_2,\Guv=\Z_4$ that exhibits the surface anomaly of a notional $4d$ $\Gir$-SPT bulk, but when realized as an igSPT, the anomaly emerges from a gapped $\N$-sector.
We note that, as for the $1d$ example, the extending degrees of freedom could either be bosons (e.g. spins) or fermions. The latter is perhaps more relevant for physical realizations, as the $\Gir=\Z_2$ symmetry could arise as an Ising spin rotation symmetry such as $\pi$ rotations around a fixed spin-axis, say $x$: $R_x(\pi)$ for a (possibly doped) Mott insulator of spin-1/2 electrons, which would transform under a $\Guv=\Z_4^F$ extension with $R_x(\pi)^2 = (-1)^{N_F}$.

An explicit $5$ cocycle for this state is:
\begin{align}
\omega_5(g,h,k,l,m) = (-1)^{ghklm}
\end{align}
The igSPT lattice model consists of $\hat{n}\in \widehat{\N}/g\in G$-rotors sitting on sites/simplicies of a $3d$ simplicial complex (which, can actually be simplified to a $3d$ cubic lattice following the DDW picture described previously) with onsite $\Guv=\Z_4$ symmetry generated by:
\begin{align}
U^\text{os}_1 = \prod_{i}(-1)^{g_{i_4}\hat{n}_i}\prod_a(-1)^{\hat{g}_a},
\end{align}
 anomaly imprinting Hamiltonian
\begin{align}
H_\Delta = -\Delta \sum_{i} \delta_{\hat{n}_i,g_{i_1i_2}g_{i_2i_3}g_{i_3i_4}},
\end{align}
which gives rise to an emergent anomalous $\Gir=\Z_2$ symmetry in the IR generated by:
\begin{align}
U^A_1 = \prod_{i}(-1)^{g_{i_1i_2}g_{i_2i_3}g_{i_3i_4}g_{i_4}} \prod_a (-1)^{\hat{g}_a},
\end{align}
which is related to the original on-site symmetry by the corresponding anomaly-lift unitary:
\begin{align}
V = \prod_{i}(-1)^{g_{i_1i_2}g_{i_2i_3}g_{i_3i_4}n_{i}}.
\end{align}

 We note that just as in $2d$, we can view this model as a DDW construction with a larger $\Z_2^3=\{1,a\}\times\{1,b\}\times \{1,c\}$ symmetry group, which we then break down its diagonal subgroup generated by $abc$. As illustrated above in $2d$, in this extended DDW picture, we can again pull apart the simplicial complex into three inter-penetrating sub-lattices, one each for a,b,c links, with the $b$ sublattice residing on the dual lattice of $a$ and $c$ dual to $b$. This phase can then be viewed as a DDW phase in which we decorate the $1d$ intersection of $a$ and $b$ DWs with $1d$ $\Guv=\Z_4$ igSPTs. The original $\Guv=\Z_4$ igSPT with just a single symmetry factor can then be recovered by locking the $a$, $b$, and $c$ rotors together at each site of the original lattice.

\subsection{Edge phase(s)}
The extended group structure, $\N=\Z_2$ follows from noting that $$(U^A_1)^2 = \prod_{i\in \d \Sigma_3} (-1)^{g_{i_1i_2}g_{i_2i_3}g_{i_3}}$$ pumps a $2d$ $\Z_2$ SPT onto the surface, which has a $H^3(\Z_2,U(1))=\Z_2$ classification. This tunes the surface to a self-dual point between a $\Z_2$ SPT (Levin Gu phase~\cite{levin2012braiding}) and a trivial $\Z_2$ symmetry paramagnet. 

\subsubsection{Spontaneous symmetry breaking edge}
A particular instance of (a gauged version) of a $2d$ self-dual Hamiltonian satisfying this pumping symmetry was studied numerically in~\cite{morampudi2014numerical} which found that it realized a first order phase transition, which in the present context corresponds to spontaneously breaking of the $\Guv=\Z_4$ symmetry. It remains an open question how stable a putative gapless symmetric edge termination would be, which would correspond to asking whether there was a continuous (multi)critical point between the Levin-Gu phase and trivial PM.

\subsubsection{QSET edge and bulk}
A new possibility that arises in $3d$ is that the $2d$ edge of an igSPT can also exhibit anomalous QSET order that can only arise at the edge of a system with an emergent $3d$ anomalous symmetry. We focus on the simplest theoretical instance where both the $3d$ bulk and $2d$ edge have QSET order. An explicit lattice model can be readily constructed as in $1d$ and $2d$, but we instead simply deduce a candidate QSET from the general principles identified in the previous sections. 

\begin{figure}[]
\begin{centering}
	\includegraphics[width=0.5\columnwidth]{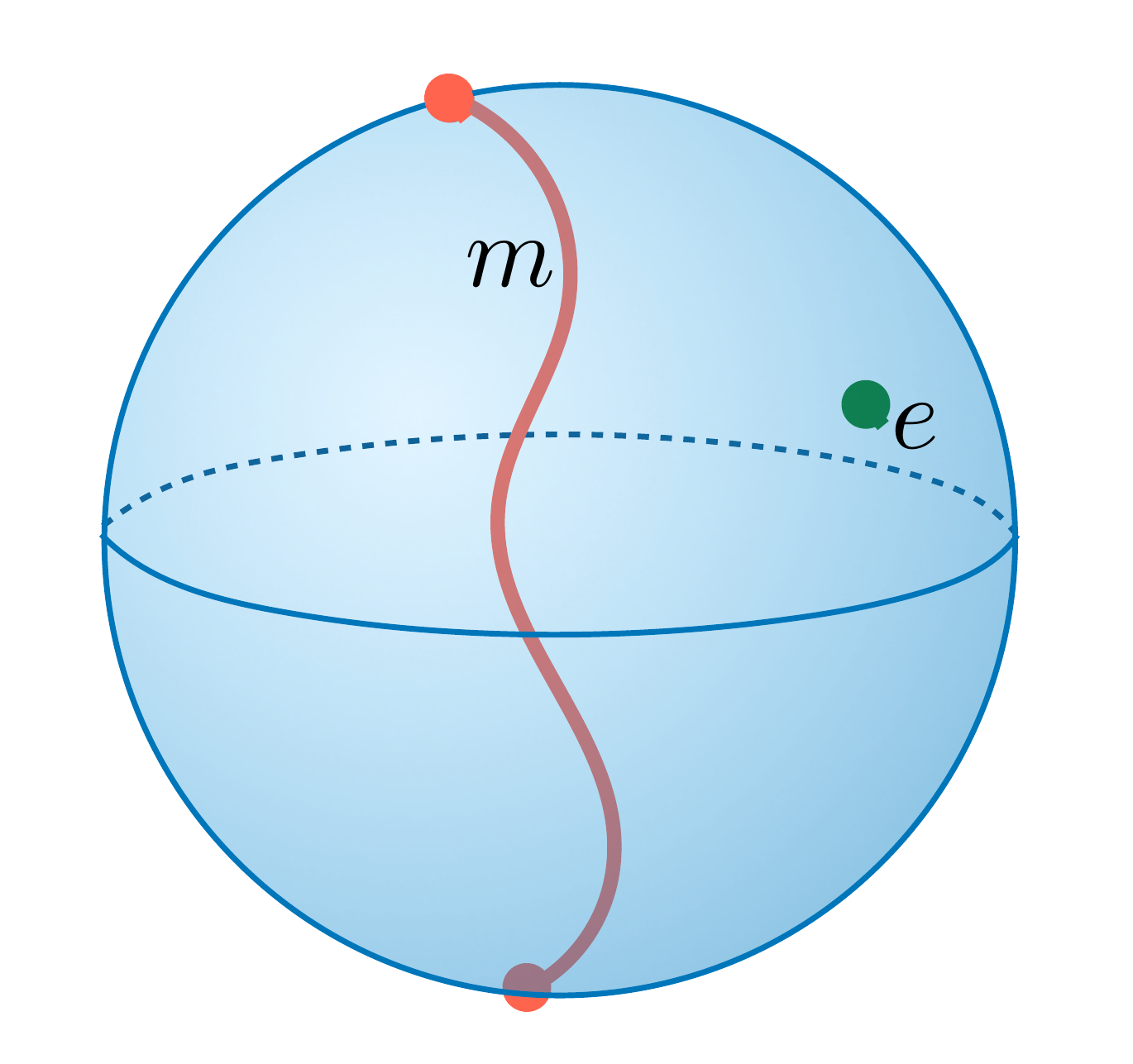}
\end{centering}
	\caption{{\bf Fully gapped $3d$ QSET}  A possible fate of the $3d$ $\Guv=\Z_4$ igSPT model is that the emergent anomaly is satisfied by a fully-gapped anomalous $\Z_2$ symmetry enriched $3d$ Toric code with point-like $e$ excitations that carry $\Z_4$ charge and string-like $m$ excitations that carry $1d$ $\Z_4$ igSPTs. The intersection of the $m$ line and the surface hosts the edge zero mode of the $1d$ igSPT.}
	\label{fig:3dqset}
\end{figure}

The QSET order can have the structure of an emergent $\N$ gauge theory, in this case a $3d$ toric code topological order with $e$ particles, and vortex-line like $m$ string excitations. In our fractionalization construction, the $e$ particle carries a half-charge of $\Gir$, i.e. transforms under $\Guv=\Z_4$. In $2d$, the string operator whose ends create $m$ particles acted locally like the lower-d SPT pumping operation. Here, the $3d$ analog of this is that the $2d$ membrane operator whose edges create $m$ line excitations carries a $2d$ $\Z_2$ gapped SPT. The edge of this membrane, which is the $m$ line therefore carries the (gapless or SSB) edge of this $2d$ SPT. In an open geometry, for example the 3-ball pictured in Fig.~\ref{fig:3dqset}, the $m$ line excitations can also intersect the edge of the system. From our study of $1d$ igSPTs we know that terminating the anomalous $1d$ edge of a $2d$ SPT requires the $\N$ extending DOF in the UV and results in a global 2-fold ground-space degeneracy of the $m$ line when it intersects the system's edge.
A succinct summary of these properties is that the anomalous feature of the $3d$ QSET is that the $m$ line excitations carry $1d$ $\Z_4$ igSPTs.


\section{Discussion}
Our group cohomology based constructions and framework, build on previous (predominately $1d$) examples~\cite{thorngren2021intrinsically,li2022symmetry}, to give a clear picture of how emergent anomalous symmetries lead to igSPT and QSET states, with SPT edge states protected by SPT-pumping symmetries.
A natural question for future exploration is whether or not igSPTs can be constructed with other types of anomalies, including i) fermionic anomalies, i.e. anomalies involving fermion parity which is not quite a symmetry because it cannot be broken by any local operator and also cannot be sensibly extended~\cite{thorngren2021intrinsically} to a larger group, and  ii) beyond group-cohomology anomalies that involve gravitational or mixed-symmetry/gravitational anomalies.
To facilitate potential experimental realizations, it would also be interesting to generate examples where the extended symmetry group was naturally realized by spin-1/2 electrons, which can imprint the anomalous IR symmetry on the low-energy spin-model of a Mott- or Cooper-pair insulator.
As we argue next, this naturally leads one to consider beyond cohomology anomalies.

\subsection{Selected boson igSPT candidates with beyond group-cohomology anomalies}
Before closing we briefly comment on some candidate $2d$ igSPTs with emergent anomalies beyond group cohomology, highlighting ones that could arise in electronic systems (i.e. emerging from unit charged, spin-1/2, fermions).
We focus on systems with physically relevant symmetries: spin-component-conservation and time-reversal $\Gir=U(1)\times \Z_2^T$, that are robust to disorder.
In a system of electrons, bosonic excitations are composites of an even number of electrons, and will have charge that is an even multiple of the electron charge, for example spins have charge $0$ and Cooper pairs charge $2$.

In $2d$, anomalous $\Gir=U(1)\times \Z_2^T$ symmetry actions are governed by three different $\Z_2$-topological invariants. These are conveniently expressed in terms of the properties of an anomalous toric-code-like SETs with anyons $\{1,e,m,f=e\times m\}$. We can distinguish different SET phases by the symmetry properties of these anyons. Following standard notation~\cite{vishwanath2013physics,wang2013boson}, we denote a half-integer charge by a $C$ subscript (note that a half-integer boson charge corresponds to odd number of units of electron charge), a time-reversal Kramers-doublet by a $T$ subscript, and fermionic self-statistics by an $F$ subscript. The 7 non-trivial anomalous $\Gir$-SETs with can be generated by taking combinations of the three root phases $e_Cm_C$, $e_Tm_T$, and $e_Fm_F$, where the symmetry properties of the fermion follow from $f=e\times m$.
Anomalies involving only $C$ or $T$ are captured by the group-cohomology classification and are included in our results. Anomalies involving $e_Fm_F$ are beyond the group-cohomology as they involve a mixed time-reversal/graviational anomaly. Namely, any pure $2d$ bosonic realization of $e_Fm_F$ must have chiral central charge $\pm 4$, which is half of the minimal ``integer" chiral phase of bosons (the $\mathbb{E}_8$ state~\cite{lu2012theory}), and hence breaks time-reversal symmetry.

On general grounds, one can lift the $e_Fm_F$ anomaly by introducing a (charge-neutral) local fermion excitation, $c$, in the system. This can be seen in either of two ways. First, working directly in the topological order, we can relabel $e_F$ and $m_F$ by binding them to a local $c$ fermion, resulting in bosons with trivial quantum numbers that can be condensed to trivially confine the SET order. Second, fermions have invertible chiral phases with any integer Hall conductance, which can be used to cancel the chiral central charge $\pm 4$ of the $2d$ realization. Hence, we expect that one can obtain an emergent $e_Fm_F$ starting from a system of charge-neutral fermions, and driving these neutral-fermions into a gapped state that imprints the anomaly on the IR boson DOF.

A potentially more interesting example is the $e_{CTF}m_{CTF}$ anomaly (i.e. with all three root anomalies). In this case, the anomalous properties can be lifted by introducing fermions with unit charge (i.e. half a boson charge) and that are Kramers doublets (spin-1/2) under time-reversal -- i.e. ordinary electrons. A concrete, albeit schematic construction of a $2d$ QSET order an emergent $e_{CTF}m_{CTF}$ anomaly can be obtained by the slab construction of~\cite{thorngren2021intrinsically}. Start from a thin slab of a $3d$ boson $\Gir$-SPT with this surface anomaly, in which we view the bosons as emerging from a Mott insulator of electrons. Then, close the electron gap on the bottom surface to bind electrons to the $m_{CTF}$ particles, and condense the resulting bound state which is now a boson with trivial quantum numbers. This confines the DOF on the bottom surface, but leaves the anomalous $e_{CTF}m_{CTF}$ SET order on the top surface.
An interesting question, that we leave for future work, is to investigate the edge of this $2d$ igSPT/QSET, i.e. the side-interface between the top and bottom surfaces in this slab construction.
While we have cited the slab construction as a proof-of-principle, the same phase should be realizable in a strictly $2d$ lattice model of electrons, and it would be interesting to see whether this could be accomplished with realistic interactions.

\subsection{Relation to half-filled topological flat bands}
Another context in which beyond cohomology anomalies emerge in the IR is half-filled topological flat bands with a non-local particle hole ($\C\T$) symmetry, including the 1/2 filled Landau level of ordinary electrons~\cite{son2015composite,wang2016half,metlitski2016particle}, and various multi-component generalizations thereof~\cite{potter2017realizing,sodemann2017composite}. Here, the role of $H_\Delta$ is played by the orbital magnetic field, which breaks the space into Landau levels (LLs). Half-filling the lowest LL results in an anomalous $\C\T$ symmetry that is emergent, in the sense that it is only operative within the LL and does not extend to the entire state Hilbert space. 
This $\C\T$ symmetry realizes the same anomaly as the surface of a $3d$ $U(1)\times \C\T$ symmetric topological insulator. This was notably used to argue~\cite{son2015composite} and numerically demonstrate~\cite{geraedts2016half} that $\C\T$-symmetric composite Fermi surface has a $\pi$ Berry phase.

In this context, the particle-hole transformation is unusual in that it involves both performing both a $\C\T$ transformation and then adding filled LLs. The latter operation produces a non-standard notion of $\C\T$ symmetry at the edge, and for example is compatible with non-vanishing thermal and/or electrical Hall conductance that is half of the allowed amount for integer states. The non-locality of this symmetry follows from the Wannier obstruction to forming a local tensor product structure to the lowest LL, which is also a key aspect of the beyond-group-cohomology anomalies that emerge.

If we make peace with the non-local symmetry, we might regard these examples as the first known examples of igSPTs. In fact, this half-filled flat band construction provides an example of a system with an emergent $e_{F}m_{F}$ anomaly discussed above. Namely, consider a four-flavor half-filled Landau levels of fermions with $\C\T$ symmetry, which realizes the same anomaly as a four Dirac cone surface of a $3d$ topological insulator in class AIII~\footnote{This system naturally has another $U(1)$ symmetry (charge conservation) which will not play an important role in the following discussion.}.
In particular, DWs between opposite $\C\T$-breaking domans in this surface carry chiral central charge $8$ -- the hallmark of the $e_Fm_F$ anomaly.
Then, consider inducing $\C\T$-symmetry interactions that gap out the fermion excitations (for example introduce fermion pairing and disorder the resulting superconductor through vortex proliferation without closing the gap to unpaired fermions as detailed in~\cite{metlitski2014interaction}). The combination of the orbital field and symmetrically-gapping interactions results in a low-energy boson-only model with an emergent anomalous anti-unitary $\Z_2^{\C\T}$ symmetry that exhibits an $e_Fm_F$ anomaly at low energies. 
We remark that this example was previously discussed as a realization of an $e_Tm_T$ anomaly in~\cite{potter2017realizing}. If we ignore the hierarchy of scales $\Delta$, the distinction between the $e_Tm_T$ and $e_Fm_F$ anomaly is lost in a system with microscopic Kramers-doublet fermions. Interpreted as an igSPT, this distinction becomes important so long as we demand that the fermion gap remains open.

It remains an open question whether one can realize such beyond-cohomology anomalies with a \emph{local} symmetry, (e.g. one that enforces vanishing thermal Hall conductance).

\subsection{Open directions}
While our focus on topological properties protected by the gapped sector has let us sharply deduce certain anomaly constraints on igSPTs and elucidate the origin of their edge states, it does not permit a detailed description of the bulk phase diagram for realistic interactions, nor does it characterize boundary-critical phenomena arising from igSPT edge states. Developing controlled theory and numerical techniques to address these open issues, and identify potential experimentally testable signatures of igSPT bulk and edge physics, and explore the importance of disorder and other non-idealities remain important targets for future work. 

It may also be interesting to explore the connections of gapless SPTs and quantum information. For example, given the importance of gapped SPT phases as resource states for measurement-based quantum computing (MBQC), it is tempting to ask whether gapless systems can have computational power, and whether they offer advantages or limitations over their gapped counterparts.

\vspace{10pt}\noindent {\it Acknowledgements -- } We thank Dominic Else, Lukasz Fidkowski, Michael Hermele, Ajesh Kumar, Ruochen Ma, Gaurav Tenkila, Ryan Thorngren,  Ruben Verresen, and Chong Wang for insightful conversations.
ACP was supported by the Alfred P. Sloan Foundation through a Sloan Research Fellowship.
This work was performed in part at the Aspen Center for Physics, which is supported by National Science Foundation grant PHY-1607611.
This research was undertaken thanks, in part, to funding from the Max Planck-UBC-UTokyo Center for Quantum Materials and the Canada First Research Excellence Fund, Quantum Materials and Future Technologies Program.

\bibliography{igSPTbib}

\appendix

\section{Additional $1d$ igSPT examples \label{app:1d}}
In this section we develop a few other examples of $1d$ igSPTs. Data for these as well as additional examples are listed in Table.~\ref{tab:results}. 

\subsection{$\Gir=\Z_{N}$}
We represent elements of $\Z_N$ by operators $g_i = \{0,1,\dots N-1\}$, and conjugate operators $\hat{g}_i$ satisfying $\[\hat{g}_i,g_i\] = 1$. These define the $\Z_N$ rotor operators: $e^{\frac{2\pi i}{N}g}$ and their dual operators $e^{\frac{2\pi i}{N}\hat{g}}$ satisfying: $\[e^{\frac{2\pi i}{N}\hat{g}},e^{\frac{2\pi i}{N}g}\]_g=e^{\frac{2\pi i}{N}}$ where $[a,b]_g = aba^{-1}b^{-1}$ is the group commutator. These are the usual representations for a $\Z_N$ quantum clock model, that are an $N$-level generalization of the Pauli algebra for a qubit.

The anomaly classification is $H^2(\Z_N,U(1)) = \Z_N$, with an explicit root cocycle: $\omega_3(a,b,c) = e^{\frac{2\pi i}{N}a\cdot v_N(b,c)}$ (this cocycle generates all the other anomalies). The corresponding anomalous symmetry action is
\begin{align}U^A_{g} &= \prod_{(ij)} e^{\frac{2\pi i}{N}g_{i,j}v_N(-g_j-g,g)} \prod_i e^{\frac{2\pi i}{N} g\hat{g}_i}\\
	&=\prod_{(ij)} e^{\frac{2\pi i}{N}g_{i,j}v_N(g_j,g)} \prod_i e^{\frac{2\pi i}{N} g\hat{g}_i}
\end{align}
 We can read from the cocycle that $\N=\widehat{\N}=\Z_N$ and $b_1(a)=a, e_2(a,b)=v_N(a,b)$. Therefore we can onsite this symmetry by extending the group with $\N=\Z_N$ rotors, $n_i$, and locking $\hat{n}_j \ir g_{i,j}$ on each link $(ij)$ with onsite symmetry 
 \begin{align}
	U^\text{os}_{(n,g)} = \prod_{i} e^{\frac{2\pi i}{N}\hat{n}_i(n+v(g_i,g))}e^{\frac{2\pi i}{N}g\hat{g}_i},
 \end{align}
which satisfies the $\Guv=\Z_{N^2}$ symmetry multiplication rule: $U^\text{os}_{(n,g)}U^\text{os}_{(n',g')} = U^\text{os}_{(n+n'+v_N(g,g'),g+g')}$.

In the IR the SPT pumping operations are generated by 
$$U_{(1,0)}\ir \prod_{(ij)} e^{\frac{2\pi i}{N}g_{i,j}} = e^{\frac{2\pi i}{N} (g_L-g_1)}$$ 
which pumps the root class of 0d $\Gir=\Z_N$-SPT onto the edge(endpoints) of the chain.

\subsection{ $\Gir=U(1)$}
$2d$ SPTs with $\Gir=U(1)$ are classified by an even-integer Hall conductance $\sigma^{xy}=2n$ with $n\in \mathbb{Z}$. Unlike their fermionic counterparts these bosonic integer quantum Hall (bIQH)~\cite{senthil2013integer} states have vanishing chiral central charge i.e. vanishing thermal Hall conductance, since each charged chiral edge mode is accompanied by a neutral counter-propagating one. The result is an SPT phase since absent $U(1)$ symmetry one could backscatter charge modes into neutral ones gapping the edge. 

Realizing an igSPT with the same edge anomaly would be quite striking, as its bulk would have a quantized conductance just as for the SPT edge. Unfortunately, our construction has some pathalogical features when applied to this case. First, one needs to introduce continuous $U(1)$ rotors $e^{ig_i}$ with $g_i\in [0,2\pi)$ and $g_i\simeq g_i+2\pi$. This is often a price that one must pay to get exactly solvable models for bIQH-type states. Ordinarily one can hope that the same topological phase could arise in a finite-dimensional truncation of these rotors, e.g. to spin-$S$ spins with maximum $S^z = \pm S$. Here, there is a more serious obstruction.

Namely it turns out the minimal extending group that can lift the anomaly is $\Z$. One way to see this is that the classification of $0d$ $U(1)$ SPTs that can get pumped onto the edge have $H^1(U(1),U(1)) = \text{Rep}\(U(1)\) = \Z$ classification. In our construction, this lower-$d$ SPT classification dictates the structure of the symmetry extension.

We can also verify by explicit construction. A cocycle for the root phase with $\sigma^{xy}=2$ is: $\omega_3(a,b,c) = e^{ia\cdot v_{2\pi}(b,c)}$. The anomalous symmetry implementation is then 
\begin{align}
	U_{g}^\text{A} = \prod_{(ij)} e^{ig_{i,j}v(g_j,g)}\prod_i e^{ig\hat{g}_i}
\end{align} where $\hat{g}_i$ is the conjugate ``angular momentum" to the rotor phase $g_i$ satisfying $\[\hat{g}_i,g_i\]=1$, which has $\Z$ eigenvalues. We read from the cocycle that $\N=\Z$, $\widehat{\N}=\widehat{\Z}=U(1)$ and $b_1(g)=g, e_2(g_1,g_2)=v_{2\pi}(g_1,g_2)$.
To onsite this symmetry we can introduce additional $\widehat{\N}=U(1)$ rotors, $\hat{n}_i$,
and lock $\hat{n}_j\ir g_{i,j}$ on each link $(i,j)$, and defining the on-site symmetry 
\begin{align}
	U_{(n,g)}^\text{os} = \prod_i e^{i \hat{n}_i(n+v_{2\pi}(g_i,g) )} e^{ ig\hat{g}_i}
\end{align}
 One can verify that the resulting extending symmetry is actually the entire real line $\mathbb{R}$. Namely, $U(1) = \mathbb{R}/2\pi \Z$, and the content of the on-site symmetry operator is that each time the symmetry shifts the phase $g_i$ past $2\pi$ it increments $\hat{n}_i$ by one unit, so that we can identify each $(n\in N,g\in \mathbb{R}/2\pi \Z)\in \Guv$ with $2\pi n + g \in \mathbb{R}$, and the on-site symmetry is simply $U_xU_y = U_{x+y}$ for $x,y\in \mathbb{R}$.

While mathematically possible, we regard these models as unphysical. For example, while it is perfectly possible to consider $1d$ array of quantum beads on loops of wire, faithfully realizing the $\mathbb{R}$ symmetry in such a model would require having no kinetic energy for the $\N$ beads, as the symmetry under cycling the $\Gir$-rotors from $g\rightarrow g+2\pi$ would increase the $\N$-rotor's angular momentum, which must commute with the Hamiltonian.

\subsection{$\Gir=\prod_{I=1}^3 \Z_{N_I}$}
An explicit cocycle for the mixed anomaly is $\omega_3(a,b,c) = e^{\frac{2\pi i}{N_{123}}a^1b^2c^3}$ corresponding to anomalous symmetry 
\begin{align}
U_{\vec{g}}^\text{A} = \prod_{(ij)} e^{\frac{2\pi i}{N_{123}}g_{i,j}^1(-g_j^2-g^2)g^3}\prod_i e^{2\pi i \sum_{I=1}^3 \frac{\hat{g}_i^Ig^I_i}{N_I}}.
\end{align}
We can onsite the symmetry by introducing $\widehat{\N}=\Z_{N_{123}}$ rotors $\hat{n}_j \ir (g_{i,j}^1)$ and define the UV actions to be
\begin{align}
	 U^\text{os}_{(n;\vec{g})} = \prod_i e^{\frac{2\pi i}{N_{123}}\hat{n}_i(n+(-g^2_i-g^2)g^3)} \prod_i e^{2\pi i \sum_{I=1}^3 \frac{\hat{g}_i^Ig^I_i}{N_I}}.
\end{align}
which obeys a group multiplication rule:
\begin{align}
\(n,\vec{g}\)\cdot \(m,\vec{h}\) = \(n+m+g^2h^3, \vec{g}+\vec{h}\)
\end{align}
We see the extended group is $\Z_{N_2}\ltimes (\Z_{N_1}\times \Z_{N_3}\times \Z_{N_{123}})$, where the action of $\Z_{N_2}$ on $\Z_{N_1}\times \Z_{N_3}\times \Z_{N_{123}}$ is:
\begin{align}
	g^2\circ (g^1,g^3,n):=(g^1,g^3,g^2g^3+n)
\end{align}
The pumping actions are generated by
\begin{align}
	U^\text{os}_{(1,0)}\ir \prod_{(ij)} e^{\frac{2\pi i}{N_{123}}g_{i,j}^1}=e^{\frac{2\pi i}{N_{123}}(g_L^1-g_1^1)}
\end{align}
which pumps a $1d$ $\Z_{N_1}$ SPT onto the edge. 

\subsection{$\Gir=U(1)\ltimes \Z_2^T$}
$1d$ anomalies of $\Gir=U(1)\ltimes \Z_2^T$ are classified by $H^3(U(1)\ltimes \Z_2^T,U(1))=\Z_2$. 
As for the $U(1)$ example above, $0d$ SPTs are classified by $H^1(G,U(1))=\Z = \N$, so one expects that this example requires an unphysical $\Z$ extension of the symmetry to lift the anomaly. This is confirmed by direct computation from an explicit cocycle: $\omega_3(g,h,k) = e^{ig^1h^2k^2}$. We can read from this cocycle that $\N=\Z$, $\widehat{\N}=\mathbb{R}$, making the model unphysical as for the $U(1)$ example.



\section{Other $2d$ igSPTs}
\subsection{$\Gir=\Z_2^2$}

In $2d$, $\Gir=\Z_2^2$ has $H^4(\Z_2^2,U(1))=\Z_2\times \Z_2$, which can be generated by combinations of the following two representative cocycles:
\begin{align}
\omega_4(g,h,k,l) \in \{ (-1)^{g^1h^2k^2l^2},(-1)^{g^1h^1k^1l^2}\}
\end{align}
In both cases, the anomalies can be lifted by $\N=\Z_2$ extensions. However, the extension group structure and edge physics is different. For the first cocycle the anomaly can be lifted by extending to  $\Guv=\Z_2\times \Z_4$, producing an igSPT with edge states protected by a $1d$ Haldane chain ($\Z_2^2$-SPT) pumping symmetry. In the second case, the extension is to $\Guv=\Z_4\ltimes \Z_2$, and the edge states are not protected by any SPT-pumping aciton
\color{black}

\subsubsection{igSPT with SPT pumping symmetry}
Let us first consider the anomalous $\Gir$ symmetry with representative cocycle $\omega_4(g,h,k,l) = (-1)^{g^1h^2k^2l^2}$, which can be written as $\omega_4=b_2\cup e_2$ with $b_2(g,h)=(-1)^{g^1h^2}$, which is a non-trivial element of $H^2(\Z_2^2,U(1))$, corresponding to the cocycle of a Haldane/AKLT spin chain. According to the general construction above, we expect the resulting igSPT has edge states protected by a Haldane SPT pumping symmetry, in close relation with the time-reversal symmetric $2d$ igSPT example discussed in the main text. This phase can also be understood via a decorated domain wall construction.

Denote the IR symmetry group as $\Z_2^A \times \Z_2^B$ where $A,B$ label the two different $\Z_2$ factors for clarity.
This igSPT phase can be thought of as a DDW state where $\Z_2^A$ DWs are decorated with $1d$ $\Z_4^B$ igSPTs.

The symmetry can be onsited by introducing $\widehat{N}=\Z_2$ rotors that are locked to $\hat{n}_i\ir s_{(i_1i_2i_3)}g^1_{i_1i_2}g^2_{i_2i_3}$ for each triangle $i$ where we remind that $s_i=\pm 1$ for up and down triangles respectively, and writing onsite symmetry:
\begin{align}
U_{(n;\vec{g})}^\text{os} = \prod_i (-1)^{\hat{n}_i(n+v(g_{i_3}^2,g^2))} \prod_a(-1)^{\sum_{I=1,2}\hat{g}^I_ag^I}
\end{align}
which obeys the group multiplication of $\Guv=\Z_2\times \Z_4$.

The $1d$ edge of this $2d$ igSPT phase carry edge states protected by an SPT pumping symmetry $U_{(0,1)}^2\ir \prod_{(ij)\in \d\Sigma_2} (-1)^{g^1_{i,j}g^2_j}$ which pumps a $1d$ $\Z_2^2$-SPT onto the edge.

We can deduce a possible symmetry gapped QSET phase for this igSPT by the fractionalization procedure outlined in the main text. This results in a $\Z_2$ gauge theory with a $e$ particle having a unit of $\Z_4$ charge, i.e. half of the elementary $B$ charge in the IR symmetry factor $\Z_2^B$, and an $m$ particle that carries the projective representation of $\Z_2^2$ that is found at the end $1d$ SPT pumped by $U_{(0,1)}^2$. 
We refer to this state as $e_{C_B}m_P$ where $C_B$ denotes a half-$B$-charge, and $P$ denotes a projective $\Z_2^2$ representation.

The structure above can be readily generalized to products of two cyclic groups, $\Gir=\Z_{N_1}\times \Z_{N_2}$, with the relevant data listed in Table~\ref{tab:results}. These results follow readily from interpreting these phases in a DDW construction where $\Z_{N_1}$ DWs carry $1d$ $\Z_{N_2N_{12}}$ igSPTs.

\subsubsection{igSPT without SPT pumping symmetry}
Now let us turn to the other representative cocycle $\omega_4(g,h,k,l) = (-1)^{g^1h^1k^1l^2} = b_2\cup e_2$ with $b_2(g,h) = (-1)^{g^1h^1}$ and $e_2(g,h) = (-1)^{g^1h^2}$. We see that the $b_2$ cocycle involves only the first $\Z_2$ symmetry factor. Since there are no $1d$ SPT phases protected by a single $\Z_2$ symmetry, when realized as an emergent anomaly, $\omega_4$, will not have a lower-dimensional SPT pumping symmetry, in contrast to all the previously constructed examples. Nevertheless, we will find that the igSPT constructed from this anomaly via our construction has SPT edge states, which are protected by a long-range string order where the condensed string has half-charge of the first $\Z_2$ symmetry factor.

Following the general procedure outlined in the main text, define the anomalous IR symmetry action associated with this anomaly cocycle as:
\begin{align}
U_\alpha^A &= \prod_{i}(-1)^{g^1_{i_1i_2}g^1_{i_2i_3}g^1_{i_3}}\prod_a(-1)^{\hat{g}^1_{a}+\hat{g}^2_{a}} \nonumber\\
U_\beta &= \prod_a (-1)^{\hat{g}^1_a}.
\end{align}
where for future convenience we have chosen to generate $\Z_2^A\times \Z_2^B = \{1,a\}\times \{1,b\}$ by $\{(\alpha=a+b),\beta=b\}$.
This symmetry can be on-sited by defining an $\N=\Z_2$ rotor on each plaquette
defining the on-site symmetry action:
\begin{align}
U_\alpha^\text{o.s.} &= \prod_i (-1)^{\hat{n}_i g^1_{i_3}}\prod_a(-1)^{\hat{g}^1_a+\hat{g}^2_a},
\nonumber\\
U_\beta^\text{o.s.} &= \prod_a (-1)^{\hat{g}^1_a}
\nonumber\\
U_n^\text{o.s.} &= U_\alpha^2 = \prod_i (-1)^{\hat{n}_i},
\end{align}
which satisfy: $U_\alpha^2 = \prod_i (-1)^{\hat{n}_i}$, $U_\alpha^4=1$, $U_\beta^2=1$, $U_\beta^\dagger U_\alpha U_\beta = U_\alpha^{-1}$, corresponding to UV symmetry: $\Guv = \Z_4^\alpha \ltimes \Z_2^\beta$.
Then, lock $\hat{n}_i \ir g^1_{i_1i_2}g^1_{i_2i_3}$ with Hamiltonian:
\begin{align}
H_\Delta = -\Delta \sum_{i} \delta_{\hat{n}_i,g^1_{i_1i_2}g^1_{i_2i_3}}.
\end{align}

Note that, in the IR, the $\N$ symmetry acts non-trivially only on the boundary of the system. Defining the system on an open spatial 2-manifold, $\Sigma_2$: 
\begin{align}
U_n = U_\alpha^2 \ir \prod_{i\in \d\Sigma_2} (-1)^{a_{i-1,i}a_i}
\end{align}
where we denote $g^1_i$ as $a_i$. Restricting this symmetry to an open region $R=1<i\leq L$ of the boundary, one obtains $(U_n)|_R = \prod_{1<i\leq L} (-1)^{a_{i-1,i}a_i}$, which transforms under the $\Gir$ symmetry as: $U_{\beta}: (U_n)|_R\mapsto (-1)^{a_1+a_L}(U_n)|_R$. Examining this expression, one finds that while the bulk the $(U_n)|_R$ string operator is invariant under $\Gir$, its ends transform as if there is a local ``1/2-charge" of $\Z_2^a$. 
Namely, acting again with $U_\beta$ on the string end transforms the end near site $1$ from $(-1)^{a_1}\mapsto (-1)^{a_1+1}$, and the transformation of the bulk string adds another factor of $(-1)^{a_1}$, leaving a local phase of $(-1)$ indicating the the string ends locally have $U_\beta^2=(-1)$.

A more precise version of this argument involves gauging the $\N$ symmetry, and the action of $\Gir$ on the ends of the restricted $\N$ symmetry action $(U_n)|_R$ translate to the $\Gir$ transformation properties of the $\N$-flux, which are well-defined modulo redefining the transformation by a $\widehat{\N}=\Z_2$ magnetic gauge transformation, i.e. correspond to an element of a projective symmetry action (PSG): $H^2(\Z_2^A,\widehat{\N})=\Z_2$. The non-trivial element of the PSG corresponds to the flux having a half-charge of $a$, which we expect to arise due to the above argument. We can verify this by confirming that the emergent anomaly cocycle decomposes as $\omega_4 =b_2\cup e_2$ where $b_2(g,h) = (-1)^{g^1h^1}$ is the non-trivial element of the PSG for $\N$-fluxes, and $e_2 = (-1)^{g^1h^2}$ specifies the group-extension that lifts the anomaly. 

As we've shwon in general in Section~\ref{sec:edge states} of the main text, the non-trivial transformation rules for the $\N$ flux ensure that the igSPT edge cannot be symmetric and gapped.
Specifically, the edge symmetry action of $U_n$ and $U_\beta$ has the same anomaly as a gapped SPT phase with $\Z^\beta_2\ltimes \Z^n_2$ symmetry with cocycle $\omega_3(g,h,k) = (-1)^{g^1h^1k^2}$.
However, as with previous examples, there is no gapped SPT with this edge anomaly when the $\N=\Z_2$ is the normal subgroup of a larger $\Z^\alpha_4$ symmetry, as follows from the general proof in Appendix~\ref{app:intrinsic}.

One can explicitly construct a symmetric gapless edge of this model as follows. Define spin Pauli operators: $\sigma^z_i = (-1)^{g^1_i}$, $\sigma^x_i = (-1)^{\hat{g}^1_i}$ for each site of the edge. Then, examine the action of $U_n$ on these operators: $U_n: \sigma^x_i \mapsto -\sigma^z_{i-1}\sigma^x_i\sigma^z_{i+1}$. I.e. the $U_n$-symmetry forbids a trivial paramagnetic interaction for the edge spins, but does allow a critical edge Hamiltonian such as:
\begin{align}
H_\text{edge} = -K\sum_i \(\sigma^x_i-\sigma^z_{i-1}\sigma^x_i\sigma^z_{i+1}\),
\end{align}
which by maps via a standard Jordan-Wigner transformation to a pair of decoupled critical Ising models in the even and odd sub-lattice.
In fact, this Hamiltonian has an enlarged emergent $U(1)$ symmetry generated by $\widetilde{N} = \sum_j \widetilde{N}_j$ with $\widetilde{N}_j = (-1)^j\frac14\(1-\sigma^z_j\)\(1+\sigma^z_{j+1}\) = (-1)^j a_{j,j+1}a_{j+1}$ measuring the number of $\down\up$ domain walls, and the $U_N$ symmetry corresponds to $(-1)^{\widetilde{N}}$.
As for the other cases described in the main text, this Luttinger liquid edge Hamiltonian needs to be symmetrized over $\Gir$, which will potentially tangle it up with bulk modes in a manner that is not precisely understood in general, but in select cases (e.g. the bulk is a gapped QSET) this edge Luttinger liquid can remain decoupled from the bulk modes.

\subsection{$\Gir=U(1)\times \Z_2^T$}
In this section, we denote: $U(1)$ rotor states by $\alpha_i\in [0,2\pi)$, the phases of $U(1)$ symmetry rotations by Greek letters without site subscripts, and $\Z_2$ time-reversal rotor states by $\tau_i\in \{0,1\}$, in order to clearly distinguish the role of each in formulas. Moreover, we denote the conjugate angular momentum to the rotor phase as $\ell_i$, with $\[\ell_i,\alpha_i\] = 1$.

This group is physically relevant for spin systems with a conserved spin-component, say $S^z$, which is odd under time-reversal (indicated by the direct product with $\T$).
The $3d$ group-cohomology anomaly classification for $\Gir$ is $\Z_2^3$. One of these $\Z_2$ factors is the pure time-reversal anomaly discussed above. The remaining two $\Z_2$ factors correspond to mixed $U(1),\T$ anomalies, and can be thought of as being generated by two different root phases and their combinations.

Following~\cite{wang2013boson}, we can label the anomalies by a representative anomalous SET $\Z_2$-gauge-theory (toric code topological order), labeled by specifying whether or not the $\Z_2$ gauge charge $e$ and flux $m$ have a half-integer charge ($C$-subscript) and/or transform as a Kramers doublet under $\T$ ($T$ subscript). The pure $\T$-anomaly is labeled by $e_Tm_T$, and the mixed anomalies can then be labeled by different stacking combinations of the two root phases $e_Cm_C$ and $e_Cm_T$.

\subsubsection{A $2d$ igSPT with $e_Cm_C$ anomaly}
The $e_Cm_C$ anomaly arises at the surface of a $3d$ $\Gir$-SPT discussed extensively in~\cite{vishwanath2013physics}. Its gapless states are closely related to field theories of deconfined quantum critical points~\cite{wang2017deconfined}. The physics of this anomalous state can be understood through a DDW perspective in which time-reversal DWs are decorated with boson integer quantum Hall (bIQH) states with Hall conductance $2$, i.e. where the time-reversal breaking DW in the anomalous theory carries gapless edge modes of the root $2d$ $U(1)$ SPT (analogous to how magnetic DWs in the surface of a $3d$ electron $\Gir$-topological insulator carry electron IQH edge modes~\cite{senthil2014symmetry}). An equivalent statement is that the bulk response theory of the $3d$ boson SPT to a background $U(1)$ electromagnetic field has a quantized magnetoelectric response:
\begin{align}
\mathcal{Z}_{3d}[A] = e^{\frac{i\theta}{8\pi^2}\int dAdA}
\end{align}
Physically this response function is invariant under $\theta\rightarrow \theta+4\pi$ (twice the periodi for fermion systems). Since $\T: dAdA\rightarrow -dAdA$, this periodicity leaves $\theta=0,2\pi$ as the distinct time-reversal invariant possibilities, with the latter corresponding to a non-trivial SPT.

\paragraph*{An explicit cocycle}
To construct an explicit cocycle for this phase, let us warm-up with a lower-dimensional example. In $1d$ the axion response theory is $\mathcal{Z}_{1d} = e^{i\frac{\theta}{4\pi}\int dA}$, with time-reversal symmetric  allowing for $\theta=0,2\pi$ corresponding to trivial and SPT phases respectively. The $1d$ SPT phase has edge states that transform under a projective representation of $U(1)\times \Z_2^T$ defined by: $U_{2\pi-\alpha}U_\alpha = (-1)$, corresponding to effectively having half-charge. A suitable cocycle which implements this projective representation is:
\begin{align}
\omega_2[(\alpha_1,\tau_1),(\alpha_2,\tau_2)] = e^{i\frac{\theta}{2}v_{2\pi}(\alpha_1,\alpha_2)}
\end{align}
with $\theta=2\pi$, since $U_{\alpha}U_\beta = \omega_2 U_{\alpha+\beta}$. 
To check this expression, we note that time-reversal restricts $\theta$ to be a multiple of $2\pi$.
Namely, since $\tau$ does not appear in $\omega_2$, $\T U_\alpha = U_\alpha \T$ in this projective representation. Therefore, we must have $\T U_{2\pi-\alpha} U_{\alpha}\T^\dagger = U_{2\pi-\alpha}U_\alpha = e^{i\theta/4\pi}$. The left hand side evaluates to $\T e^{i\theta/4\pi} \T^\dagger = e^{-i\theta/4\pi}$. Comparing these requires $\theta$ to be an integer multiple of $2\pi$. 
The cocycle is obviously trivial for $\theta$ an integer multiple of $4\pi$, since it evaluates to $1$ for all arguments.
Finally, we note that, while time reversal does not directly play a role in this cocycle (there is no explicit $\tau$ dependence), it is crucial to quantizing $\theta$ (as for the field theory). 
Namely, with just a $U(1)$ symmetry alone, the cocycle is a pure boundary for $\theta \notin 4\pi \Z$: $e^{i\frac{\theta}{4\pi}v_{2\pi}(\alpha_1,\alpha_2)} = [d\omega_1](\alpha_1,\alpha_2)$ with $\omega_1(\alpha) = e^{i\frac{\theta}{4\pi}[\alpha]}$.
However, $d\omega_1\neq \omega_2$ when the group is extended to include time-reversal. Namely, for $\Gir=U(1)\times \Z_2^T$: denoting $g_i = (\alpha_i,\tau_i) \in U(1)\times \Z_2$, the coboundary condition is modified to:
\begin{align}
df(g_1,g_2) = \frac{\[f(g_2)\]^{1-2\tau_1}f(g_1)}{f(g_1g_2)}
\end{align}
and we no longer have $d\omega_1=\omega_2$ for arguments with non-vanishing $\tau_1$. Note, however, that for $\theta\in 4\pi \Z$, $\omega_2=1$ for all arguments and the cocycle is explicitly trivial.

From this $1d$ picture, we can analogously construct a 3-cocycle for the $2d$ anomaly:
\begin{align}
\omega_3[(\alpha_1,\tau_1),\dots,(\alpha_4,\tau_4)] = e^{\frac{i\theta}{4\pi}v_{2\pi}(\alpha_1,\alpha_2)v_{2\pi}(\alpha_3,\alpha_4)}. 
\end{align}
Again time reversal quantizes $\theta$ to be an integer multiple of $2\pi$, and $\theta=4\pi$ is trivial because $\omega_3$ evaluates to $1$ for all arguments in that case.

\paragraph*{igSPT from a $U(1)$ rotor model}
Since time-reversal group elements do not explicitly appear in the cocycle, we can actually construct the $2d$ igSPT purely from a $U(1)$ rotors with no Ising spins for the $\Z_2^T$ subgroup.
This cocycle defines the anomalous $U(1)$ symmetry action:
\begin{align}
U_{\alpha}^\text{A} = \prod_{i} (-1)^{v_{2\pi}(\alpha_{i_1i_2},\alpha_{i_2i_3})v_{2\pi}(\alpha_{k},\alpha)} \prod_{a} e^{i \alpha \ell_a}
\end{align}

We can onsite this symmetry by introducing a $\N=\Z_2$ extension with onsite symmetry:
\begin{align}\label{UosU1timesZ2T}
U_{(n,\alpha)}^\text{os} = \prod_i (-1)^{\hat{n}_i(n+v_{2\pi}(\alpha_{i_3},\alpha))}\prod_a e^{i \alpha \ell_a},
\end{align}
and energetically locking $\hat{n}_i \ir v_{2\pi}(\alpha_{i_1i_2},\alpha_{i_2i_3})$ on each triangle $i$ with: 
\begin{align}
H_\Delta = -\Delta \sum_{i}\delta_{\hat{n}_i,v_{2\pi}(\alpha_{i_1i_2},\alpha_{i_2i_3})}.
\end{align}
We can identify each element of the extended symmetry group $(n\in \Z_2,\alpha\in U(1)=\mathbb{R}/2\pi \Z)\in \Guv$, as an element of $U(1)$ with \emph{half} the elementary charge, i.e. $\Guv=\mathbb{R}/4\pi \Z$. To see this map:  $(n,g) = 2\pi n+g~\text{mod}~4\pi$, and notice that the onsite group operation is simply ordinary addition in $\mathbb{R}/4\pi \Z$: $U_{(n,g)}U_{(m,h)} = U_{(n+m+v_{2\pi}(g,h),g+h)}$.
Physically, this corresponds to lifting the anomaly by adding local bosons with half of the charge of the UV degrees of freedom. Put another way, we can view this igSPT as arising in a boson-paired Mott insulator with a gap to unpaired bosons.
We note in passing that, just as for the igSPT with $e_Tm_T$ anomaly, we expect that the $e_Cm_C$ anomaly cannot be lifted by viewing the bosons as Cooper pairs of fermionic electrons with half of the bosonic charge. This would instead be a natural candidate for lifting the beyond group cohomology $e_{CF}m_{CF}$ anomaly.

\paragraph*{Edge states}
The $\N$ group is generated by $U^\text{os}_{(1,0)}$. In the IR, this operation pumps a $1d$ $U(1)\times \Z_2^T$ SPT:
\begin{align}
	U^\text{os}_{(1,0)} &\ir \prod_{(ijk)\in \Sigma_2} (-1)^{v_{2\pi}(\alpha_{ij},\alpha_{jk})}
\nonumber\\
&= \prod_{(ijk)\in \Sigma_2} d_{U(1)}\(e^{\frac{i}{2}[\alpha]}\)_{ijk}
\nonumber\\
&= \prod_{(ij)\in \d\Sigma_2} (-1)^{v_{2\pi}(g_{ij},g_{j})}
\end{align}
where $d_{U(1)}$ denotes the coboundary operator for the $U(1)$ symmetry subgroup, and $[\alpha]$ denotes $\alpha~\text{mod}~2\pi$, and in the second line we have used that $v_{2\pi}(\alpha,\beta) = \frac{1}{2\pi}\([\alpha]+[\beta]-[\alpha+\beta]\)$.
In light of our discussion above, we see that this pumps the $1d$ $U(1)\times \Z_2^T$ SPT with mixed $U(1)$ and $\T$ anomaly onto the boundary.


\paragraph*{QSET Order} To obtain a QSET order here, we can follow the fractionalization procedure described in main text. Here, we need to fractionalize the physical boson unit charge $B$ into a pair of $1/2$-charge spinons $B=b^2$ bound together in the UV by an emergent $\N=\Z_2$ gauge field. As we have seen in previous examples, in the deconfined phase of this gauge theory, the anomaly is ``lifted", from the point of view of the fractional excitations, and the gauge flux ($m$ particles) carry the projective representation of $U(1)\times \T$ that arise at the $1d$ G-SPT pumped by $U_{2\pi-\alpha}U_\alpha$. In this case, the projective representation is characterized by a half charge of the $U(1)$, and non-Kramers $\T^2$, i.e. the resulting QSET order is $e_Cm_C$, as expected from the outset.

\subsubsection{$2d$ igSPT with $e_Cm_T$ anomaly}
The other nontrivial $U(1)\times \Z_2^T$ anomaly can be realized by cocycle:
$\omega_4[(\alpha_1,\tau_1),\dots (\alpha_4,\tau_4)] = (-1)^{\tau_1\tau_2 v_{2\pi}(\alpha_3,\alpha_4)}$, corresponding to anomalous $U(1)$-symmetry:
\begin{align}
U_\alpha^\text{A} = \prod_{i}(-1)^{\tau_{i_1i_2}\tau_{i_2i_3}v_{2\pi}(\alpha_{i_3},\alpha)} \prod_a e^{i\alpha\ell_a},
\end{align}
and ordinary $\T = \prod_i (-1)^{\hat{\tau}_i}K$.
This can be on-sited by introducing $\N=\Z_2$ rotors and locking $\hat{n}_i\ir \tau_{i_1i_2}\tau_{i_2i_3}$ on each triangle $i$, and defining onsite symmetry action as in \eqref{UosU1timesZ2T} -- again resulting in an extension of $U(1)$ to include half-charges, i.e. extending from $\mathbb{R}/2\pi \Z$ to $\mathbb{R}/4\pi \Z$.

The edge states are protected by the SPT-pumping operation:
\begin{align}
U^\text{os}_{(1,0)} &\ir  \prod_{i\in \Sigma_2} (-1)^{\tau_{i_1i_2}\tau_{i_2i_3}} 
\nonumber\\
&= \prod_{(ij)\in \d\Sigma_2}(-1)^{\tau_{ij}\tau_{j}},
\end{align}
which we recognize (from the discussion of the $\Guv=\Z_4^T$ igSPT in the main text) pumps a Haldane/AKLT chain onto the boundary. Note sharing similar edge states, the $\mathbb{R}/4\pi \Z \ltimes \Z_2^T$ $2d$ igSPT is distinct from $2d$ $\Z_4^T$ one which has a pure time reversal anomaly. The $\Z_4^T$ igSPT involves a pure time-reversal anomaly, and the SPT pumping is implemented by $\T^2$. Here, instead we have a mixed $U(1)$ and $\T$ anomaly, in which it is the $U(1)$ symmetry that pumps the lower-$d$ $\Z_2^T$ SPT.

Following the pattern of previous examples a possible QSET order that comes from the fractionalization procedure is $e_Cm_T$, as expected from our label of the anomaly.

\begin{widetext}
\section{The SPT pumping action of $\N$ \label{app:sptpump}}
In this appendix, we demonstrate that the $\N$ symmetry actions in \eqref{eq:sptpumping} act only on the $(D-1)d$ edge of the igSPT and pump $(D-1)d$ SPTs onto the edge. 
Recall $b_D$ is a $D$ cocycle with coefficient $\widehat{\N}$, the cocycle condition reads:
\begin{align}
	1=\partial b_D(g_1,\cdots,g_{D+1})&=b_D(g_2,g_3,\cdots,g_{D+1})\cdot b_D(g_1g_2,g_3,\cdots,g_{D+1})^{-1}\cdots
	\nonumber\\
	&\cdot b_D(g_1,g_2,\cdots,g_Dg_{D+1})^{(-1)^D}\cdot b_D(g_1,g_2,\cdots,g_D)^{(-1)^{D+1}}
	\label{eq:cocyclecondition}
\end{align}
Replacing $g_i\rightarrow g_{i_l}^{-1}g_{i_{l+1}}$ for $I=1,2,\cdots,D$, and $g_{D+1}\rightarrow g_{i_{D+1}}^{-1}$, we obtain the relation:
\begin{align}
	1&=b_D(g^{-1}_{i_2}g_{i_3},g^{-1}_{i_3}g_{i_4},\cdots, g^{-1}_{i_{D}}g_{i_{D+1}},g^{-1}_{i_{D+1}})\cdot b_D(g^{-1}_{i_1}g_{i_3},g^{-1}_{i_3}g_{i_4},\cdots,g^{-1}_{i_D}g_{i_{D+1}},g^{-1}_{i_{D+1}})^{-1}\cdots 
	\nonumber\\
	&\cdot b_D(g^{-1}_{i_1}g_{i_2},g^{-1}_{i_2}g_{i_3},\cdots, g^{-1}_{i_{D-1}}g_{i_{D}},g^{-1}_{i_{D}})^{(-1)^D}\cdot b_D(g^{-1}_{i_1}g_{i_2},g^{-1}_{i_2}g_{i_3},\cdots, g^{-1}_{i_{D-1}}g_{i_{D}},g^{-1}_{i_{D}}g_{i_{D+1}})^{(-1)^{D+1}}
\end{align}
The action \eqref{eq:sptpumping} involves the last term, therefore we can use the above relation to replace it by the other $D$ terms, each of which depends only on a face of the $D$-complex, i.e. the first term doesn't depend on $g_{i_1}$, the second doesn't depend on $g_{i_2}$, etc. Replacing each term of \eqref{eq:sptpumping} in this way, we obtain an expression in which each term corresponds to a face of a $D$-simplex. Each face in the bulk receives contributions from its two adjacent simplices and cancel, hence we are left with only an edge action: 
\begin{align}
	U^\text{os}_{(n,0)}=\prod_{(i_1,\cdots,i_D)\in \d\Sigma}b_D^s(g^{-1}_{i_1} g_{i_2},g^{-1}_{i_2}g_{i_3},\cdots, g^{-1}_{i_{D-1}}g_{i_{D}},g^{-1}_{i_{D}})\otimes_{\Z}n,
\end{align}
where the product is now only over $(D-1)$ simplices on the edge $\d\Sigma$ of the spatial lattice $\Sigma$. These $\N$ actions pump $(D-1)d$ $\Gir$-SPTs onto the edge of the igSPT. Conjugating a trivial symmetry action $U_g^\text{trivial}|\{g_a\}\>=|\{gg_a\}\>$ by the operator $U^\text{os}_{(n,0)}$ results in:
\begin{align}
	&U^\text{os}_{(n,0)}U^\text{trivial}_g U^{\text{os}\dagger}_{(n,0)}|\{g_a\}\rangle 
    =\prod_{(i_1,\cdots,i_D) \in \d\Sigma}\frac{b_D^s(g^{-1}_{i_1} g_{i_2},g^{-1}_{i_2}g_{i_3},\cdots, g^{-1}_{i_{D-1}}g_{i_{D}},g^{-1}_{i_{D}}g^{-1})}{b_D^s(g^{-1}_{i_1} g_{i_2},g^{-1}_{i_2}g_{i_3},\cdots, g^{-1}_{i_{D-1}}g_{i_{D}},g^{-1}_{i_{D}})}\otimes_{\Z}n|\{gg_a\}\> 
    .	
    \label{eq:modsymmetryedge}
\end{align}
This is precisely the symmetry action of a $(D-1)d$ $\Gir$-SPT with anomaly specified by the $D$ cocycle $[b_D\otimes_{\Z} n]\in H^D(\Gir,U(1))$~\cite{chen2013symmetry}.
To see this, consider restricting the action of \eqref{eq:modsymmetryedge} to an open sub-region $A$ of the edge. 
Then, applying the cocycle condition (\eqref{eq:cocyclecondition}) with $g_l\rightarrow g_{i_l}^{-1}g_{i_{l+1}}, l=1,\cdots,D-1$ and $g_D\rightarrow g_{i_D}^{-1}g^{-1},g_{D+1}\rightarrow g$, the above expression can be further reduced to an action with phase factor supported only on the $(D-2)d$ boundary of $A$:
\begin{align}
    \(U^\text{os}_{(n,0)}U^\text{trivial}_g U^{\text{os}\dagger}_{(n,0)}\)_{A}|\{g_a\}\rangle &=\prod_{(i_1,\cdots,i_{D-1})\in \d A} b_D^s(g^{-1}_{i_1}g_{i_2},g^{-1}_{i_2}g_{i_3},\cdots,g^{-1}_{i_{D-1}}g^{-1},g)\otimes_{\Z} n|\{gg_a\}\rangle
    .
\end{align}
Comparing with \eqref{eq:UAgeneral} we see the action is the standard anomalous symmetry action on the edge of a $(D-1)d$ $\Gir$-SPT with anomaly specified by the $D$-cocycle $[b_D\otimes_{\Z} n]\in H^D(\Gir,U(1))$
\end{widetext}

\section{Gauging the extending symmetry\label{app:gauged}}
To gauge $\N$ explicitly in our lattice models, we introduce an $|\N|$-dimensional Hilbert space on each link of the dual lattice with basis $|a_{ij}\in \N\rangle$. We define the gauge field operator by  $\hat{n}\otimes_\Z A_{ij}|a_{ij}\>=\hat{n}\otimes_\Z a_{ij}|a_{ij}\>$ and the electric field operator $E_{ij}$ by $E_{ij}\otimes_\Z n|a_{ij}\>=|a_{ij}+n\>$. Note in general the operators $A_{ij},E_{ij}$ are not well defined since they are $\N,\widehat{\N}$-valued which are only well defined modulo $|\N|$, only the $U(1)$-valued vertex operators $\hat{n}\otimes_\Z A_{ij},E_{ij}\otimes_\Z n$ are well-defined.  Gauge transformations at $i$ is generated by 
   \begin{align}
	\Omega_i=\hat{n}_i \prod_{j\in r(i)}E_{ij},
   \end{align}
   here $r(i)$ stands for sites adjacent to $i$ on the dual lattice.  The gauge invariant subspace is then obtained by demanding Gauss's law $\Omega_i=1$ on every site.  
   \begin{figure}[t]
	\begin{centering}
		\includegraphics[width=0.9\columnwidth]{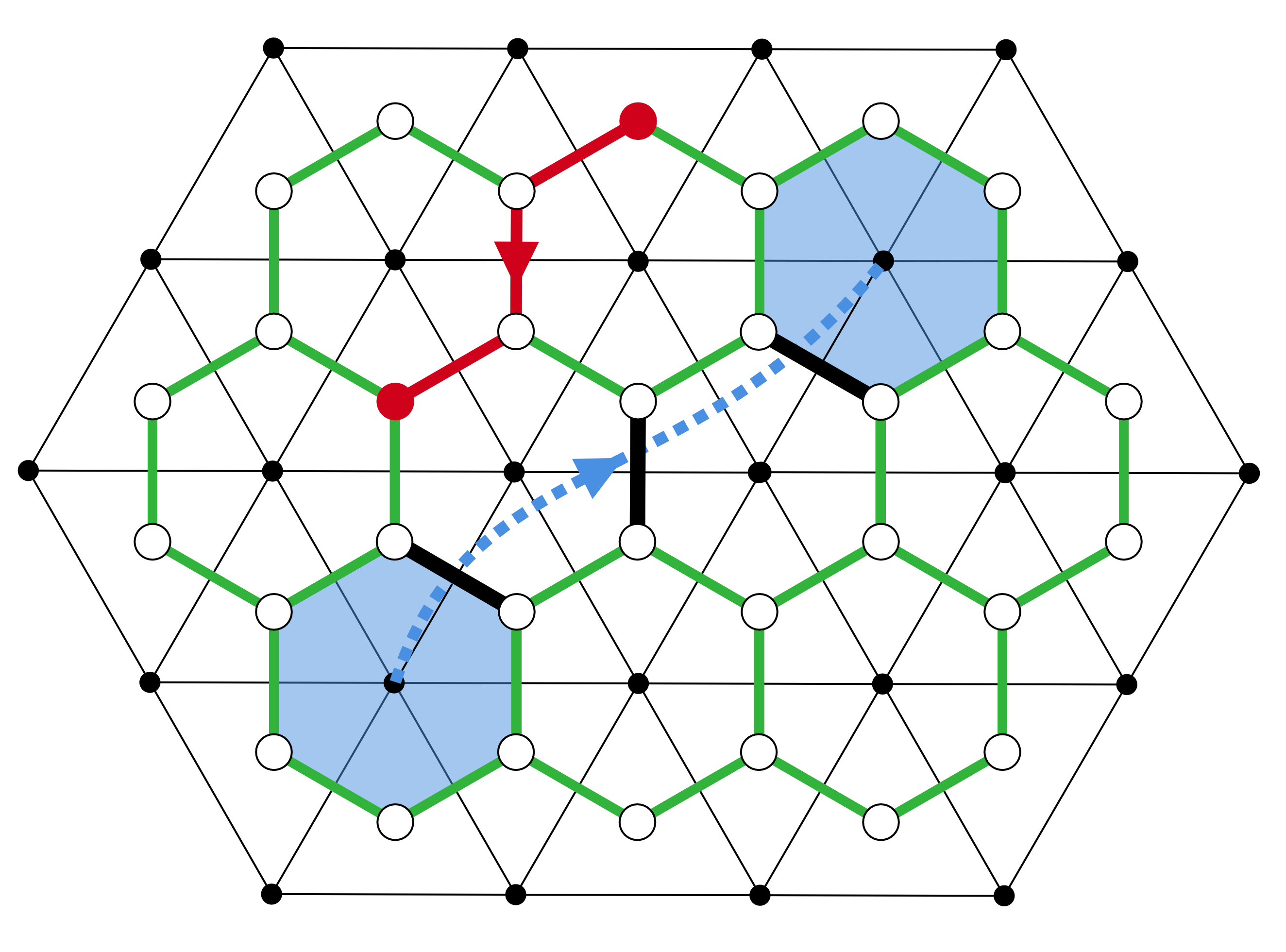}
	\end{centering}
		\caption{{\bf Gauging the extending symmetry.} Schematic of the effet of gauging the $\N$ symmetry in the igSPT lattice models models (shown in $2d$). Recall $\Gir$-rotors live on the vertices of the simplicial lattice(black dots) while $\N$ rotors live on the vertices of the dual lattice(white dots). The $\N$ gauge fields then live on the links of the dual lattice(green links). An electric string operator along the dashed blue path is a product of electric field operator over links that cross the path(thickened black links), it  creates a pair of flux/anti-flux at its end, shown as the blue hexagons. A gauge field string(Wilson line) is plotted in red and creates a pair of charge/anti-charge at its end, shown as the red dots.}
		\label{fig:gauged_lattice}
	\end{figure}

Now we consider excitations in the gauged model.
In the deconfined phase of a $\Gir$-symmetric $\N$ gauge theory, topological excitations include fluxes and charges, fluxes are $\N$-valued objects and charges are $\widehat{\N}$-valued. 
A charge is a site on the dual lattice with nonzero $\hat{n}_i$. Charge creation is generated by the gauge invariant string operator 
\begin{align}
	W_e=n_{i_1}-A_{i_1i_2}-A_{i_2i_3}-\cdots -A_{i_{L-1}i_L}-n_{i_L},
	\label{eq:charge string}
 \end{align}
 $\hat{n}\otimes_\Z W$ creates a charge $+\hat{n}$ at site $i_1$ and a charge $-\hat{n}$ at site $i_L$. 
An elementary flux is an elementary(smallest) hexagon on the dual lattice with non-zero value of Wilson loop $\sum_{(ij)\in h(a)} A_{ij}$ where $h(a)$ stands for the elementary hexagon surrounding the site $a$ of the simplicial lattice. A pair of flux/anti-flux can be created by acting on the vacuum with a string of electric field operators. Denote the path of the string operator by $C$, then flux creation is generated by 
\begin{align}
	W_m=\prod_{(pq)\bot C}E_{pq}\label{eq:flux string}.
\end{align}
$W_m\otimes_\Z n$ generates a flux $n$ and an anti-flux $-n$ on the two hexagons at the end of the path $C$, the product is over all links of the dual lattice that intersect the path $C$. The two types of excitation as well as their corresponding string operators are shown in Fig. \ref{fig:gauged_lattice}. 

\subsection{Review: symmetry fractionalization within the group cohomology framework}
Before we proceed to analyze the symmetry properties of excitations in the $\N$-gauged igSPT models, we briefly review the group cohomology (partial) classification of symmetry enriched topological orders (SETs). Since we are interested in studying the result of gauging the central extension $\N$, we restrict our attention to Abelian topological orders enriched by a general (possibly non-Abelian) global symmetry $G$.

\subsubsection{Symmetry fractionalization in 2d}
In $2d$, all excitations are point-like. Local operators cannot create topologically non-trivial anyon excitations. Instead, an isolated anyon, $a$, can only be created by non-local operators, for example by a string operator that creates an $a$ anyon and its anti-particle, $\bar{a}$ and moves the $\bar{a}$ far away. 
Therefore while the overall symmetry action on the system must form an ordinary linear representation, $U_g$, with $U_gU_h=U_{gh}\forall g,h\in G$, the action of symmetry restricted to a region with an anyon $a$, $U_g^a$, may act projectively: $U_g^aU_h^a = \omega^a(g,h)U_{gh}^a$ where $\omega^a(g,h)\in U(1)$. This is analogous to the projective action at the edge of a $1d$ G-SPT, with one key difference: the projective phases are discrete and constrained by the fusion rules of the theory. For example, given an anyon $a$ such that $N$ copies of $a$ fuse to the identity: $a^N=\underset{N\times}{\underbrace{\(a\times a\times \dots a\)}}=1$, it is possible to create $N$ $a$-excitations with a local operator. Since all local operators must transform under an ordinary linear representation of $G$, this requires $\[\omega^a(g,h)\]^N$ be trivial, i.e. is $\Z_N$ valued rather than $U(1)$ valued. Let us now formalize this structure.

For a general Abelian topological order, the set of anyon excitations $\mathcal{A}$ together with the fusion rules $\times:\mathcal{A}\times\mathcal{A}\rightarrow \mathcal{A}$ form an Abelian group. The porjective phases $\omega^a$ can be regarded as a map: $G\times G\times \mathcal{A}\rightarrow U(1), (g_1,g_2,a)\mapsto \omega^a(g_1,g_2)$. Consistency with the fusion rules requires $\omega^a\omega^b=\omega^{ab}$ which allows us to view $\omega$ as a map: $G\times G\rightarrow \widehat{\mathcal{A}}=\text{Hom}(A,U(1))$, i.e. a 2-cochain with $\hat{\mathcal{A}}$-coefficients. The associativity of the unitaries $U^{a}_g$ requires the cochain $\omega$ be closed: $\omega\in Z^2(G,\widehat{\mathcal{A}})$. 

The symmetry restriction $U_g^a$ is not uniquely defined, but rather can be modified by relabeling by $U_g^a\rightarrow \beta_g^a U_g^a$, which redefines $\omega\rightarrow d\beta \omega$, where $d\beta \in B^2(G,\widehat{\mathcal{A}})$. 
Thus, equivalence classes of symmetry fractionalization are classified by $[\omega]\in H^2(G,\widehat{\mathcal{A}}) = Z^2(G,\widehat{\mathcal{A}})/B^2(G,\widehat{\mathcal{A}})$.

As an example, in an $\N$ gauged theory the fusion rules form the group $\N\times \widehat{\N}$, where the group $\N$ is generated by gauge fluxes and $\widehat{\N}$ is generated by gauged charges. The symmetry fractionalization pattern is therefore classified by $H^2(G,\widehat{\N}\times\N)=H^2(G,\widehat{\N})\times H^2(G,\N)$. 

\subsubsection{Symmetry fractionalization in 3d}
In $3d$ Abelian topological orders, which may be regarded as $\N$ gauge theories, there are both point-like gauge-charge excitations, and loop-like gauge flux excitations. In an $\N$ gauge theory, gauge charges have fusion group $\widehat{\N}$ and gauge fluxes have fusion group $\N$.
As in $2d$, symmetry properties of point-like excitations are classified by $H^2(G,\widehat{\N})$, where $\N$ is the fusion group of point-like excitations.
For loop excitations, there are two cases for their symmetry properties. First, loops can carry an overall fractional charge. Second, loops may carry symmetry-protected gapless modes~\footnote{It is also possible that loop excitations carry ungappable chiral modes, however these cases are beyond the group cohomology formalism and will not be relevant for the in-cohomology igSPT examples here.}. 
Without loss of generality, we ignore the first possibility because one can always shrink the loop to a point to produce a fractional-charged point-excitation, or equivalently, the fractional charge can be removed by relabeling excitations by binding a point excitation to the loop.
For the second case, we follow closely the formalism of~\cite{kawagoe2021anomalies} for studying $1d$ symmetry-protected gapless modes at the edge of a $2d$ SPT. 
Specifically, in-cohomology symmetry-obstructions to gapping flux-lines arise from non-trivial fusion symbols for $G$-domain walls along an flux line. Denote the flux line anyon type as $a$, and label domain walls by the element $g\in G$ that relates the change in the symmetry breaking configuration across the domain wall. Then, the fusion symbols, $\omega^a(g,h,k)\in U(1)$ denote overall phase difference resulting from fusing domain walls $g,h,k$ within an $a$-flux line in different orders [$(gh)k$ vs. $g(hk)$].
For the $1d$ edges of $2d$ $G$-SPTs equivalence classes of $\omega$ are characterized by $H^3(G,U(1))$~\cite{kawagoe2021anomalies}.
Here, an important difference for the SET rather than the SPT, is that $\omega$ depends on the type of flux-line, $a$ and is required to be consistent with the fusion rules. Following the arguments for $2d$ above, this implies that equivalence classes of $\omega$ for $G$-SETs are characterized by $H^3(G,\widehat{\mathcal{A}})$ where $\mathcal{A}$ is the dual of the fusion group for loop excitations.

As an example, in a 3d $G$-enriched $\N$-gauge theory, fusion rules of gauged charges form the group $\widehat{\N}$ while fusion rules of flux loops form the group $\N$. Apply the general classification we see the symmetry fractionalization pattern is given by the group $H^2(G,\N)\times H^3(G,\widehat{\N})$.


\subsection{Symmetry fractionalization of the $\N$-gauged igSPT}
Let us first focus on $2d$. To determine the symmetry fractionalization pattern one looks for restriction of the symmetry action $U_g$ to a subregion $M$ that only contains a single charge or flux.  The $\Gir$ action is given by \eqref{eq:ugos}
 \begin{align}
	U_g|\{\hat{n}_i\},\{g_a\}\rangle=\prod_i \hat{n}_i\otimes_\Z e_2(g_{i_{3}}^{-1}g^{-1},g)|\{\hat{n}_i\},\{g_{a}\}\rangle
 \end{align}
Now consider creating a pair of charge $\hat{n}$ and anti-charge $-\hat{n}$ using the gauge field string \eqref{eq:charge string}
Now since only two sites $i_1,i_L$ host non-zero $\hat{n}$, we can restrict the symmetry action to $i_1$ by setting
  \begin{align}
	U_g^{e,i_1}|\{\hat{n}_i\},\{g_a\}\rangle=\hat{n}\otimes_\Z e_2(g_{(i_1)_3}^{-1}g^{-1},g)|\{\hat{n}_i\},\{g_a\}\rangle,
  \end{align}
which satisfies $U^e_gU^e_h=\hat{n}\otimes_\Z e_2(g,h)U^e_{gh}$ showing that chargs carry fractionalized transformation given by $[e_2]\in H^2(\Gir, \N)$.

In order to see the symmetry fractionalization of a gauge flux, it's useful to do a change of basis that transforms the IR space into simple $\hat{n}_i$ paramagnet as we did for the 1d Ising igSPT in \ref{sec:transintoir}. Before gauging, this is accomplished by the unitary:
\begin{align}
	V=\prod_i b_D^{s_i}(\{g_{i_\alpha}\})\otimes_\Z n_i
\end{align} 
where $b_D(\{g_{i_\alpha}\})$ stands for $b_D(g_{i_1i_2},g_{i_2i_3},\dots, g_{i_{D-1}i_D})$.  
\begin{widetext}
$V$ transforms the Hamiltonian to
\begin{align}
	V^\dagger H_{\Delta}V
	&=-\Delta \prod_i b_D^{-s}(\{g_{i_\alpha}\})\otimes_\Z n_i\left( \sum_j \delta_{\hat{n}_j,b_D^s(\{g_{j_\alpha}\})}\right) b_D^s(\{g_{i_\alpha}\})\otimes_\Z n_i=-\Delta \sum_j \delta_{\hat{n}_j+b_D^s,b_D^s}
		\nonumber\\&
	=-\Delta \sum_j \delta_{\hat{n}_j,0}
\end{align}
Such change of basis allows us to identify the symmetry fractionalization pattern for fluxes, to illustrate this, let us focus on the case of two special dimension. 
\end{widetext}

After gauging $\N$ we need to make sure the unitary $V$ used for changing basis is gauge invariant. To achieve this we first rewrite $V$ in a form that is convenient for gauging. Recall the cocycle condition for $b_2$,
\begin{align}
	1=db_2(g,h,k)=\frac{b_2(h,k)b_2(g,hk)}{b_2(gh,k)b_2(g,h)}
\end{align}
replace $g\rightarrow g_{i_1}^{-1}g_{i_2}, h\rightarrow g_{i_2}^{-1}g_{i_3},k\rightarrow g_{i_3}^{-1}$, we have:
\begin{align}
	b_2(g_{i_1}^{-1}g_{i_2},g_{i_2}^{-1}g_{i_3})=\prod_{(ab)\in \partial \Delta_i}b_2^{s^i_{ab}}(g_a^{-1}g_b,g_b^{-1}),
\end{align}
here $\Delta_i$ is the simplex $i$ and $\partial\Delta_i$ are the three edges of it. $s^i_{ab}=\pm 1$ depends on the direction of the edge $ab$ being compatible with the orientation of the triangle $i$ or not. With the help of this identity we can write
\begin{align}
	V&=\prod_i b_2^{s_i}(\{g_{i_\alpha}\})\otimes_\Z n_i
	=\prod_i\prod_{(ab)\in\partial \Delta_i}b_2^{s_i\cdot s^i_{ab}}(g_a^{-1}g_b,g_b^{-1})\otimes_\Z n_i\nonumber\\
	&=\prod_{(ab),(ij)\bot (ab)}b_2(g_a^{-1}g_b,g_b^{-1})\otimes_\Z(n_i-n_j)
\end{align} 
the last line is obtained by noticing that in the product in the second last expression, each edge $(ab)$ of the simplicial lattice receives contributions from its two adjacent triangles $i,j$ with $(ij)\bot (ab)$, and the exponets $s_is^i_{ab}$ in these two terms are always opposite of each other. Therefore  the result is a product over directed links $(ab)$, and for each fixed $(ab)$ the corresponding $(ij)$ link on the dual lattice is perpendicular to $(ab)$ with $i$ to the left and $j$ to right of $(ab)$.

With this new form we can make $V$ gauge invariant by replacing $n_i-n_j$ by the gauge invariant quantity $n_i-n_j+A_{ji}$,
\begin{align}
	\widetilde{V}&\equiv\prod_{(ab),(ij)\bot (ab)}b_2(g_a^{-1}a_b,g_b^{-1})\otimes_\Z(n_i-n_j+A_{ji})\\
	&=V \prod_{(ab),(ij)\bot (ab)}b_2(g_a^{-1}a_b,g_b^{-1})\otimes_\Z A_{ji}
\end{align}
\begin{figure}[t]
\begin{centering}
\includegraphics[width=0.8\columnwidth]{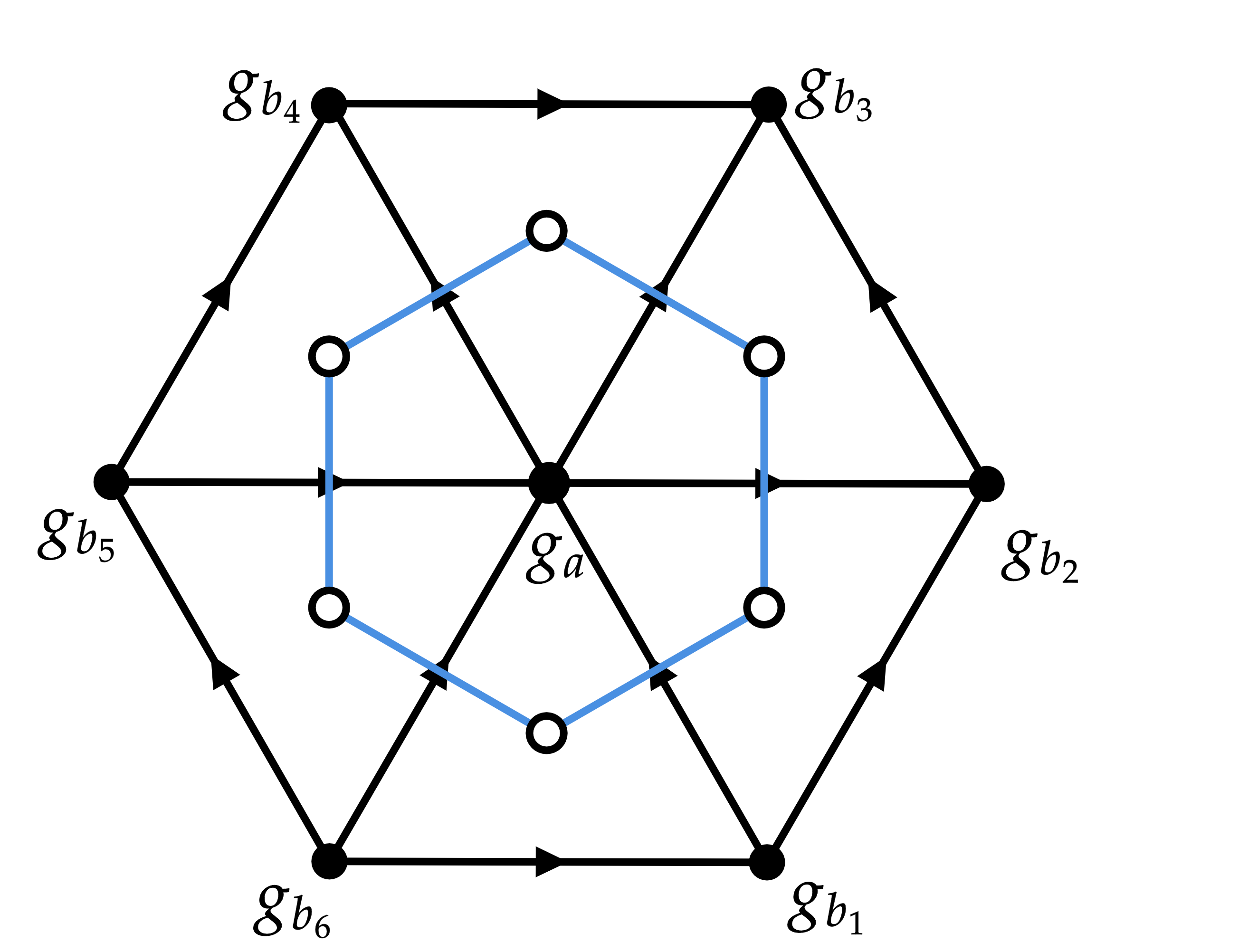}
\end{centering}
\caption{The terms in $\prod_{(ab),(ij)\bot (ab)}\frac{b_2(g_a^{-1}g^{-1},g)}{b_2(g_b^{-1}g^{-1},g)}\otimes_\Z a_{ji}$ that involve the vertex $a$ are links $(ab_k)$ shown in the figure, and the dual links $(ij)$ that are perpendicular to $(ab_k)$ form  blue hexagon that encircles the site $a$. Notice when the arrow on a link $ab_k$ is pointing away from $a$, the term is $b_2(g_a^{-1}g^{-1},g)[a_{ji}]$, while when the arrow is pointing towards $a$, the term is $b_2^{-1}(g_a^{-1}g^{-1},g)\otimes_\Z a_{ji}=b_2(g_a^{-1}g^{-1},g)\otimes_\Z a_{ij}$, either way the gauge field $a_{ij}$ that appears is pointing in the counter-clock direction, therefore the product of all 6 terms that involve the vertex $a$ is simply $b_2(g_a^{-1}g^{-1},g)\otimes_\Z(\text{flux of }h(a))$, where we $h(a)$ stands for the hexagon on the dual lattice that encircles site $a$.}\label{fig:flux}
\end{figure} 
\begin{widetext}
 The Hamiltonian is transformed by $\widetilde{V}$ to the trivial paramagnet one as before. Next we calculate the transformed $\Gir$-symmetry action. We do it by examining the action on the basis $|\{\hat{n}_i\},\{g_a\},a_{ij}\>$,
	\begin{align}
		&\widetilde{V}^\dagger U_g \widetilde{V}|\{\hat{n}_i\},\{g_a\},\{a_{ji}\}\>=\widetilde{V}^\dagger U_g\prod_{(ab),ij\bot ab}b_2(g_a^{-1}g_b,g_b^{-1})\otimes_\Z a_{ij}|\{\hat{n}_i+b_2(\{g_{i_\alpha}\})\},\{g_a\},\{a_{ij}\}\>\\
		&=\widetilde{V}^\dagger \prod_i \left(\hat{n}_i+b_2(\{g_{i_\alpha}\})\right)\otimes_\Z e_2(g_{i_{3}}^{-1}g^{-1},g)\prod_{(ab),ij\bot ab}b_2(g_a^{-1}g_b,g_b^{-1})\otimes_\Z a_{ji}|\{\hat{n}_i+b_2(\{g_{i_\alpha}\})\},\{gg_a\},\{a_{ij}\}\>\\
		&=\left(\prod_i \left(\hat{n}_i+b_2(\{g_{i_\alpha}\})\right)\otimes_\Z e_2(g_{i_{3}}^{-1}g^{-1},g)\right)\left(\prod_{(ab),ij\bot ab}\frac{b_2(g_a^{-1}g_b,g_b^{-1})}{b_2(g_a^{-1}g_b,g_b^{-1}g^{-1})}\otimes_\Z a_{ji}\right)|\{\hat{n}_i\},\{gg_a\},\{a_{ij}\}\>\\
		&\ir \left(\prod_{(ab),ij\bot ab}\frac{b_2(g_a^{-1}g_b,g_b^{-1})}{b_2(g_a^{-1}g_b,g_b^{-1}g^{-1})}\otimes_\Z a_{ji}\right)\left(\prod_i b_2(\{g_{i_\alpha}\})\otimes_\Z e_2(g_{i_{3}}^{-1}g^{-1},g)\right)|\{gg_a\},\{a_{ij}\}\>\\
		&=\prod_{(ab),ij\bot ab}\frac{b_2(g_a^{-1}g_b,g_b^{-1})}{b_2(g_a^{-1}g_b,g_b^{-1}g^{-1})}\otimes_\Z a_{ji} U^A_g|\{g_a\},\{a_{ij}\}\> \label{VtransU}
	  \end{align}\\
	  where $U_g^A$ is the anomalous symmetry action defined in \eqref{eq:UAgeneral}. To get to the second last line notice in the $\widetilde{V}$ transformed frame the IR space corresponds to $\hat{n}_i=0$. Now to further simplify the expression, we apply the cocycle condition $db_2(g,h,k)=1$ with $g=g_a^{-1}g_b,h=g_b^{-1}g^{-1},k=g$ and obtain 
	  \begin{align}
		1=\frac{b_2(g_b^{-1}g^{-1},g)b_2(g_a^{-1}g_b,g_b^{-1})}{b_2(g_a^{-1}g^{-1},g)b_2(g_a^{-1}g_b,g_b^{-1}g^{-1})}
	  \end{align}
	apply it to \eqref{VtransU} we obtain
	  \begin{align}
		\widetilde{V}^{\dagger} U_g \widetilde{V}\ir \prod_{(ab),(ij)\bot (ab)}\frac{b_2(g_a^{-1}g^{-1},g)}{b_2(g_b^{-1}g^{-1},g)}\otimes_\Z a_{ji} U^A_g
	  \end{align}
	  Now we can change the product from being over links to being over sites, for each fixed site $a$ there is one term $b_2(g_a^{-1}g^{-1},g)[a_{ji}]$ from each edge $ij$ of the dual lattce the encircles the the site $a$, as shown in  Fig. \ref{fig:flux}.
	  Therefore we have 
	  \begin{align}
		\widetilde{U}_g=\widetilde{V}^\dagger U_g \widetilde{V}\ir \prod_a b_2(g_a^{-1}g^{-1},g)\otimes_\Z\left[\sum_{(ij)\in h(a)}a_{ij}\right] U^A_g=\prod_a b_2(g_a^{-1}g^{-1},g)\otimes_\Z\left(\text{flux of } h(a)\right) U^A_g
	  \end{align}
	  where $h(a)$ stands for the hexagon in the dual lattice that surrounds the site $a$. 
\end{widetext} 
	  Now it is clear how to localize the symmetry $\widetilde{U}_g$ to a single flux. If we start with  a zero flux and $\Gir$-symmetric vacuum and create a pair of flux $+n$ and $-n$ at site $a$ and $b$  using a string of electric field operators, then we can localize the symmetry action $\widetilde{U}_g$ to a region $M$ that contains only the $+n$ flux by setting $\widetilde{U}_g^m=b_2(g_a^{-1}g^{-1},g)\otimes_\Z n U^{A,M}_g$, here $U^{A,M}_g$ stands for the restriction of the anomalous action to the region $M$. Now it's clear the restricted symmetry action $\widetilde{U}^m_g$ forms a projective representation $\widetilde{U}^m_g\widetilde{U}^m_h=b_2(g,h)\otimes_\Z n\widetilde{U}^m_{gh}$, confirming that the symmetry properties of fluxes are determined by the cocycle $b_2$.

An analogous derivation to that above shows that in $3d$ the local symmetry action of $\Gir$ on $\N$-flux lines is governed by the cocycle $b_3\in H^3(G,\N)$ appearing in the emergent-anomaly cocycle decomposition $\omega_5 = b_3\cup e_2$. Physically, we can interpret $b_3$ as the $\N$-valued fusion-symbol for fusing the $0d$ endpoints of $1d$ $\Gir$ domain wall defects within an $\N$-flux line~\cite{kawagoe2021anomalies}. 

\section{Proof of Intrinsicness of igSPT with group cohomology anomalies \label{app:intrinsic}}
In this appendix we show that the gapless SPTs in our construction are in fact intrinsic, i.e. that their edge states can not be realized as edge states of gapped SPTs with the same symmetry, $\Guv$.
This can be straightforwardly verified for the examples in the main text: for $\Guv=\Z_4$ in $1d$ and $3d$, and $\Guv=\Z_4^T$ in $2d$ where there are no non-trivial gapped SPTs.
In other cases, such as $\Guv=\Z_2\times \Z_4$, $\Gir=\Z_2^2$ in $2d$, there is not only a single $2d$ igSPT but also various $2d$ gapped SPTs with anomalies in $\Z_2$, $\Z_4$ and one with mixed $\Z_2,\Z_4$ anomaly. However one can check explicitly that none of these gapped SPTs have the same SPT pumping property of the igSPT, which has the property that acting twice with the $\Z_4$ generator in the igSPT pumps a $1d$ $\Gir=\Z_2^2$ SPT onto the boundary. In a gapped system, consistency would then require that each $\Gir$-DW carry ``half" of the minimal allowed $1d$ $\Gir=\Z_2^2$ SPT. However, this is not possible in a gapped system.

In this appendix, we give a general argument that such obstructions to realizing the SPT pumping behavior in a gapped system are intrinsic, i.e that an igSPT with emergent group cohomology anomaly is intrinsically gapless.
We proceed by assuming the converse and deriving a contradiction.  Let $\Sigma$ be the $Dd$ bulk of a igSPT, $\partial \Sigma$ be the $(D-1)d$ edge. 
The igSPT edge is characterized by two key features: i) an SPT pumping symmetry: $U_n,~n\in \N$ pumps a lower dimensional $\Gir$ SPT, and ii) an $\N$-gap: there are no gapless $\N$ DOF in the bulk or at the edge.
Suppose now that the same two characteristics were possible with a gapped $\Guv$-symmetric bulk.
Then there would be an action of $\Guv$ on $\partial\Sigma$, denoted as $V_{\gamma},\gamma\in\Guv$, that contains the pumping actions: $V_n=U_{n},~\forall n\in\N$ of the form:
\begin{align}
    V_{\gamma}|\{g_i\}\rangle=\Omega(\gamma,\{g_i\})|\{p(\Guv)g_i\}\rangle
\end{align}
where $p$ is the projection $\Guv\xrightarrow{p}G$ and $\Omega(\gamma,\{g_i\})$ is a general phase factor that depends on the spin configuration as well as the group element $\gamma$. 
Crucially, if the bulk is gapped, the global action of $V_{\gamma\in\Guv}$ on the entire boundary must form an ordinary linear representation of $\Guv$. We will now show that this fail to be the case for systems with SPT pumping symmetries.
For $n\in\N$, $V_n$ is a $\{g_i\}$-dependent phase $\Omega(n,\{g_i\})$ since $p(n)=1_G$.
For nontrivial extensions $\Guv\neq G\times\N$ and there exist $\gamma_1,\gamma_2\notin \N$, but $n\equiv\gamma_1\gamma_2\in \N$. Since $\N$ is central, $n=\gamma_1^{-1}n\gamma_1=\gamma_2\gamma_1$. 
Hence for an ordinary linear representation we should have: $ V_{\gamma_1}V_{\gamma_2} = V_{\gamma_1\gamma_2} =V_{\gamma_2\gamma_1}=V_{\gamma_2}V_{\gamma_1}$.
Instead, considering the action of these operations on $\d\Sigma$, one finds:
\begin{align}
    V_{\gamma_1}V_{\gamma_2}|\{g_i\}\>&=\Omega(\gamma_1,\{p(\gamma_2)g_i\})\Omega(\gamma_2,\{g_i\})|\{g_i\}\rangle\label{gamma12},
    \\
    V_{\gamma_2}V_{\gamma_1}|\{g_i\}\>&=\Omega(\gamma_2,\{p(\gamma_1)g_i\})\Omega(\gamma_1,\{g_i\})|\{g_i\}\>\label{gamma21},
    \\
    V_{n=\gamma_1\cdot\gamma_2}|\{g_i\}\rangle&=U_{n}(\{g_i\})|\{g_i\}\rangle\label{W}
    .
\end{align}
Denote by a $\Gir$-rotor configuration with a domain wall  between $g$ and $h$ configurations by $\text{DW}(g,h)$ (we will not need to specify the detailed form of the DW configuration just that we have a region $A$ such that $g_{i\in A}=g$ and $g_{i\in A^c}=h$ where $A^c$ denotes the complement of $A$). Let $\{g_i\}=\text{DW}(1_G,p(\gamma_1))$, which is a nontrival domain wall since $\gamma_1\notin\N$. Notice $p(\gamma_2)\cdot\text{DW}(1_G,p(\gamma_1))=\text{DW}(p(\gamma_2),1_G)$, thus from \eqref{gamma12}=\eqref{W} we have 
\begin{align}
    \Omega(\gamma_1,\text{DW}(p(\gamma_2),1_G))&\Omega(\gamma_2,\text{DW}(1_G,p(\gamma_1)))\nonumber\\
	&=U_{n}(\text{DW}(1_G,p(\gamma_1)))\label{Omega12}
\end{align}
Next set $\{g_i\}=\text{DW}(p(\gamma_2),1_G)$, then $p(\gamma_1)\cdot \text{DW}(p(\gamma_2),1_G)=\text{DW}(1_G,p(\gamma_1))$, therefore from \eqref{gamma21}=\eqref{W}, we have
\begin{align}
    \Omega(\gamma_2,\text{DW}(1_G,p(\gamma_1)))&\Omega(\gamma_1,\text{DW}(p(\gamma_2),1_G))\nonumber\\
	&=U_{n}(\text{DW}(p(\gamma_2),1_G))\label{Omega21}
\end{align}
Notice \eqref{Omega12} and \eqref{Omega21} have the same LHS,while the RHS are in general different, because the phase factor $U_{n}$ takes the form: 
\begin{align}
	U_n(\{g_i\})= \prod_{(i_1,\cdots,i_D)}b_D^s(g^{-1}_{i_1} g_{i_2},\cdots, g^{-1}_{i_{D-1}}g_{i_{D}},g^{-1}_{i_{D}})\otimes_{\mathbb{Z}}n,
	\label{Unapp}
\end{align}
which is not a function of domain wall classes. That is to say, although $\text{DW}(p(\gamma_2),1_G)$ and $\text{DW}(1_G,p(\gamma_1))$ are equivalent domain walls in the sense that the $\text{DW}(1_G,p(\gamma_1))=p(\gamma_1)\cdot \text{DW}(p(\gamma_2),1_G)$, the last argument of $b^s_D$ in $U_n$ is $g_{i_{D}}^{-1}$, which is not a function of domain wall classes. Because of this, $U_n$ in \eqref{Unapp}, gives different phases depending on the order of application of $\gamma_1$ and $\gamma_2$. This raises a contradiction with the assumption that the bulk is gapped since the edge-restriction of symmetry on the edge of a gapped bulk would necessarily have commuting action of $\gamma_{1,2}$. As a simple example, consider $2d$ igSPT with $\Guv=\Z_{2}\times \Z_4$ and $\Gir=\Z_2^2$. The sole non-trivial SPT pumping operation is $U=(-1)^{(g_{i+1}^1-g_{i}^1)g_{i+1}^2}$, and the $g^2_{i+1}$ factor is what makes $U$ not a function of domain wall classes. Apply the construction to this example by setting $\gamma_1=\gamma_2=(1,1)\in \Z_2\times \Z_4$, which satisfy $\gamma_1+\gamma_2=(0,2)\in\mathcal{N}$. Then the phase factor $W$ receives a single contribution in the configuration $\text{DW}((0,0),(1,1))$: $(-1)^{1\cdot 1}=-1$, while it is one in the configuration $\text{DW}((1,1),(0,0))$.

\section{Relating anomaly cancellation and SPT pumping pictures of igSPT edge states \label{app:anomalycancel}}
While we have focused on the SPT pumping as the origin of igSPT edge states, TVV~\cite{thorngren2021intrinsically} instead deduce the existence of edge states from an anomaly cancellation argument. Here, we recount this argument and relate these two mechanisms.

First recall the anomaly cancellation argument. Consider gauging the $\Guv$ symmetry by coupling the igSPT on a $(D+1)d$ spacetime manifold $X_{D+1}$ to a background $\Guv$-gauge field $A^{\Guv} = (A^{\N},A^{\Gir})$, where $A^{\N}$ is an $\N$ gauge field that acts only on the gapped sector and $A^\Gir$ a $\Gir$-gauge field that acts on the IR DOF. Then, consider integrating out the gapped $\N$-DOF, which results in the partition function $\mathcal{Z}[A^{\Guv}] = \mathcal{Z}_\text{IR}[A^{\Gir}] e^{i\int_{X_{D+2}} \omega(A^{\Guv})}$ where $\omega_{D+2}$ is the anomaly cocycle, and $X_{D+2}$ is a Wess-Zumino-Witten (WZW) type extension of spacetime whose boundary is the physical spacetime: $X_{D+1} = \d X_{D+2}$, and $\mathcal{Z}_\text{IR}$ is the partition function for the gapless IR DOF, which couple only to the IR part of the gauge field.
When considered as a function of $g\in \Gir$, $\omega_{D+2}$ is a non-trivial cocycle, and there is no way to define the above action on a spatial manifold with a boundary, $\d X_{D+1} \neq \emptyset$. Rather, in this case, the anomalous $\Gir$-action can only arise as the boundary of a gapped $\Gir$-SPT on $X_{D+2}$. However, the group extension lifts the anomaly such that interpreted as a $\Guv$-cocycles one writes $\omega_{D+2}=\d\alpha_{D+1}$~\cite{tachikawa2020gauging}, and with the extended DOF it is possible to define the action on an open spacetime as: $\mathcal{Z} = \mathcal{Z}_\text{IR}\[A^G\]e^{i\int_{X_{D+1}} \alpha(A^{\Guv})}$. 
As argued in~\cite{thorngren2021intrinsically}, $\alpha$ cannot be invariant under $\N$ gauge transformations: $A^{\N}\rightarrow A^{\N} + d\chi^{\N}$, otherwise one would have $\omega= d\alpha$ as $\Gir$-cocycles, and there would be no emergent $\Gir$-anomaly. Rather, $\N$ gauge transforms change $\alpha$ by an exact but nonzero form: $\alpha\rightarrow \alpha + d\lambda(A^{\Guv},\chi^{\N})$, which results in a transformation of the partition function: $\mathcal{Z}\rightarrow e^{i\int_{\d X_D} \lambda}\mathcal{Z}$.  From this, one deduces that additional gapless edge states are required to cure this lack of gauge invariance. 

We now show that the SPT pumping picture reproduces the anomaly cancellation picture as well as clarifying the physical properties of the gapless edge states that produce the cancellation. Consider instead the the decomposition $\omega_{D+2}=b_D\cup e_2$ given by the anomaly-lifting extension. 
Due to the non-trivial group extension structure, the merger of two $A^{\Gir}$ fluxes with flux values $g_i1,g_2$ with $0\neq e_2(g_1,g_2)\in \N$, acts as a source for $A^{\N}$: $\d A^{\N} = e_2(A^{\Gir})$~\cite{tachikawa2020gauging}.
This gives a solution to the anomaly lifting equation $\omega_{D+2}=\partial\alpha_{D+1}$ with $\alpha_{D+1}=b_D\cup A^{\N}$~\footnote{Note that from $\omega \in H^*(\Gir,M) = H^*(B\Gir,M)$ and gauge field $A^\Gir:X\rightarrow B\Gir$, we can define an action $\int_X \omega[A]$ by the pull back $\omega[A](x) \equiv \omega\(A(x)\)$ where we use the same symbol for both interpretations of $\omega$.},
from which we can deduce explicitly the lack of gauge invariance of the partition function $\mathcal{Z}$ on open manifolds and see the connection to our SPT pumping picture.
Specifically: consider an $\N$ gauge transformation $A^{\N} \rightarrow A^{\N}+d\chi^{\N}$  where $\chi^{\N}(x\in \d X) = n\in \N$ is constant on the boundary. This changes the partition function by: $\mathcal{Z} \rightarrow e^{i\int_{\d X_D} b_D(A^{\Gir})\otimes_\Z n}\mathcal{Z}$. The prefactor of this change is precisely the partition function for the $(D-1)d$ SPT pumped by the $n\in \N$ symmetry.
From this, we see that the SPT-pumping picture reproduces the anomaly cancellation argument, but additionally reveals a further substructure: $\alpha = b\cup A^{\N}$ for the solution to the the anomaly lifting equation $\d\alpha = \omega$, and gives a simple physical interpretation in terms of SPT pumping for the relevant class of igSPTs.

\section{Perspective from Gauge-Response Theory from Gapped SPTs}
The emergent $\Gir$-symmetry anomaly underlying the topological feature of a $Dd$ igSPT arises from an anomaly-free UV symmetry extension $\Guv$. A more conventional context for this anomaly is to arise at the $Dd$ surface of a $(D+1)d$ gapped $\Gir$-SPT with no extended DOF. A useful way to characterize gapped $\Gir$-SPTs is to consider their response to a background $\Gir$-gauge field, which will be described by a TQFT. In this section, we review some TQFTs for various $\Gir$-SPTs relevant to the igSPT examples we study, recall the construction of Ref.~\cite{wang2015field} which enables one to deduce explicit cocycles from these TQFTs. We then argue that the structure of these TQFTs and corresponding cocycles implies the decomposition form $\omega_{D+2}=b_D\cup e_2$ needed for our igSPT construction, and show how the lower-$d$ SPT edge pumping symmetry that protects the igSPT edge states arises from a dimensional reduction procedure from the TQFT describing the higher-d $\Gir$-SPT with the same surface anomaly.

In this section we consider only Abelian, unitary symmetries.

\subsection*{Topological terms in discrete gauge theories}
We follow the notation of Ref.~\cite{wang2015field}, where we define a $\Z_N$ gauge field $A$ such that its holonomies take values $\oint A \in \frac{2\pi}{N}$ and its gauge transformations satisfy $\oint \delta A = 0$ where $\oint$ is short-hand for a discrete lattice sum around a closed loop.
We abbreviate $N_{12} = \text{GCD}(N_1,N_2)$ where GCD denotes the greatest common divisor.

We define the response field theory as the partition function for the system in the presence of a background (nondynamical) gauge field $A$. This is well defined as long as the system has a unique gapped ground-state, which requires that we consider an SPT with periodic boundary conditions. This gapped requirement may fail in the presence of certain gauge flux configurations, which must then be excluded.

For example, in $1d$ the $\Gir=\Z_{N_1}\times \Z_{N_2}$ SPTs, which can be interpreted as $\Z_{N_1}$ domain walls (DWs) being decorated by $p\in N_{12}$ charges of the $Z_{N_2}$ symmetry. As a result, the configurations with $dA_1\neq 0$, i.e. with spacetime instantons where the $A_1$ flux changes, induce a change in the $\Z_{N_2}$ charge of the system causing a zero of the partition function, $\text{tr} e^{-\beta H[A]}$ since the trace forces the starting and ending states to have the same charge. Hence, we only expect to get a well-defined partition function if $A_{1,2}$ are flat, i.e. $dA_{1,2}=0$.
The corresponding gauge response theory is~\cite{wang2015field}:
\begin{align}
\mathcal{Z}[A^{I=1,2}] = e^{\frac{ipN_1N_2}{N_{12}}\int_{M_{2}} A^1A^2}
\end{align}
where $A^{I=1,2}$ is a $\Z_{N_I}$ gauge field, $M_2$ is a (closed) $1+1d$ spacetime, and we have suppressed $\wedge$ products between each term for simplicity. This term is gauge invariant under $A^I\rightarrow A^I + dg^I$ with $I=1,2$ only if we demand that the gauge fields are flat: $dA^{1,2}=0$. 
The restrictions on gauge configurations depend on dimensionality and topological term in a way detailed in Ref.~\cite{wang2015field}.

\subsection*{Deducing Cocycles from Response Theory}
In general, for finitely generated Abelian symmetry groups, $\Gir=\prod_{I=1}^Q \Z_{N_{I}}$ the topological terms will have Lagriangian proportional to products of $k$ factors of $A^I$ distinct $I$ and $\lfloor(D-k)/2\rfloor$ factors of $dA^J$. For example in $2d$, one could have either mixed Chern-Simons terms type terms $2d$, $\mathcal{L}_{2d}\sim A^IdA^J$, or discrete terms such as $A^1A^2A^3$. In $3d$ one can have either $\mathcal{L} = A^1A^2dA^3$ or $A^1A^2A^3A^4$. In $4d$ (which is relevant to $3d$ IGSPTs) there are various possibilities such as $\mathcal{L}\sim AdAdA, AAAdA, AAAAA$ with various combinations of flavor indices.

As shown in Ref.~\cite{wang2015field}, from these topological terms, one can readily read off an explicit cocycle, $\omega_{D+2}$ for the corresponding SPT, by taking $e^{i\mathcal{L}[A]}$ and replacing each factor of $A^I \rightarrow \frac{2\pi}{N_I} g^I$ and each factor of $dA^I$ by $2\pi v_{N_I}(g,h)$ where we remind that the the vorticity function is defined by $v_N(a,b) = 1$ if $a+b>N$ (with addition taken in $\Z$ i.e. not modulo $N$) and $0$ otherwise. 
For example for $\Gir=\Z_{N_1}\times \Z_{N_2}$ in $3d$,  
\begin{align}
\mathcal{Z}[A] &= e^{i\frac{N_1N_2}{(2\pi)^2 N_{12}}\int A^1A^2dA^2} 
\nonumber\\
&\downarrow
\nonumber\\
 \omega_4(g,h,k,l) &= e^{\frac{2\pi i}{N_{12}}g^1h^2 v_{N_2}(k^2,l^2)}.
\end{align}
We note that, the decomposition $\omega_{D+2}=b_D\cup e_2$ follows naturally from the exterior product structure of the topological terms in the response theory in this case.

\subsection*{igSPT edge-SPT pumping symmetry from dimensional reduction}
We also note that there seems to be a rather general prescription to deduce the $(D-1)d$ SPT pumping symmetry of a $Dd$ igSPT with emergent anomalous $\Gir$ symmetry, from the response theory for the $(D+1)d$ gapped SPT whose surface has the same anomaly.

We illustrate this first with a $D=1$ $\Gir=\Z_N$ example, with $\mathcal{Z}[A] = e^{\frac{i}{2\pi}\int_{M_2} AdA}$, with corresponding cocycle $\omega_3(g,h,k) = e^{\frac{2\pi i}{N}gv_N(h,k)}$. 
We say that the $0d$ SPT pumping symmetry arose from just the first factor $e^{2\pi ig/N}$, which we could obtain by composing symmetry transformations $(N-j)$ and $j$ which set the $v_N$ argument to $1$ in the cocycle term.

Acting with a symmetry $j$ is like considering $\text{tr}\[U_je^{-\beta H}\]$ which is equivalent to a gauge configuration with a flux $\oint_\tau A = \frac{2\pi j}{N}$ in the time cycle of the partition function. Composing symmetries that add up to $N$ is then like performing a large gauge transformation around the time-cycle. In the igSPT context, we would like to explore the effect of applying this transformation only to the anomalous $1d$ surface of the $2d$ SPT (in fact, we would like to go further and ask about the effect on the $0d$ edges of the $1d$ igSPT which do not exist when realized as the surface of a $2d$ SPT). 

Considering the $2d$ SPT response theory on a three-torus with coordinates $(\tau,x,y)$, we can emulate the effect of the SPT-pumping transformation restricted to the $y=0$ line at time $\tau=0$ by taking $A = A_{1d}(x)+A_F$ with $dA_F=-2\pi d\tau dy \delta(\tau)\delta(y)$, which inserts a $2\pi$ flux in the $\tau$-cycle of the torus only at $y=0$. Inserting this form into the response theory results in a $1d$ partition function $\mathcal{Z}_{1d}[A_{1d}] = e^{i\int A^x_{1d} dx}$, which we can identify as the response field theory for the $(0+1)d$ $\Gir$-SPT that gets pumped onto the igSPT boundary by the $\N$ symmetry.

This dimensional reduction procedure works for all of the examples listed in Table~\ref{tab:results}. Selected additional examples are summarized in Table.~\ref{tab:response} below.

\begin{table*}[]
\setlength{\tabcolsep}{6pt}
\renewcommand{\arraystretch}{2.5}
\begin{tabular}{C{1in}|C{1.2in}|C{1.8in}|C{1.4in}}
{igSPT} & {Response Theory $\mathcal{Z}_{D}[A]$} & {Reduction Ansazt $A=A_{D}+A_{F}$} & {Reduced Response Theory $\mathcal{Z}_{D-2}[A]$}\\
\hline
{1d $\Z_N$} & {$e^{\frac{i}{2\pi}\int_{M_2}AdA }$} & {$dA_F=2\pi \delta(\tau)\delta(y)d\tau dy$} & {$e^{i\int A^x_{1}dx}$}\\
\hline 
{2d $\Z_{N_1}\times\Z_{N_2}$} & {$e^{i\frac{N_1N_2}{(2\pi)^2N_{12}}\int A^1A^2dA^2}$} & {$dA_F = 2\pi \delta(\tau)\delta(x^3) d\tau dx^3$} & {$e^{\frac{iN_1N_2}{2\pi N_{12}}\int_{M_2}A^1A^2}$}\\
\hline
{3d $\Z_{N}$} & {$e^{\frac{i}{(2\pi)^2} \int_{M_5} AdAdA}$} & {$dA_F = 2\pi \delta(\tau)\delta(x^4)d\tau dx^4$} & {$e^{\frac{i}{2\pi} \int AdA}$}
\end{tabular}
\caption{{\bf edge-SPT pumping symmetry from
dimensional reduction} }

\label{tab:response}
\end{table*}
\end{document}